%% file: fbqc_4cat.tex
\documentclass[aps,reprint,superscriptaddress,prx,floatfix]{revtex4-2}
\usepackage{template}

\begin{document}
\title{Fault-tolerant Fusion-based Quantum Computing with the Four-legged Cat Code}

\author{Harshvardhan K. Babla}
\thanks{Corresponding author: harsh.babla@yale.edu}

\author{James D. Teoh}
\altaffiliation[Present address:~]{Quantum Circuits Inc.}

\author{Jahan Claes}
\altaffiliation[Present address:~]{Logiqal Inc.}

\author{Daniel K. Weiss}
\altaffiliation[Present address:~]{Quantum Circuits Inc.}

\author{Shraddha Singh}
\altaffiliation[Present address:~]{IBM Quantum}

\author{Robert J. Schoelkopf}
\author{Shruti Puri}
\affiliation{Departments of Applied Physics and Physics, Yale University, New Haven, CT, USA}
\affiliation{Yale Quantum Institute, Yale University, New Haven, CT, USA}

\date{\today}

\begin{abstract}
    The four-legged cat code is a quantum error-correcting code designed to address the predominant error in bosonic modes: single-photon loss. It was the first such code to surpass the break-even point, thereby demonstrating the practical utility of quantum error correction. In this work, we propose a planar fault-tolerant architecture for this code by concatenating it with the $XZZX$ code via fusion-based error-correction. To the best of our knowledge, this is the first 2D nearest-neighbor architecture for fault-tolerant fusion-based error-correction. We demonstrate how all the required operations, namely resource state preparation and Bell measurements, can be carried out using standard circuit-QED techniques, such as intercavity beam-splitter coupling, cavity displacements, cavity-transmon dispersive coupling, and transmon drives.
    We show analytically and numerically that all dominant hardware errors in the bosonic modes and control ancillae are corrected, to first-order, at the hardware level. Consequently, the outer $XZZX$ code only needs to address smaller residual errors, which are quadratically suppressed, effectively doubling the architecture's fault-distance. Moreover, the performance of our architecture is not limited by unwanted nonlinearities such as cavity self-Kerr, and it avoids demanding coupling techniques like $\chi$-matching or high-order coupling. Overall, our architecture substantially reduces the hardware complexity needed to achieve fault tolerance with the four-legged cat code.
\end{abstract}
\maketitle

\input{tex/1.intro.tex}
\input{tex/2.0.preliminaries.tex}
\input{tex/3.0.this_work.tex}
\input{tex/4.discussion.tex}

\section*{Acknowledgements}
We thank Steven M. Girvin, Liang Jiang, Luigi Frunzio, Kaavya Sahay, Takahiro Tsunoda, and Patrick Winkel for helpful discussions. We thank the Yale Center
for Research Computing for technical support and high-performance computing resources. 

This research was supported by the U.S. Army Research Office (ARO) under grant W911NF-23-1-0051. The views and conclusions contained in this document are those of the authors and should not be interpreted as representing the official policies, either expressed or implied, of the ARO or the US Government. The US Government is authorized to reproduce and distribute reprints for Government purposes notwithstanding any copyright notation herein.

R.J.S. is a founder and shareholder of Quantum Circuits Inc. (QCI).

\bibliographystyle{unsrt}
\bibliography{citations}

\clearpage
\input{appendix.tex}
\bibliographystyleapp{unsrt}
\bibliographyapp{citations}

\end{document}

%% file: tex/1.intro.tex
\section{Introduction}

To enable practical quantum computation, it is essential to protect quantum information from decoherence. Quantum error correction (QEC) realizes this by redundantly encoding a logical qubit(s) in a higher-dimensional Hilbert space \cite{nielsen_chuang_2010, knill_laflamme_1997}. The four-legged cat (4C) encodes a single logical qubit in the Hilbert space of a harmonic oscillator (also known as a bosonic mode) \cite{leghtas_2013, mirrahimi_2014}. This code is particularly well suited to correct single-photon loss, the predominant error in bosonic systems \cite{leghtas_2013,mirrahimi_2014}. It stands as a landmark in the field as the basis for the first experiment to demonstrate ``break-even'' QEC, where the logical qubit’s lifetime exceeds that of the best unprotected component \cite{ofek_2016}. Since then, advances in quantum hardware and control techniques have driven further progress, with other oscillator codes \cite{sivak_2023, ni_2023, brock_2025} and a qubit-based code \cite{google_2024, paetznick_2024} achieving similar breakthroughs.

\begin{figure*}[ht!]
	\includegraphics[width=\linewidth]{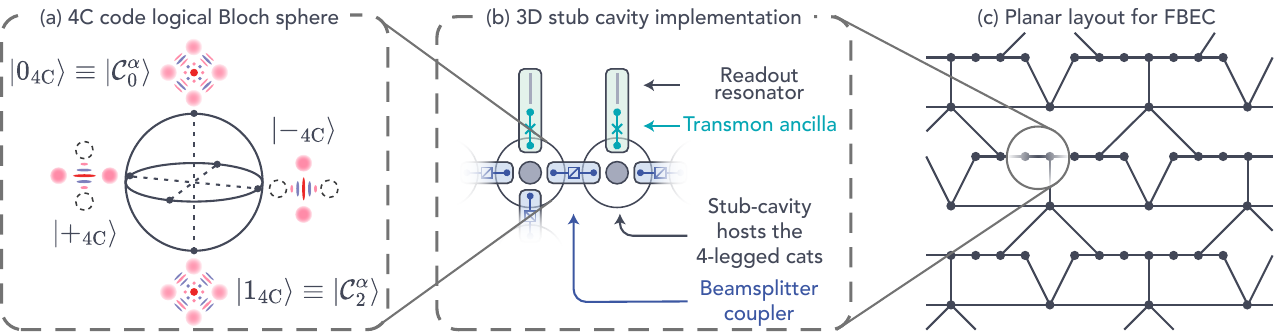}
    	\caption{(a) The logical Bloch sphere of the 4C code, supported on even photon-number parity states (see Eq.~\ref{eqn:4cats}). The logical $Z$-($X$-) states are illustrated by the Wigner-function cartoons at the poles (equators) of the Bloch sphere. (b) The cat states can be encoded within electromagnetic modes of 3D microwave stub cavities. Neighboring stub cavities are coupled via beamsplitter couplers (blue). Each stub cavity is dispersively coupled to a transmon ancilla (green) with a dedicated readout resonator (gray). (c) The stub-cavities ($\sbullet$\,) are tiled on a planar layout with nearest-neighbor connectivity (\textemdash), to implement the fusion-based $XZZX$ code. Here we show the most qubit-efficient layout, with degree-3, degree-4, and degree-5 connections. Other layouts which tradeoff qubit count for simpler connectivity are given in App. \ref{app:planar_fbec}.}
	\label{fig:overview}
\end{figure*}

Although achieving break-even QEC for memory is a pivotal milestone, it is merely the first step toward realizing practical quantum computation. To enable indefinitely long quantum computations, QEC codes must support arbitrarily low logical error probabilities across a universal set of operations \cite{nielsen_chuang_2010, shor_1996, divincenzo_1996, kitaev_1997a, kitaev_1997b, gottesman_1998, gidney_2021}. While qubit-based codes, such as the surface code \cite{bravyi_1998, fowler_2012a, fowler_2012b, tomita_2014}, guarantee this error suppression, bosonic codes do not. However, bosonic codes offer a key advantage: they suppress errors directly at the hardware level. By concatenating a bosonic code with a qubit-based code, it is possible to harness the strengths of both approaches, resulting in a hardware-efficient fault-tolerant (FT) architecture \cite{guillaud_2019, puri_2020, darmawan_2021, guillaud_2021, chamberland_2022, regent_2023, hann_2024, ruiz_2024, messinger_2024}. Crucially, the operations used for concatenation must be carefully designed to preserve the intrinsic error suppressing properties of the bosonic code. This ensures that the outer qubit-based code only addresses smaller residual errors, potentially reducing the overall hardware overhead required for fault tolerance.

In this work, we concatenate the 4C code with a fusion-based \cite{bartolucci_2023} version of the $XZZX$ code \cite{claes_2023, sahay_2023a} to achieve full fault-tolerance. We design a novel planar layout to implement fusion-based error correction (FBEC) \cite{bartolucci_2023, sahay_2023a,bombin_2023}. Our approach for concatenation relies on two key operations: (1) error-detected preparation of 6-body resource states and (2) destructive FT Bell measurements, also known as fusions. These operations are mediated by dedicated three-level ancillae coupled to each bosonic mode and beamsplitter couplers across the bosonic modes. We tailor these operations to retain the cat code's resilience to single-photon loss while ensuring first-order FT against other dominant hardware errors. These errors include decay, dephasing, and readout errors in the ancillae, as well as unwanted nonlinearities in the bosonic mode, such as self-Kerr \cite{zhu_2013, kirchmair_2013, blais_2021}. Addressing these ancilla errors and unwanted nonlinearities has been a prevailing challenge for bosonic qubits \cite{ofek_2016, sivak_2023, xu_2023b, ni_2023}. We highlight that this is a core advantage of our scheme. Overall, the outer code only needs to address rarer errors such as double-photon loss, cavity dephasing, the occurrence of more than one dominant error in the ancilla, and ancilla heating. With each operation providing quadratic suppression of the dominant physical error rates, the concatenation is expected to effectively double the fault distance of the $XZZX$ code to all major hardware errors.

Furthermore, our architecture also overcomes two core challenges faced by linear-optical FBEC, the platform where FBEC was first proposed \cite{bartolucci_2023, sahay_2023a, grice_2011}. First, unlike linear optics, our protocol recycles measured qubits, reducing the qubit count to be independent of algorithm length. To the best of our knowledge, this is the first 2D nearest-neighbor architecture for fault-tolerant measurement-based quantum computing. Second, in the absence of errors, all operations in our protocol proceed deterministically. Failures are only caused by single hardware errors during resource state preparation, which occur with probability $\sim 1\%$ for realistic hardware coherences. Fusion measurements are even more robust, as they remain unaffected by single hardware errors. This contrasts sharply with linear optics, where all operations are probabilistic. In particular, the intrinsic success rate of resource state preparation is only $\sim 1.5\%$ \footnote{If no additional qubits are used, the resource state preparation succeeds with probability $~ 1/(2^6) \approx 1.5 \%$.}, and $XX$ measurement outcomes are recovered by bare fusions only $50\%$ of the time \cite{sahay_2023a}. A large qubit overhead is needed to improve upon these success rates \cite{grice_2011}.

We propose to encode the 4C within an electromagnetic mode of a superconducting microwave resonator, such as a 3D stub cavity. These cavities demonstrate lifetimes comparable to, and often exceeding, those of Josephson-junction-based qubits \cite{reagor_2013, reagor_2016, ofek_2016, rosenblum_2018, reinhold_2020, chakram_2021, milul_2023, ni_2023, sivak_2023, goldblatt_2024}. Notably, recent experiments have achieved single-photon lifetimes beyond 30 ms \cite{romanenko_2020, milul_2023}. Furthermore, these cavities feature intrinsically low dephasing and heating rates, making photon loss the primary physical error mechanism \cite{milul_2023, goldblatt_2024}. Neighboring cavities are coupled via beamsplitter interactions \cite{sirois_2015, gao_2018, gao_2019, zhang_2019, chapman_2023, yao_lu_2023, zhou_2023}, with each cavity controlled by a dispersively coupled transmon qutrit \cite{rosenblum_2018, reinhold_2020, tsunoda_2023}.

Previous proposals for scaling bosonic codes have relied on multi-photon dissipation and drives \cite{mirrahimi_2014, mundhada_2019, grimsmo_2020, xu_2023a, steve_isa_2024, vanselow_2025}, or intricate coupling schemes like $\chi$-matching \cite{rosenblum_2018, reinhold_2020, xu_2023b}; both of which have proven experimentally challenging. These approaches were required to stabilize the 4C code against no‑jump evolution and to implement entangling operations. In contrast, our fusion‑based scheme inherently suppresses no‑jump evolution by repeatedly teleporting the logical 4C state. Furthermore, our entangling operations rely solely on well‑established techniques in circuit quantum electrodynamics (cQED) \cite{blais_2004, blais_2007, schuster_2007, blais_2021}.

We numerically evaluate the infidelity in the fusion measurements and resource state preparation circuits due to hardware errors. We show that the infidelities of preparing the resource states and of the $ZZ$ fusion scale quadratically with hardware coherence times, confirming first-order fault tolerance against all dominant hardware errors in both the bosonic mode and the ancillae. With the coherence levels of the currently feasible cQED hardware, we anticipate resource state preparation infidelity below 0.01\% with a failure probability of 1\%. During fusion measurements, we estimate a 1\% (0.1\%) likelihood of an incorrect $ZZ$ ($XX$) measurement outcome. This bias in the measurement infidelity motivates our choice of the $XZZX$ code, which is known to have higher thresholds under biased noise \cite{claes_2023}. Overall, we expect all errors to be below the threshold of the $XZZX$ code \cite{claes_2023}, suggesting that the 4C-FBEC architecture may be a favorable candidate for subthreshold FT QEC. However, more rigorous simulations are needed to confirm this.

This article is organized as follows. We start by reviewing the preliminary concepts needed for our protocol (Sec. \ref{sec:preliminaries}). This section gives an overview of the 4C code error correction and the relevant operations in the cQED platform. Our contribution, the FT protocol for concatenating the 4C code with the $XZZX$ code, is presented in Sec. \ref{sec:this_work}. Here, we introduce a novel planar architecture for the FBEC implementation and FT protocols for all components required to perform universal quantum computing with 4C-FBEC: (1) resource state preparation, (2) fusions, and (3) non-Clifford operations. We evaluate the errors in (1) and (2) numerically in Sec. \ref{sec:numerical_sims} and confirm the first-order fault-tolerance of most FBEC operations to dominant hardware errors. Finally, in Sec. \ref{sec:discussion}, we discuss further opportunities for optimization and compare to previous architectures for scaling up the 4C code. 

%% file: tex/2.0.preliminaries.tex
\section{Preliminaries}
\label{sec:preliminaries}
In this section, we cover the background information needed to understand our protocol. We begin by describing the 4C code and its QEC properties \cite{leghtas_2013, mirrahimi_2014}. Next, we present an overview of the relevant operations within the cQED \cite{blais_2021} platform. In particular, we discuss standard techniques to control single and pairs of bosonic modes, as well as the effect of common hardware errors during these operations. 

\input{tex/2.1.four_cats.tex}
\input{tex/2.2.cqed.tex}

\begin{table*}[ht]
    \centering
    \begin{tblr}{
        colspec = {X[c,m, wd=5.5cm]X[c,m]X[c,m]X[c,m]X[c,m]X[c,m]},
        stretch = 0,
        rowsep = 6pt,
        vlines = {black, .6pt},  
        cell{4,5,6,7,9,10,12,14}{1}={mode=dmath},
      }
        \hline
        \SetCell[r=2]{c,m}{Primitive \\ operation} & \SetCell[r=2]{c,m}{Application \\ within our \\ architecture} & \SetCell[c=4]{c,m}{Effect of dominant error mechanisms during the operation} \\ 
        \hline[dashed]
        & & {Photon loss \\ $\Big( \opa$ or $\opb \Big)$} & {Ancilla decay \\ $\Big(\dyad{g}{e}$ or $\dyad{e}{f}\Big)$} & {Ancilla dephasing \\ $\Big(\dyad{e}{e}$ or $\dyad{f}{f}\Big)$} & {Spurious non-linearities \\ $\Big(\opa^{\dagger \, 2} \opa^2$ etc. $\Big)$} \\
        \hline \hline
        {\textbf{Photon-number msmts. (PNM)} \\ (selective rotation below followed by an ancilla measurement)} & & \SetCell[r=5]{c,m}{parity check \\[15pt] $\Big\Downarrow$ \\[15pt] \errordetected{\underline{state prep.}: \\reset \& retry} \\[8pt] \errorcorrected{\underline{fusions}: \\repeat measurement}} & & \SetCell[r=5]{c,m}{\errorcorrected{repeat measurement. \\[15pt] $\Big\Downarrow$ \\[15pt] majority vote on outcomes} } & \SetCell[r=8]{c,m}{no effect \\ (they commute with the photon-num.)} \\

        \text{PNM}^{ge}(\mcN) = & \SetCell[r=2]{c,m}{Fusions} & & \SetCell[r=2]{c,m}{\uncorrectable{incorrect measurement \& dephased bosonic mode}} & & \\
        \sum_{n \in \mcN} \dyad{n} \otimes \op{\sigma}_x^{ge} + \sum_{n \notin \mcN} \dyad{n} \otimes \opI_{\text{anc}} & & & & & \\
        
        \hline[dashed]
        \text{PNM}^{gf}(\mcN) = & \SetCell[r=2]{c,m}{GHZ state \\prep. \\[10pt] Fusions} & &  \SetCell[r=5]{c,m}{\errordetected{flagged by $\ket{e}$  \\[15pt] $\Big\Downarrow$ \\[15pt] reset \& retry}}& & \\
        \sum_{n \in \mcN} \dyad{n} \otimes \op{\sigma}_x^{gf} + \sum_{n \notin \mcN} \dyad{n} \otimes \opI_{\text{anc}} & & & & & \\
        
        \hline
        \textbf{SNAP gates} & \SetCell[r=3]{c,m}{GHZ state \\prep. \\[10pt] Non-Clifford $Z_\fC(\theta)$} & \SetCell[r=3]{c,m}{\errordetected{parity check \\[5pt] $\Big\Downarrow$ \\[5pt] reset \& retry}} & & \SetCell[r=3]{c,m}{\errordetected{flagged by $\ket{f}$  \\[5pt] $\Big\Downarrow$ \\[5pt] reset \& retry}}  & \\[0pt]
        S(\vec{\phi}) \otimes \dyad{g} + S(- \vec{\phi}) \otimes \dyad{f} & & & & & \\
        \text{where } S(\vec{\phi}) = \sum_n e^{i \phi_n}\dyad{n} & & & & & \\
        
        \hline \hline
        \textbf{Beamsplitter (BS) gates} & \SetCell[r=2]{c,m}{GHZ state \\prep. \\[10pt] Fusions} &  \SetCell[r=4]{c,m}{parity check \\[15pt] $\Big\Downarrow$ \\[15pt] \errordetected{\underline{state prep}: \\reset \& retry} \\[8pt] \errorcorrected{\underline{fusions}: \\(see Sec. \ref{sec:fusions})}} &  \SetCell[r=2]{c,m}{no effect (both ancillae in $\ket{g}$)} &  \SetCell[r=2]{c,m}{no effect (both ancillae in $\ket{g}$)} & \SetCell[r=4]{c,m}{\errordetected{parity check \\[5pt] $\Big\Downarrow$ \\[5pt] if no photons lost, then correct with SNAP; \\else, reset \& retry}} \\[-5pt]
        \rmB \rmS(\theta) = \exp \left[ \frac{\theta}{2} (\opaDag \opb - \opbDag \opa) \right] & & & & & \\
        
        \hline
        \textbf{Ancilla-controlled logical gates} & \SetCell[r=2]{c,m}{$\text{CZ}_\fC$ for 6-ring prep. \\[10pt] Non-Clifford $ZZ_\fC(\theta)$} & &  \SetCell[r=2]{c,m}{\errordetected{flagged by $\ket{e}$  \\[5pt] $\Big\Downarrow$ \\[5pt] reset \& retry}} &  \SetCell[r=2]{c, m}{\errordetected{flagged by $\ket{f}$  \\[5pt] $\Big\Downarrow$ \\[5pt] reset \& retry}} &  \\[-5pt]
        cZZ_\fC = \opI \, \otimes \, \left| g \right>_1 \hspace*{-0.3em} \left< g \right|_1 + ZZ_\fC \otimes \left| f \right>_1 \hspace*{-0.3em} \left< f \right|_1  & & & & & \\
        \hline
    \end{tblr}
    \caption{Summary of dominant hardware errors affecting primitive operations for a single bosonic mode (PNM, SNAP) and a pair of modes (BS, ancilla-controlled logical gates). These operations are natively implemented in cQED using a dispersively-coupled non-linear ancilla (e.g., transmon) and beamsplitter couplers. The primary errors in such a platform include photon loss in the bosonic mode, ancilla decay and dephasing, as well as non-linearities inherited by the bosonic mode (e.g., self-Kerr). Operations can: inherently correct a single error (\errorcorrectednobox{green}), detect errors (\errordetectednobox{blue}), or fail undetectably (\uncorrectablenobox{red}), causing uncorrectable errors on the 4C code. For errors that can be detected or corrected, we detail the detection process (before the arrow) and the corresponding response (after the arrow). Errors during unselective operations and bosonic-mode resets are excluded, as these operations are faster than any physical coherence time and they minimally impact the protocol.}
    \label{tbl:primitive_ops}
\end{table*}

\input{tex/2.3.primitive_ops.tex}    

%% file: tex/2.1.four_cats.tex
\subsection{The four-legged cat code}
\label{sec:4cats}

\begin{figure}[!ht]
	\includegraphics[width=1\linewidth]{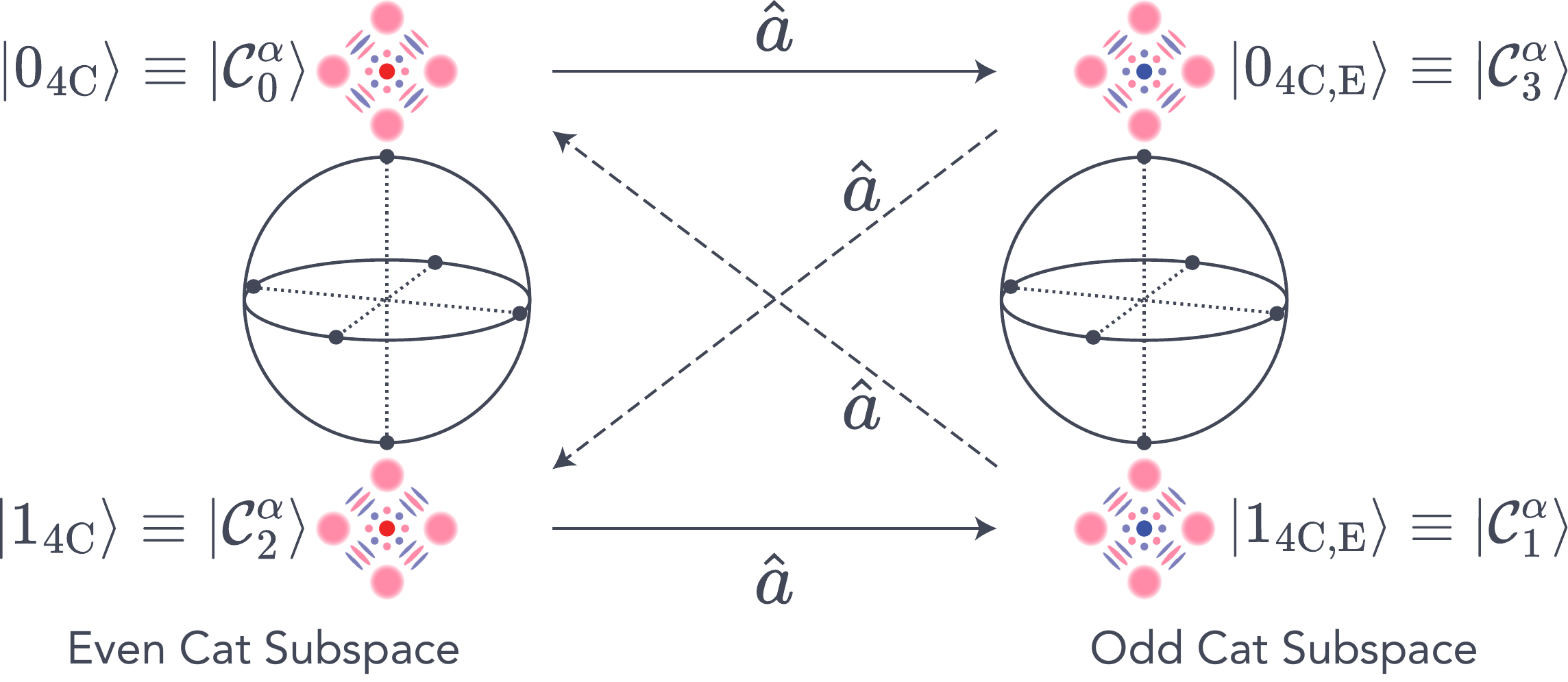}
	\caption{Left: Bloch sphere representing the codespace of the four-legged cat (4C) code, spanned by even-parity states $\cat{0}$ and $\cat{2}$. Right: Bloch sphere for the error space, spanned by odd-parity states $\cat{3}$ and $\cat{1}$. The error-space is formed by the action of single-photon loss $\op{a}$ (solid arrows) on the codewords. A second photon loss (dashed arrows) returns the system to the codespace with an uncorrectable bit flip. The even and odd 4C states sit on the poles of these Bloch spheres, illustrated by cartoons of their respective Wigner functions.}
	\label{fig:4cats}
\end{figure}

The \textit{4C states} $\{ \cat{n} | n = 0,1,2,3 \}$ are defined as an equal superposition of four coherent states $\{ \ket{i^m \al} | m = 0,1,2,3 \}$, each displaced to a vertex of a square centered at the origin in the oscillator's phase space, as illustrated in Fig.~\ref{fig:4cats} \cite{leghtas_2013}. The relative phases of these superpositions are chosen such that these states have disjoint support over the Fock states modulo four, 
\allowdisplaybreaks{
\begin{subequations} 
    \fontsize{9}{12}\selectfont
    \begin{alignat}{3}
        \cat{0} &\propto \coh{0} + \coh{1} + \coh{2} + \coh{3}     &&= \sum_k a_k \ket{4k}, \\
        \cat{1} &\propto \coh{0} - i \coh{1} - \coh{2} + i \coh{3} &&= \sum_k b_k \ket{4k + 1}, \\
        \cat{2} &\propto \coh{0} - \coh{1} + \coh{2} - \coh{3}     &&= \sum_k c_k \ket{4k + 2}, \\
        \cat{3} &\propto \coh{0} + i \coh{1} - \coh{2} - i \coh{3} &&= \sum_k d_k \ket{4k + 3}.
    \end{alignat}
    \label{eqn:4cats}
\end{subequations}
}

The prefactors $\{a_k, b_k, c_k, d_k\}$ depend on the ``size of the cat'' $\abs{\al}$ (see App. \ref{app:four_cat_Z}) \cite{li_2017}. Without loss of generality, we assume that $\alpha$ is real, such that the four coherent states are aligned horizontally or vertically in the oscillator's phase space. 

As indicated in Eq. \ref{eqn:4cats}, two of the 4C states have support on even Fock states, whereas the other are supported on odd Fock states. As shown in Fig. \ref{fig:4cats}, we may form a logical qubit by spanning the two even states (herein referred to as the ``codespace''): $\ket{0_\fC} \equiv \cat{0}, \ket{1_\fC} \equiv \cat{2}$ \cite{leghtas_2013, mirrahimi_2014, ofek_2016}. Under this convention, the two odd states span the ``error-space'', i.e. the subspace formed by the action of the single-photon loss on the codespace: $\ket{0_\fCE} \equiv \cat{3} \propto \opa \cat{0}, \ket{1_\fCE} \equiv \cat{1} \propto \opa \cat{2}$, where $\op{a}$ is the bosonic mode's annihilation operator. 

Notably, adopting the opposite convention, where the codespace is defined by odd parity and the error space by even parity, yields a set of logical states with equivalent QEC capabilities. In fact, our preferred state preparation scheme randomly prepares any one of the 4C states (see App. \ref{app:prep_4cats}) \cite{ofek_2016}. Therefore, we must ensure that each operation is agnostic to the prepared 4C state.

Since the 4C codewords have well-defined parity, a single-photon loss can be detected by measuring the photon-number parity, $\op{\Pi} = \exp(-i \pi \opaDag \opa)$, of the mode. Thus, the parity must be measured frequently enough to make sure that double-photon loss is rare, as it is an undetectable error that leaves the parity unchanged. The photon loss rate is proportional to the number of photons stored in the mode $\bar{n} \approx |\al|^2$, implying that parity must be measured more frequently for larger cats. Upon detecting a change in the parity, we correct for photon loss by updating our knowledge of the code basis in software \cite{ofek_2016}. This approach is readily implemented in cQED systems \cite{vlastakis_2013,sun_2014,rosenblum_2018,ofek_2016} (see Sec. \ref{sec:core_ops_single}). Thus, each of our proposed operations must work independently of the basis.  Alternatively, photon loss can be corrected using measurement-free or autonomous QEC \cite{mirrahimi_2014}. In this approach the basis remains fixed, and the entropy accumulated in the bosonic mode (due to photon loss) is transferred to an auxiliary system. Then, this auxiliary system is reset to evacuate the entropy, restoring the system to the codespace \cite{leghtas_2013}. We do not consider autonomous QEC as it demands for the challenging high-order coupling techniques that we seek to circumvent. 

A subsequent photon loss (or two simultaneous photon losses) results in an uncorrectable bit-flip in both the code and error spaces, i.e. $\ket{0_\fC} \equiv \cat{0} \rightleftharpoons \ket{1_\fC} \equiv \cat{2}$ and $\ket{0_\fCE} \equiv \cat{3} \rightleftharpoons \ket{1_\fCE} \equiv \cat{1}$. Likewise, a photon gain (heating) is also an uncorrectable error; since it may be incorrectly decoded as a photon-loss. However, the probability of two simultaneous photon losses is quadratically suppressed \cite{michael_2016, girvin_2023}, and heating errors are typically about two orders of magnitude less likely than a single-photon loss \cite{goldblatt_2024}.

In the absence of photon losses, the bosonic mode undergoes \textit{no-jump evolution}, $\rho(t) = \op{K}_0(t) \rho(0) \op{K}_0^\dagger(t)$, where $\op{K}_0(t) = \exp(-\half \kappa t \, \opaDaga)$, $\rho(t)$ is the density matrix at time $t$, and $\kappa$ is the photon-loss rate \cite{michael_2016, girvin_2023}. Unfortunately, $\op{K}_0$ commutes with $\op{\Pi}$ rendering the no-jump evolution undetectable by the parity checks. In order to preserve logical information, $\op{K}_0$ must not distinguish between the logical-$Z$ states. More precisely, $\bra{0_\fC} \op{K}_0^\dagger(t) \op{K}_0(t) \ket{0_\fC} = \bra{1_\fC} \op{K}_0^\dagger(t) \op{K}_0(t) \ket{1_\fC}$. This is achieved when $\tan |\al|^2 = \tanh |\al|^2$, such that the logical-$Z$ states have equal average photon number (see App. \ref{app:four_cat_nbar}) \cite{li_2017}. Even when this condition is satisfied, the no-jump evolution causes the size of the 4C states to shrink exponentially, $|\al(t)| = e^{-\half \kappa t} |\al(0)|$.  Our FBEC architecture naturally counteracts this damping by frequently teleporting the 4C states into freshly initialized bosonic modes via fusions (see Sec. \ref{sec:fbqc_overview}). 

Another factor in choosing $|\alpha|$ concerns the logical-$X$ eigenstates. These states are approximately described as superpositions of two coherent states aligned horizontally (for $\ket{+_\fC}$) or vertically (for $\ket{-_\fC}$), known as ``two-legged cat (2C) states'' \cite{leghtas_2013} and shown in Fig. \ref{fig:4cats}. While the approximation improves exponentially as $|\alpha|$ increases, there are specific small values of $|\alpha|$ where the 2C states exactly correspond to the logical-$X$ eigenstates (see App. \ref{app:four_cat_X}). For our numerical simulations, we use $|\alpha| = \sqrt{8}$, as it closely satisfies both conditions (see App. \ref{app:alpha_choice}). This corresponds to an average of $\bar{n} \approx 8$ photons in the bosonic mode, well within experimental feasibility \cite{vlastakis_2013, sivak_2023}.

The 4C code can also correct for a continuous set of phase rotations with an angle smaller than $\pi/4$~\cite{leghtas_2013,mirrahimi_2014,ofek_2016,xu_2023b}. In fact, the 4C code was numerically shown to have optimum performance for certain bosonic channels with both photon-loss and dephasing errors~\cite{leviant_2022}. However, none of the operations we propose explicitly detect or correct for dephasing errors. This choice is motivated by two reasons. First, correcting these errors in practice requires complex homodyne measurements of the bosonic mode, which are not readily implemented in cQED. Second, such dephasing errors are relatively rare and can be addressed by the outer $XZZX$ code. In particular, the microwave resonators (used to host the 4C) exhibit an inherent noise bias, with dephasing rates that are at least an order of magnitude smaller than decay rates~\cite{chakram_2021,rosenblum_2018, reinhold_2020,milul_2023,goldblatt_2024,teoh_2023}. While other codes~\cite{leviant_2022, gkp_2001, steve_isa_2024} may be optimal under a pure photon-loss channel, however we prefer the 4C code for its practical scalability and its resilience to ancilla-induced errors (see Sec.~\ref{sec:this_work}).

The logical-$Z$ operator in the 4C basis, defined as $\op{Z}_\fC = \exp \parens{-i \frac{\pi}{2} \opaDaga}$, captures the mode's photon-number modulo 4. In contrast, the logical-$X$ operator $\op{X}_\fC$, determines whether a 2C state is oriented vertically or horizontally in the mode's phase space. 

Fusion operations, required for FBEC, entail measuring the joint logical operators $\op{Z}_\fC \otimes \hat{Z}_\fC$ and $\op{X}_\fC \otimes \hat{X}_\fC$ on two 4C states, each within a bosonic mode with operators $\op{a}$ and $\op{b}$ respectively. The logical-$ZZ$ operator, $\op{Z}_\fC \otimes \hat{Z}_\fC = \exp \left[-i \frac{\pi}{2} (\opaDaga + \opbDagb) \right]$, measures the joint photon-number of both modes modulo 4, whereas the logical-$XX$ operator identifies if a pair of 2C states are aligned or anti-aligned. As we shall see in Sec. \ref{sec:fusions}, these fusions can be implemented by decoupling the joint information and encoding them in the local photon-number basis of the individual modes. 

%% file: tex/2.2.cqed.tex
\subsection{The cQED platform}
\label{sec:cqed_platform}

We leverage the remarkable experimental advancements in superconducting couplers \cite{sirois_2015, gao_2018, gao_2019, zhang_2019, chapman_2023, yao_lu_2023, zhou_2023} to connect neighboring bosonic modes. These couplers effectively admit a programmable interaction between modes, given by (setting $\hbar = 1$)
\begin{align}
    \opH_{\rmB \rmS} = \half \left[g(t) \, \opaDag \opb + g^*(t) \, \opa \opbDag \right] + \Delta(t) \opaDaga,
    \label{eqn:H_bs}
\end{align}
where $\opa$ and $\opb$ are the annihilation operators of the two modes, $g$ is the interaction strength of the beamsplitter interaction, and $\Delta$ is an effective detuning between the two modes. In general, the beamsplitter parameters, $g$ and $\Delta$ are time-dependent and can be controlled by microwave drives.


To create useful non-classical superpositions, each bosonic mode needs a source of nonlinearity. As illustrated in Fig. \ref{fig:overview}(b), our proposal couples each bosonic mode to a dedicated auxiliary transmon \cite{koch_2007}. The resulting Hamiltonian for each pair of bosonic modes and their respective ancillae combines the inter-mode beamsplitter coupling with an always-on dispersive interaction. In the rotating frame, this Hamiltonian is given by \cite{blais_2004,blais_2007,blais_2021, tsunoda_2023},
\begin{align}
    \begin{split}
        \op{H} &= \opH_{\rmB \rmS}  - \chi_{e,a} \opaDaga \left| e \right>_a \hspace*{-0.3em} \left< e \right|_a - \chi_{f,a} \opaDaga \left| f \right>_a \hspace*{-0.3em} \left< f \right|_a \\
        & \quad - \chi_{e,b} \opbDagb \left| e \right>_b \hspace*{-0.3em} \left< e \right|_b - \chi_{f,b} \opbDagb \left| f \right>_b \hspace*{-0.3em} \left< f \right|_b,
    \end{split} \label{eqn:H_bs_disp}
\end{align}
where $\ket{e}_j$ and $\ket{f}_j$ are the first and second excited states of the three-level ancilla coupled to the $j^{\text{th}}$ mode. The ground state of these ancillae is denoted $\ket{g}_j$ (not shown). $\chi_{i,j}$ represents the strength of the dispersive interaction between the $i^{\text{th}}$ level of the ancilla and its corresponding mode, $j$. For simplicity, we enforce that the second ancilla is always in its ground state $\ket{g}_b$ during all two-mode operations and ignore the second line of Eq. \ref{eqn:H_bs_disp}. 

In the previous section, we discussed the effect of photon loss (within the bosonic modes) on the 4C states. In addition to this, we must also consider the effect of the dominant errors in the ancilla: decay and dephasing of the $\ket{f}$ and $\ket{e}$ states. In fact, the coherence time of the ancilla is typically two orders of magnitude shorter compared to the bosonic mode \cite{reagor_2013, reagor_2016, rosenblum_2018, reinhold_2020, chakram_2021, milul_2023, goldblatt_2024}. Therefore, to retain the advantages of bosonic QEC, it is crucial to prevent ancilla errors from propagating to the bosonic modes \cite{rosenblum_2018, xu_2023b}. 

Decay in the transmon ancilla can be modeled as cascaded transitions: first from $\ket{f}$ to $\ket{e}$, and then from $\ket{e}$ to $\ket{g}$, with respective lifetimes $T_1^{fe}$ and $T_1^{eg}$. Since neither decay commutes with the dispersive interaction (Eq. \ref{eqn:H_bs_disp}), they dephase the bosonic mode. To protect the bosonic mode from this loss channel, we can operate the ancilla exclusively in its $g$-$f$ subspace, and reserve the $\ket{e}$ state as a flag to detect single decay events \cite{reinhold_2020,rosenblum_2018,tsunoda_2023}. If the ancilla is measured in $\ket{e}$, a failure is declared and the gate is retried. Errors in these gates are detected with a probability $p_{\text{fail}}$ that scales linearly with the decay rate $1/T_1^{fe}$. Therefore, the infidelity of a gate that passes the check, $\veps_{\text{pass}}$, is suppressed quadratically since it takes two decay events to cause an undetected error: $\veps_{\text{pass}} \propto 1/(T_1^{fe} \, T_1^{eg}) \propto (1/T_1^{fe})^2$ \cite{tsunoda_2023, goldblatt_2024}. 

We use this error-detection scheme only during resource state preparation. This enables us to retry the preparation until success, without disrupting the other parts of the computation, a technique known as preselection \cite{paetznick_2024}. Preselection substantially reduces the hardware requirements for subthreshold QEC, as $\varepsilon_{\text{pass}}$ can be orders of magnitude smaller than $p_{\text{fail}}$ for realistic hardware coherences (see Sec. \ref{sec:numerical_sims}). On the other hand, the fusion operations are designed to be fully FT to this error. This means that the dephasing induced on the bosonic mode has does not have any impact on the logical information, without any preselection (see Sec. \ref{sec:fusions}).

Ancilla dephasing is modeled by the jump operators $\dyad{e}$ and $\dyad{f}$, with coherence times $T_\phi^{ee}$ and $T_\phi^{ff}$, respectively. Unlike decay, dephasing commutes with the static Hamiltonian (Eq. \ref{eqn:H_bs_disp}). However, as we shall soon see, gates and measurements on the bosonic mode are implemented by driving the ancilla. In this case, a dephasing event may result in an incorrect operation. To enable preselection, we design gates to detect single dephasing events. This results in a quadratically suppressed gate infidelity $\veps_{\text{pass}} \propto (1/T_\phi^{ee})^2 \propto (1/T_\phi^{ff})^2$, at the expense of a failure probability that scales linearly $p_{\text{fail}} \propto (1/T_\phi^{ee}) \propto (1/T_\phi^{ff})$ \cite{tsunoda_2023}. For fusion operations, repeated measurements are employed to achieve a comparable quadratic suppression of the measurement infidelity due to ancilla dephasing.

Lastly, we need to consider how our scheme performs against undesired non-linearities introduced by the dispersive coupling. These effects include self-Kerr $\half K \opa^{\dagger \, 2} \opa^2$, corrections to the dispersive interaction $ \chi'_e \opa^{\dagger \, 2} \opa^2 \dyad{e} + \chi'_f \opa^{\dagger \, 2} \opa^2 \dyad{f} $, and other photon-number-conserving non-linearities \cite{zhu_2013, kirchmair_2013, blais_2021}. If no photons are lost, these non-linearities induce a deterministic, photon-number-dependent phase. This phase can be corrected using SNAP gates \cite{heeres_2015} and other control techniques (see Appendix \ref{app:chi-prime}). However, if a photon is lost during a gate, the mode accrues different phases before and after the gate. Since we cannot predict when a photon will be lost, the overall evolution becomes stochastic. To address this, our state-preparation circuits include single-photon-loss detection to enable preselection, while the fusion operations are engineered so that their nonlinearities have negligible impact on the logical information.

With the static coupling between the ancilla and bosonic mode established, let us now explore how to make the ancilla-mediated operations on one or two modes fault-tolerant.

%% file: tex/2.3.primitive_ops.tex
\subsection{Single-mode primitives}
\label{sec:core_ops_single}

\textbf{Unselective operations}: To prepare the 4C states, we need to displace a bosonic mode or rotate an ancilla qutrit, independent of one another. A strong linear drive on the bosonic mode, of magnitude $\Omega_d \gg \chi_e, \chi_f$, can displace the bosonic mode approximately independent of the ancilla's state. Likewise, a strong drive on the ancilla at the frequency $\omega_{ge}$ realizes a mode-independent rotation $\op{R}_\phi^{ge}(\theta)$, about an arbitrary axis $\cos(\phi) \op{\sigma}_x^{ge} + \sin(\phi) \op{\sigma}_y^{ge}$ on the equator of the Bloch sphere for the $g$-$e$ subspace. Here, $\op{\sigma}_x^{ge} = \dyad{g}{e} + \dyad{e}{g}$ and $\op{\sigma}_y^{ge} = i \dyad{g}{e} - i \dyad{e}{g}$ are the $X$ and $Y$ Pauli matrices in this subspace. Rotations in the ancilla's $g$-$f$ subspace $\op{R}_\phi^{gf}(\theta)$, can be implemented similarly. 

While ancilla decay can be flagged by the $\ket{e}$ state, the unselective operations are not FT against ancilla dephasing and photon-loss in the bosonic mode. However, these operations can be performed several orders of magnitude faster than any coherence time in the system \cite{blais_2004, blais_2007}. As a result, errors occurring during these operations negligibly impact the protocol.

\textbf{Bosonic-mode reset (Q-switch)}: Since bosonic modes are linear and long-lived, it is challenging to initialize or reset the system to a known state. In the proposed cQED setup (Fig. \ref{fig:overview}), this can be addressed by leveraging the fast decay rate of the readout mode. Specifically, the coupled transmon can be driven to parametrically induce a swap interaction between the readout resonator and the storage cavity \cite{pfaff_2017}. This interaction effectively enhances the decay rate of the bosonic mode, rapidly cooling it to its ground state. Notably, this operation is robust to ancilla decay and dephasing errors, as the ancilla only acts as a coupler. Additionally, single-photon loss assists this parametric reset.

\textbf{Photon-number measurements (PNMs)}: PNMs are binary-outcome measurements of the photon number in the bosonic mode, such as parity, $\Pi$ or the 4C $Z$-logical operator $Z_\fC = \sqrt{\Pi}$. These measurements are crucial during state preparation as well as fusions. 

To implement PNMs, we first note that the dispersive interaction (see Eq.~\ref{eqn:H_bs_disp}) can be interpreted as a shift of each ancilla's transition frequency dependent on the number of photons in its corresponding bosonic mode. This effect is particularly perceptible in the \textit{number-split regime}, where the shift in the ancilla's frequency per photon, $\chi_{i,j}$ is larger than the linewidths of both the ancilla and the bosonic mode \cite{schuster_2007,gambetta_2006,johnson_2010}.

In this regime a weak drive on the ancilla at the frequency $\omega_{ge} + n \chi_e$ selectively induces Rabi oscillations in the ancilla's $g$-$e$ manifold, conditioned on $n$ photons in the bosonic mode \cite{johnson_2010}. In contrast to the unselective ancilla pulses, it is important that this Rabi drive is weak (Rabi frequency $\Omega_e \ll \chi_e$) such that the neighboring transitions [at $\omega_{ge} + (n \pm 1) \chi_e$] are not triggered. Thus, the pulse duration $T$, for a full $\pi$-rotation must be longer than $1/\chi_e$. Similarly, a weak drive at frequency $\omega_{gf} + n \chi_f$ selectively rotates the ancilla within the $g$-$f$ subspace. We denote these photon-number selective ancilla rotations as $\op{R}_{n, \phi}^{ge/gf}(\theta) = \dyad{n} \otimes \op{R}_\phi^{ge/gf}(\theta) + (\opI_{\text{bm}} - \dyad{n}) \otimes \opI_{\text{anc}}$, where ${R}_\phi^{ge/gf}(\theta)$ is an unselective rotation in the ancilla's $g$-$e$ or $g$-$f$ subspace, and $\opI_{\text{bm}}$ ($\opI_{\text{anc}}$) is the identity operator for the bosonic mode (ancilla).

Such selective ancilla rotations may be superposed by driving the ancilla at multiple frequencies. If the respective drive strengths are all weaker than $\chi_{i,j}$, the drives minimally interfere with one another and selectively drive the ancilla for an arbitrary set of Fock states. \footnote{These pulses can be further improved using optimal control techniques \cite{heeres_2017, reinhold_2020}}. PNMs are implemented by initializing the ancilla in $\ket{g}$, selectively excite it to $\ket{e}$ for a set of Fock states $\mcN$, and then measure it. Importantly, these are quantum non-demolition (QND) measurements \cite{heeres_2017, reinhold_2020, blais_2021}, allowing successive PNMs to extract multiple bits of information.

As mentioned previously, these measurements can be made robust to a single ancilla decay by encoding the information within the $g$-$f$ manifold. On the other hand, ancilla dephasing results in a faulty rotation. Consequently, the photon-number information may be incorrectly mapped onto the ancilla, leading to a measurement error upon readout. This can be mitigated by repeating the PNMs to verify the result and suppress measurement infidelity. Note that ancilla dephasing commutes with the dispersive interaction and therefore does not propagate onto the bosonic mode. 

In Sec. \ref{sec:resource_states} we describe how to realize error-detected parity measurements $\op{\Pi}^{gf}$, by exciting the ancilla from $\ket{g}$ to $\ket{f}$ selectively for odd Fock states (up to some reasonable photon-number cutoff dependent on the size of the cat). In Sec. \ref{sec:fusions} we explore a more optimized measurement scheme where all three levels of the ancilla are used to encode different information. The $\ket{e}$ state asks whether the bosonic mode is in the vacuum state and the $\ket{f}$ state encodes the photon-number parity or photon-number modulo 4 information. As we shall see, we can exploit the end-of-the-line nature of fusions to make these optimized measurements robust to ancilla decay and dephasing. 

\textbf{Selective Number-dependent Arbitrary Phase (SNAP) gates}: SNAP imparts an arbitrary phase on each Fock state in the bosonic mode, $S(\vec{\phi}) = \sum_n e^{i \phi_n} \dyad{n}$ \cite{heeres_2015,reinhold_2020}. SNAP has been experimentally demonstrated for several photon-number preserving applications \cite{heeres_2015}. In this work, we use SNAP to undo the parasitic evolution due to the bosonic mode's self-Kerr effect (Sec. \ref{sec:resource_states}) \cite{heeres_2015} and to implement $Z_\fC(\theta)$ gates (Sec. \ref{sec:non_clifford}) \cite{reinhold_2020}.

We can apply a geometric phase, $\phi$ on any single Fock state via two consecutive photon-number selective rotations on the ancilla, along different axes, $\op{R}_{\phi,n}^{gf}(-\pi) \op{R}_{0,n}^{gf}(\pi)$ \cite{heeres_2015}. As before, these photon-number selective operations can be chained together, allowing each Fock state $\ket{n}$ in the bosonic mode to acquire a distinct phase $\phi_n$. 

As with PNMs, we operate within the $g$-$f$ manifold and reserve the $\ket{e}$ state to detect single ancilla decay events \cite{reinhold_2020}. Since the ancilla ideally returns to the $\ket{g}$ state at the end of the operation, the $\ket{f}$ state can be used as a flag for a single dephasing event. If the ancilla is measured in $\ket{f}$, the bosonic mode is not disturbed and the gate may be retried \cite{reinhold_2020}. This fact may not be immediately obvious; therefore, we point the reader to detailed derivation in Ref. \cite{reinhold_2020}. In summary, the gate-infidelity for this error-detected construction of SNAP scales quadratically with both ancilla decoherence times, $\veps_{\text{pass}} \propto (1/T_1^{\dyad{e}{f}})^2$ , $(1/T_\phi^{\dyad{f}})^2$.

\subsection{Two-mode primitives}
\label{sec:core_ops_two}

\textbf{Beamsplitter (BS) unitary}: The BS is an indispensable tool for coupling bosonic modes, and has been experimentally realized to high-fidelity \cite{sirois_2015, gao_2018, gao_2019, chapman_2023, yao_lu_2023, zhou_2023}. In our protocol, the BS is used during both resource state preparation and fusions. 

It can be realized by initializing both ancillae in their ground states such that the bilinear interaction $\opH_{\rmB \rmS}$ remains the only active term in Eq. \ref{eqn:app_dispersive}. This Hamiltonian facilitates the implementation of any beamsplitter unitary, $\rmB\rmS(\theta)$, as given by Table \ref{tbl:primitive_ops} \cite{zhang_2019}. This operation is not affected by the dominant ancilla decay and dephasing errors since neither ancilla is excited out of its ground state. As for single-photon loss in the bosonic mode, we employ a different correction strategies for resource-state preparation and fusions. 

\textbf{Ancilla-controlled logical gates}: Lastly, we consider logical-$ZZ$ gates between two 4C states. To realize these, we first reinterpret the dispersive interaction as an ancilla-state-dependent beamsplitter detuning $\Delta' = \Delta - \chi_{f,a} \left| f \right>_a \hspace*{-0.3em} \left< f \right|_a$ \cite{tsunoda_2023}. This implies that the two modes evolve conditioned on the ancilla's state. The evolutions can be designed so that the bosonic modes return to their initial states at the end of the operation, but with different geometric phases that are kicked back onto the ancilla. In particular, choosing $g=\sqrt{15}/2 \, \chi_{f,a}$, $\Delta = -\chi_{f,a}/2$, and a gate time $T= \pi/\chi_{f,a}$ we can implement an ancilla-controlled $ZZ_\fC$ on the bosonic modes, $cZZ_\fC = \opI_{\text{bm}} \, \otimes \, \opI_{\text{bm}} \, \otimes \, \left| g \right>_a \hspace*{-0.3em} \left< g \right|_a + Z_\fC \otimes Z_\fC \otimes \left| f \right>_a \hspace*{-0.3em} \left< f \right|_a$~\cite{tsunoda_2023}.

Coupled with unselective ancilla rotations $\op{R}_{\phi}^{gf}(\theta)$, the $cZZ_\fC$ operation can be exponentiated to form a family of parametrized entangling gates on the bosonic modes, \cite{tsunoda_2023},
\begin{align}
    \fontsize{8}{12}\selectfont
    ZZ_\fC(\theta) = \op{R}_{\pi}^{gf}\parens{-\frac{\pi}{2}} \cdot  cZZ_\fC \cdot \op{R}_{0}^{gf}(\theta) \cdot  cZZ_\fC \cdot  \op{R}_{\pi}^{gf}\parens{\frac{\pi}{2}}. 
    \label{eqn:ZZ_theta}
\end{align}

Similar to the error-detected SNAP gate, a successful $ZZ_\fC(\theta)$ gate returns the ancilla to $\ket{g}_a$, while the $\ket{e}_a$ state flags a single ancilla decay event \cite{reinhold_2020,rosenblum_2018} and $\ket{f}_a$ flags a single dephasing event \cite{tsunoda_2023}. These errors are only detectable, as an unknown unitary is applied to the bosonic modes in the presence of an error.

Lastly, single-photon loss during the operation can be detected by measuring the stabilizers of the 4C code (i.e. local photon number parity) after the gate. However, single-photon loss is no longer correctable since such an event dephases the ancilla resulting in an unknown overall unitary \cite{tsunoda_2023}.

%% file: tex/3.0.this_work.tex
\section{FBEC with the Four-Legged Cat Code}
\label{sec:this_work}

In this section, we present our novel protocol to concatenate the four-legged cat code with the fusion-based version of the $XZZX$ code \cite{claes_2023, sahay_2023a}. First, we review FBEC and introduce a novel fully-planar architecture for its implementation. To the best of our knowledge, this is the first 2D nearest-neighbor architecture for FT FBEC \cite{raussendorf_2001, raussendorf_2002, raussendorf_2003, raussendorf_2006,claes_2023}. Next, we examine how to implement the two key FBEC operations, entangled resource state preparation and fusion measurements, using the 4C code. Importantly, these operations are first-order FT, ensuring that no single dominant hardware error can propagate from the 4C to the $XZZX$ code. We support our analysis through numerical simulations. Lastly, we explain the implementation of a non-Clifford operation, $Z_\fC(\theta)$, that is essential for universal quantum computation. 

\input{tex/3.1.planar_fbec.tex}
\input{tex/3.2.resource_states.tex}
\input{tex/3.3.bell_msmts.tex}
\input{tex/3.4.non_clifford.tex}

\input{tex/3.5.numerical_sims.tex}

%% file: tex/3.1.planar_fbec.tex
\subsection{Planar FBEC}
\label{sec:fbqc_overview}

\begin{figure*}[ht!]
	\includegraphics[width=\linewidth]{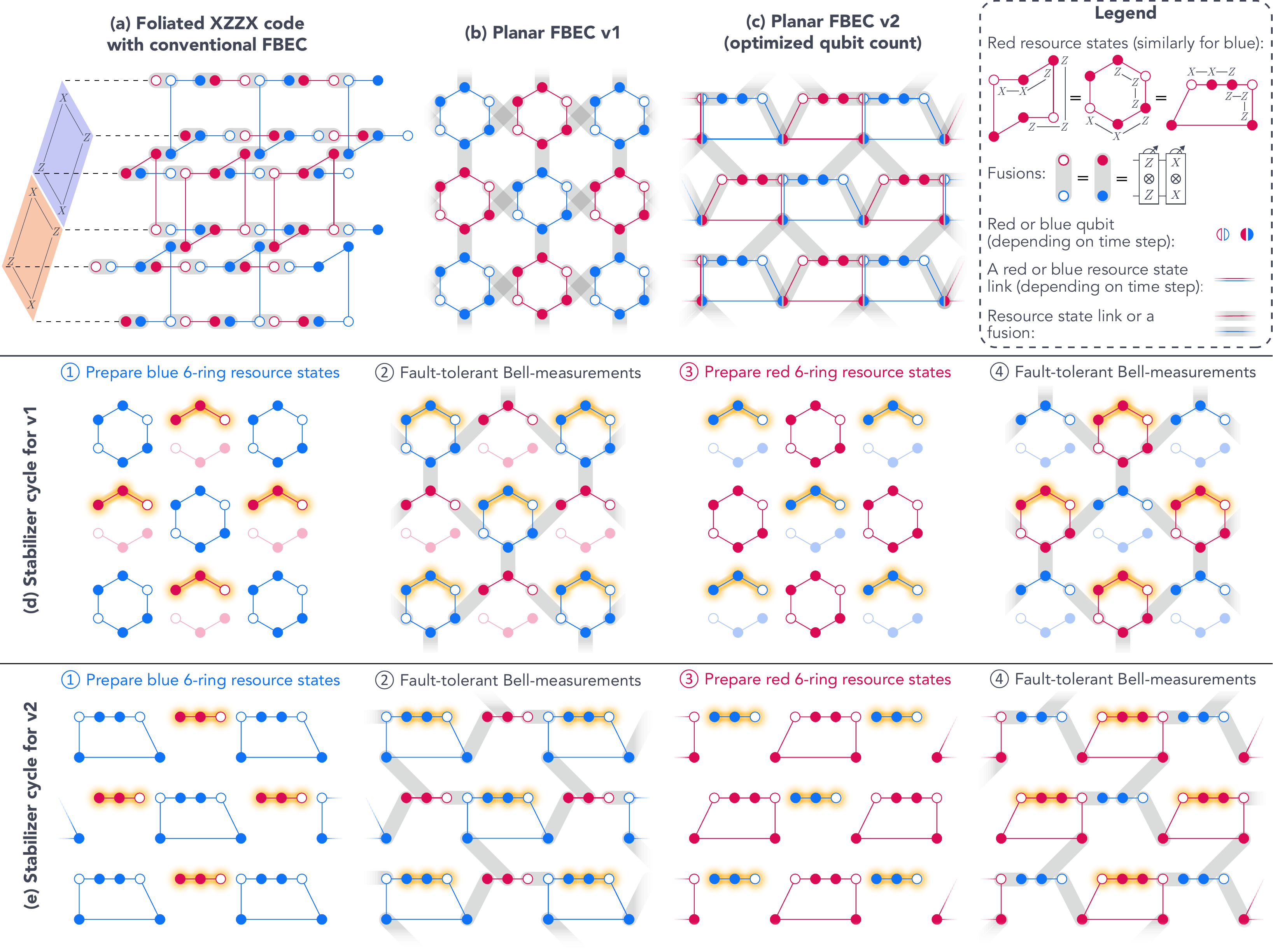}
	\caption{(a) In conventional implementations, the $XZZX$ code in FBEC uses a 3D grid of resource states \cite{claes_2023, sahay_2023a}. Here we show, six data qubits (in the bulk of the code) that are repeatedly teleported while two stabilizers (purple and orange plaquettes) are simultaneously measured. This is achieved by alternating layers of blue and red 6-ring resource states and fusing them together with Bell measurements (gray ovals). (b) Our initial planar architecture (Planar FBEC v1) simplifies the 3D design by retaining one layer of blue and red resource states, each. However, overlapping fusions prevent a fully planar configuration. (c) The optimized design (Planar FBEC v2) achieves a truly planar layout with 25\% fewer qubits. Here, some qubits are dynamically repurposed between blue and red resource states depending on the time step. Consequently, connections between qubits behave as either links between resource states or fusions, depending on the time step. (d, e) The four-step sequence for Planar FBEC v1 and v2, shows how the logical information (yellow highlights) is teleported back-and-forth between the blue and red qubits. v2 is obtained from v1 by noting that some qubits (light-red and light-blue) do not actively participate in every time-step, and can thus be repurposed. }
	\label{fig:qec_schedule}
\end{figure*}

Fusion-based quantum computing \cite{bartolucci_2023, sahay_2023a, bombin_2023} is a quantum computing paradigm tailored for systems that lack a universal set of deterministic unitary gates. In this approach, computation relies on two fundamental operations: generating few-body entangled resource states and performing destructive Bell measurements ($X \otimes X$ and $Z \otimes Z$), known as fusions.

In this work, we focus on the recently proposed FBEC protocol \cite{claes_2023, sahay_2023a} based on the $XZZX$ code \cite{ataides_2021}. This protocol achieves QEC by sequentially teleporting an encoded planar $XZZX$ state, from existing resource states to newly prepared ones \cite{raussendorf_2001, raussendorf_2002, raussendorf_2003, raussendorf_2006} [see Fig. \ref{fig:qec_schedule}(a)]. Both the teleportation and stabilizer measurements are performed through fusions. We choose the $XZZX$ version of FBEC for two key reasons. Firstly, the $XZZX$ is notable for achieving higher thresholds than the standard surface code in the presence of biased noise \cite{ataides_2021, claes_2023, sahay_2023a}. Our fusion measurements manifest this bias, as hardware errors primarily affect the $ZZ$ outcomes rather than the $XX$ ones. Second, the preparation of resource states required for the $XZZX$-FBEC protocol is simpler for the 4C code. 

The resource states needed for FBEC are composed of four ``X-type'' (depicted by $\sbullet \,$) and two ``Z-type'' (depicted by {\large $\circ$}) qubits \cite{claes_2023, sahay_2023a, sahay_2023b}. Conventionally, this distinction is based on the basis in which the qubits are initialized \cite{claes_2023, sahay_2023a, sahay_2023b}. However, our protocol often initializes bosonic modes outside the 4C codespace (see Sec. \ref{sec:resource_states}). Therefore, we categorize X- and Z- type qubits by the type of three-body stabilizer centered on each qubit. For an X-type qubit, the stabilizer is $Z \otimes X \otimes X$, with the X-type qubit placed to the left and the Z-type qubit to the right. For a Z-type qubit, the stabilizer is $Z \otimes Z \otimes Z$ with X-type qubits on each side. 

By reusing previously measured qubits, FBEC can implement an arbitrarily long computation using a limited number of qubits \cite{sahay_2023b}. This is accomplished by repeatedly teleporting the logical state back and forth between two planes of resource states. These two planes, represented in red and blue in Fig. \ref{fig:qec_schedule}, can be organized into a 2D layout as shown in Fig. \ref{fig:qec_schedule}(b). While this layout only features nearest-neighbor connections, some of these connections overlap, breaking planarity. To address this, we present an optimized layout in Fig. \ref{fig:qec_schedule}(c) that eliminates overlapping connections and further reduces the qubit footprint. Before discussing this optimization in detail, let us first how a single round of stabilizer measurements is implemented in the initial architecture. 

The four-step stabilizer schedule, shown in Fig. \ref{fig:qec_schedule}(d) works as follows. Let us assume that the $XZZX$ code is initially encoded in the red resource states, and the logical information (yellow highlights) resides in the top half of each red 6-ring (yellow highlights), which were carried over from the previous cycle. The bottom halves (light red) and all blue qubits have been measured previously. This cycle begins by reinitializing the blue qubits as fresh 6-ring resource states (step 1). Next, the blue 6-rings are fused (gray ovals) with the carried-over red states (step 2), completely measuring the red qubits and teleporting the $XZZX$ code to the blue sublattice. Now, the information (yellow highlights) resides in the top half of each blue 6-ring. The red qubits are then reinitialized as fresh 6-rings (step 3), and the code is teleported back to the red sublattice via further fusions (step 4). This leaves the information encoded in the top half of each red 6-ring, ready for the next stabilizer cycle. The $XZZX$ stabilizers are reconstructed by multiplying the fusion measurement outcomes \cite{claes_2023,sahay_2023a}. This layout can be further optimized (Fig. \ref{fig:qec_schedule}(c)), reducing the qubit count by 25\%. The light red and light blue qubits, which are measured but not immediately reinitialized, are recolored and reused, ensuring that each qubit actively participates in every cycle.

Our proposed implementations of these steps with the 4C code ensure that no single dominant hardware error can propagate to the $XZZX$ code. The resource state preparation scheme can detect and discard states with a single hardware error. This quadratically suppresses the infidelity of the prepared states, as two hardware errors are needed to bypass error detection. Furthermore, the fusions are inherently first-order FT. Now, let us look at how to implement these operations. 

%% file: tex/3.2.resource_states.tex
\subsection{Preparing the resource states}
\label{sec:resource_states}

\begin{figure}[ht]
	\includegraphics[width=1\linewidth]{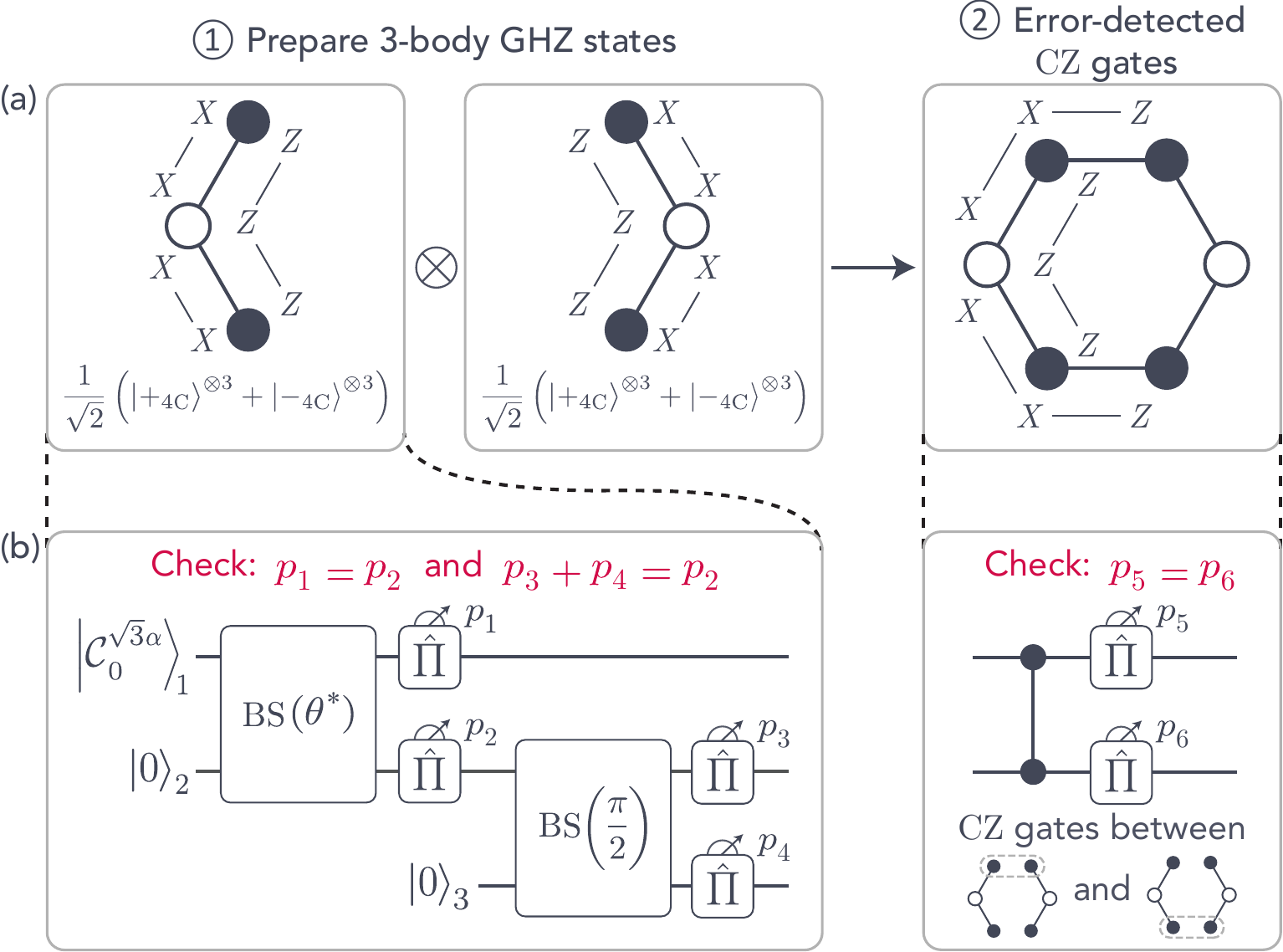}
	\caption{(a) The 6-ring FBEC resource states are fault-tolerantly constructed in two steps. First, two 3-body GHZ states are prepared. Then, these GHZ states are entangled using CZ gates. We show all three stabilizers that define each GHZ state and some stabilizers that define the 6-ring state. (b) To form a GHZ state in the 4C code, the first mode in each set is initialized in the $\Bigcat{0}$ state (three times the usual energy of $\ket{0_\fC}$), with all other modes in vacuum, $\ket{0}$. These modes are entangled using beamsplitters, BS$(\theta)$, and mid-circuit photon-number parity measurements, $\hat{\Pi}$, to form $\frac{1}{\sqrt{2}}(\ket{+_\fC}^{\otimes 3} + \ket{-_\fC}^{\otimes 3})$. Photon loss is detected via parity checks (marked in red), where $p_i \in {0,1}$ are parity results summed modulo 2. If any check fails, the state is discarded and the circuit is retried. As a result the infidelity of the resulting state is suppressed quadratically.}
	\label{fig:resource_prep}
\end{figure}

As shown in Fig. \ref{fig:resource_prep}(a), the 6-ring resource states are constructed in two steps. First, we prepare two copies of the 3-body GHZ state, $\frac{1}{\sqrt{2}}(\ket{+_\fC}^{\otimes 3} + \ket{-_\fC}^{\otimes 3})$. Then, these GHZ states are entangled using CZ gates, to form the 6-ring state. 

We emphasize that, in the absence of errors, these resource states can be prepared deterministically. A single dominant hardware error leads to a detectable failure, which occurs with a probability of about $1\%$ for practical coherence times (see Sec.~\ref{sec:numerical_sims}). This contrasts with conventional linear-optical FBEC, which requires an increasing number of qubits to reduce the failure rate below $25\%$ \cite{grice_2011}.

To prepare the GHZ state, the first mode is initialized in $\Bigcat{0}$, while others are set to vacuum. The cat state is initialized with three times the usual energy, because the circuit, in Fig. \ref{fig:resource_prep}(b), distributes the energy evenly across the three modes through beamsplitters. The first beamsplitter transfers two-thirds of the energy to the second mode, and the second transfers half of this to the third. 

This energy transfer can be understood by looking at the action of a beamsplitter on a generic coherent state $\ket{\beta}$ and vacuum: $\rmB \rmS(\theta) \big[ \ket{\beta} \otimes \ket{0} \big] = \ket{\cos(\frac{\theta}{2}) \beta} \otimes \ket{-\sin(\frac{\theta}{2}) \beta}$. By tuning the beamsplitter interaction time, we can set $\theta$ to $\theta^* \equiv 2\arccos(\frac{1}{\sqrt{3}})$ and transfer two-thirds of the energy to the second mode. Thus, the first beamsplitter acts on the 4C code as,
\begin{align}
	\rmB \rmS (\theta^*) \left[ \Bigcat{0} \otimes \ket{0} \right] = \ket{\alpha} \otimes \ket*{- \sqrt{2}\alpha} + \ket{- \alpha} \otimes \ket*{\sqrt{2}\alpha} \nonumber \\[5pt]
	  + \ket{i \alpha} \otimes \ket*{-i \sqrt{2}\alpha}  + \ket{- i \alpha} \otimes \ket*{i\sqrt{2}\alpha}.
\end{align}
Note that we have temporarily disregarded the third mode as it remains unentangled and inactive. 

Next, we measure photon-number parity on both modes. This measurement is represented by the projector $\mcM_{\Pi_i} \propto \opI + (-1)^{p_i}\op{\Pi}_i$, where $p_i \in \{0, 1\}$ is the result of the parity measurement on the $i^{\text{th}}$ mode. Both measurement outcomes result in a Bell state in the 4C basis, albeit with opposing parity. Since the parity of the input states is even and the beamsplitter conserves joint parity, the parity measurement outcomes must match, $p_1 = p_2$. A mismatch signals a measurement error or photon loss, prompting a restart.

As an example, let us consider the post-measurement state when both parity outcomes are even, i.e. $p \equiv p_1 = p_2 = 0$. This yields,
\begin{subequations}
	\begin{align}
		& \Big[ \mcM_{\Pi_1} \otimes \mcM_{\Pi_2} \Big] \rmB \rmS (\theta^*) \left[ \Bigcat{0} \otimes \ket{0} \right] \nonumber \\[5pt]
            &\propto \Big[ \ket{\alpha} + \ket{-\alpha} \Big] \otimes \left[ \ket*{\sqrt{2}\alpha} + \ket*{- \sqrt{2}\alpha} \right] \nonumber \\
		&\qquad + \big[ \ket{i\alpha} + \ket{-i\alpha} \big] \otimes \left[ \ket*{i\sqrt{2}\alpha} + \ket*{-i \sqrt{2}\alpha} \right] \\[5pt]
            &\approx \ket{+_{\fC}} \otimes \ket{+_{\fC}} + \ket{-_{\fC}} \otimes \ket{-_{\fC}}.
	\end{align}
\end{subequations}
In the last line, we have used the fact that the logical-X eigenstates of the four-legged cat code are exponentially well approximated by the two-legged cat states (see Sec. \ref{sec:4cats} and App. \ref{app:four_cat_X}). Importantly, note that the odd parity outcome $p \equiv p_1 = p_2 = 1$, is equally probable.  For this outcome, the post-measurement state is an equally valid Bell pair of odd 4C states, given by $\ket{+_{\text{4C,E}}} \otimes \ket{+_{\text{4C,E}}} - \ket{-_{\text{4C,E}}} \otimes \ket{-_{\text{4C,E}}}$. In either case, parity can simply be tracked in software.

After entangling the first two modes, we continue by entangling the third mode. As before, we do so with a beamsplitter followed by a pair of parity measurements, this time between the second and third modes. Unlike the previous step, a beamsplitter angle of $\frac{\pi}{2}$ is used, transferring half of the energy from the second mode to the third. In this step, the joint-parity checks can be cross-referenced with earlier parity measurements to detect single-photon loss. Specifically, $p_3 + p_4 = p$, where $p_3$ and $p_4$ are the parity outcomes for the second and third modes, obtained during the second round of measurements (see Appendix \ref{app:general_resource_prep} for a complete derivation).

The parity measurements in this circuit are implemented by selectively exciting the ancilla from $\ket{g}$ to $\ket{f}$ for odd states in the bosonic mode, i.e., $\text{PNM}^{gf}({1, \dots, 2n+1, \dots })$. As discussed in Sec. \ref{sec:core_ops_single}, these measurements inherently detect a single ancilla decay. By repeating the measurements twice, we can also detect a single dephasing event or SPAM error. Together, error-detected PNMs and joint-parity checks ensure that all single-jump errors are flagged. The circuit is retried until all checks are successfully passed. Notably, the probability of failure for these checks scales linearly with hardware error rates, meaning the infidelity of the prepared GHZ state is primarily limited by rare double-jump errors. This approach ensures a quadratic suppression of infidelity, albeit with the tradeoff of occasional circuit retries. Crucially, retrying the resource state preparation should not interrupt the other parts of the ongoing computation. 

With the three-body GHZ states prepared, we couple two copies of these states using CZ gates to form the desired 6-ring resource state. This gate can be implemented with the ancilla-controlled $ZZ_{\fC}\parens{\frac{\pi}{2}}$ (see Eq. \ref{eqn:ZZ_theta}), which is equivalent to CZ up to single-qubit Z-rotations. Like the GHZ state circuit, the CZ gates are designed to detect single decay or dephasing errors in the ancilla (see Sec. \ref{sec:core_ops_two}). Additionally, parity measurements performed after the gate detect single-photon losses, keeping the infidelity of the CZ gates quadratically suppressed \cite{tsunoda_2023}.

Furthermore, the 6-ring preparation protocol can be made robust to parasitic nonlinearities, such as self-Kerr. Since single-photon losses are detected and discarded, successful attempts ideally proceed without photon loss. In these cases, the Kerr evolution remains unitary and can be effectively reversed using the SNAP gate, $S\parens{\mqty[0 & 0 & Kt & \dots & \half(n^2 - n)Kt & \dots]^\rmT} = \parens{e^{-i \half K {\opa}^{\dagger 2} \opa^2 t}}^\dagger$ \cite{heeres_2015}. Since the SNAP gate can detect single hardware errors, it is consistent with the error-detection capabilities within the rest of the 6-ring circuit \cite{reinhold_2020}.
t

%% file: tex/3.3.bell_msmts.tex
\subsection{Fault-tolerant fusions}
\label{sec:fusions}

\begin{figure}[t]
	\includegraphics[width=1\linewidth]{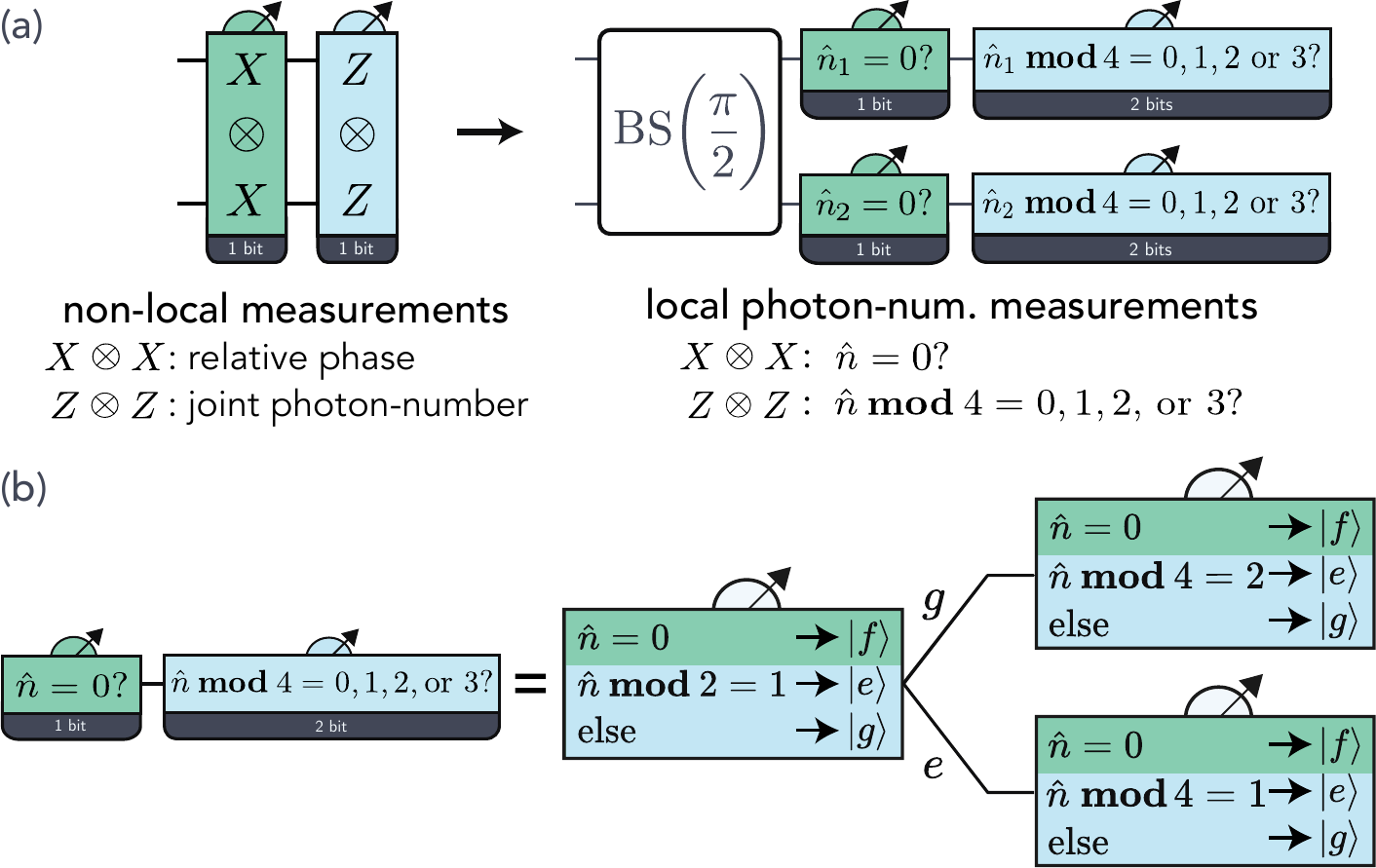}
	\caption{(a) Fusions measure $\op{X}_\fC \otimes \op{X}_\fC$ and $\op{Z}_\fC \otimes \op{Z}_\fC$ for a pair of qubits. In the 4C code, these correspond to measuring the relative phase and joint photon number between modes. A 50:50 beamsplitter, $\text{BS}(\frac{\pi}{2})$, transforms this non-local Bell-state information into properties of the local photon number $\op{n}$, which are then measured via PNMs. The $\op{X}_\fC \otimes \op{X}_\fC$ information (green) is encoded in whether the local photon number is zero $(\op{n} = 0)$. The $\op{Z}_\fC \otimes \op{Z}_\fC$ information (blue) is encoded in the photon number modulo 4 $(\op{n} \mod 4 = 0, 1, 2, 3)$. (b) The $\op{X}_\fC \otimes \op{X}_\fC$ information is extracted by selectively exciting the ancilla from $\ket{g}$ to $\ket{f}$ if the bosonic mode has zero photons. The 4-parity is determined in two steps: first, a parity measurement $\op{\Pi}$; then, the 4-parity is obtained using $\sqrt{\Pi}$ or $\sqrt{\Pi}'$, depending on the outcome of the parity measurement.}
        \label{fig:bell}
\end{figure}

The key to our protocol relies on the ability to use destructive fusions in FBEC. Hence, the bosonic modes are not required to remain in the 4C codespace, provided the measurement outcomes are accurately recovered. As shown in Fig. \ref{fig:bell}(a), this is achieved by evolving the 4C states through a 50:50 beamsplitter, $\text{BS}(\frac{\pi}{2})$. The crucial insight is that the beamsplitter encodes the Bell-state information in the photon number of each mode \cite{hastrup_2020,daiqin_su_2022,hastrup_2022}. This encoded information can then be extracted by measuring the photon numbers individually using PNMs. We highlight that our fusions are deterministic, unlike linear optical FBEC, where the $XX$ outcome is lost $50\%$ of the time. 

Table \ref{tbl:even_cat_bell} illustrates how the beamsplitter transforms Bell states within the 4C basis \cite{hastrup_2020, daiqin_su_2022, hastrup_2022}. These states are characterized by the eigenvalues $\lambda_{xx}$ and $\lambda_{zz}$ of the operators $X_\fC \otimes X_\fC$ and $Z_\fC \otimes Z_\fC$, respectively. We abbreviate these measurements $XX$ and $ZZ$ respectively. 

\begin{table*}[t]
    \centering
    \begin{tabular}{||c | c || c | c | c | c ||} 
    \hline 
    $\lambda_{xx}$ & $\lambda_{zz}$ & Bell-states & After $\rmB \rmS(\frac{\pi}{2})$ & After $\rmB \rmS(\frac{\pi}{2}) \& \ \opa$ & After $\rmB \rmS(\frac{\pi}{2}) \& \ \opb$ \rule{0pt}{3ex} \\[3pt]
    \hline\hline 
    $+1$ & $+1$ & $\cat{0}\cat{0} + \cat{2}\cat{2}$ & $\ket{0}\bigcat{0} + \bigcat{0}\ket{0}$ & $\bigcat{3}\ket{0}$ & $\ket{0}\bigcat{3}$ \rule{0pt}{4ex} \\[15pt]
    $+1$ & $-1$ & $\cat{0}\cat{2} + \cat{2}\cat{0}$ & $\ket{0}\bigcat{2} + \bigcat{2}\ket{0}$ & $\bigcat{1}\ket{0}$ & $\ket{0}\bigcat{1}$ \\[15pt]
    $-1$ & $+1$ & $\cat{0}\cat{0} - \cat{2}\cat{2}$ & $\catB{0}\catB{0} - \catB{2}\catB{2}$ & $\catB{3}\catB{0} - \catB{1}\catB{2}$ & $\catB{0}\catB{3} - \catB{2}\catB{1}$\\[15pt]
    $-1$ & $-1$ & $\cat{0}\cat{2} - \cat{2}\cat{0}$ & $\catB{1}\catB{1} - \catB{3}\catB{3}$ & $\catB{0}\catB{1} - \catB{2}\catB{3}$ & $\catB{1}\catB{0} - \catB{3}\catB{2}$ \\[10pt]
    \hline
    \end{tabular}
    \caption{Bell measurements in the 4C codespace. The four Bell states (column 3) are labeled by their eigenvalues $\lambda_{xx}$ and $\lambda_{zz}$ (columns 1 and 2), corresponding to the operators $\hat{X}_\fC \otimes \hat{X}_\fC$ and $\hat{Z}_\fC \otimes \hat{Z}_\fC$, respectively. Our protocol encodes Bell-state information into local photon-number properties using a 50:50 beamsplitter, $\mathrm{BS}(\frac{\pi}{2})$ \cite{hastrup_2020,daiqin_su_2022,hastrup_2022}. Column 4 shows the resulting states after the beamsplitter, up to an exponentially small error $\mcO(e^{-2|\alpha|^2})$. These states can be distinguished by two photon-number properties: $(\op{n} = 0)$ and $(\op{n} \mod 4 = 0,1,2,$ or $3)$, where $\op{n}$ is the photon-number operator of either mode. Columns 5 and 6 present the states after a single photon is lost from the first ($\opa$) or second ($\opb$) mode. Each half of the superposition in columns 3, 4, and 5 is distinct from one another. Therefore, we may use the same measurements to reliably recover the $\lambda_{xx}$ and $\lambda_{zz}$ eigenvalues even after a photon has been lost. Normalization constants and global phases have been ignored for all states. See Appendix \ref{app:bell_msmt} for an exact expression of the Bell-states after the beamsplitter.}
    \label{tbl:even_cat_bell}
\end{table*}

We begin by describing the $XX$ measurement. For $\lambda_{xx} = +1$ states, interference through the beamsplitter produces a superposition where one mode is in the vacuum state, while the other contains a cat state with twice the original number of photons. Conversely, the $\lambda_{xx} = -1$ states result in a superposition where both modes contain cat states. Thus, the beamsplitter maps the $XX$ information to whether one of the modes has zero photons or not (see column 4 of Table \ref{tbl:even_cat_bell}). This can be determined using PNMs on each mode, conditioned on detecting the vacuum state. This is represented by the unitary $V = \text{PNM}^{gf}(\{0\})$ followed by an ancilla measurement.

Even with perfect measurements, an unavoidable error can arise in the $XX$ measurement. Since the cat state $\cat{0}$ has a small but non-zero overlap with the vacuum, the $\ket{\lambda_{xx} = -1, \lambda_{zz} = +1}$ state has an exponentially small probability, proportional to $e^{-|\al|^2}$, of being misidentified as the $\ket{\lambda_{xx} = +1, \lambda_{zz} = +1}$ state. Our numerical simulations (see Sec. \ref{sec:numerical_sims}) suggest that this error dominates over the hardware errors. Minimizing the error requires a larger $|\alpha|^2$, which competes with the photon loss rate, $|\alpha|^2/T_1^{\text{loss}}$, that can make other measurements worse.

Once $\lambda_{xx}$ is determined, we can proceed to the $ZZ$ measurement. Importantly, this measurement depends on $\lambda_{xx}$. If one mode is found to be in the vacuum state ($\lambda_{xx} = +1$), $\lambda_{zz}$ can be deduced by measuring the photon number modulo four (4-parity) in the other mode (see column 4 of Table \ref{tbl:even_cat_bell}). This is accomplished with a straightforward binary search. First, photon-number parity is measured using a PNM conditioned on an odd number of photons, $\Pi = \text{PNM}^{ge}(\{1, 3, \dots, 2n+1\})$. Depending on whether this initial measurement indicates even or odd parity, the subsequent measurement uses either $\sqrt{\Pi} = \text{PNM}^{ge}(\{2, 6, \dots, 4n+2\})$ or $\sqrt{\Pi}{'} = \text{PNM}^{ge}(\{1, 5, \dots, 4n+1\})$. Alternatively, if neither mode is in the vacuum state ($\lambda_{xx} = -1$), the states can be directly distinguished by measuring the photon-number parity. 

Remarkably, these entangling measurements preserve the code's resilience against single-photon loss, regardless of whether the loss occurs before, during, or after the beamsplitter. If the photon is lost after the beamsplitter, the resulting states correspond to those shown in columns 5 and 6 of Table \ref{tbl:even_cat_bell}. These states can be distinguished from each other and from the lossless states (column 4) using the same local PNMs: $V$, $\Pi$, $\sqrt{\Pi}$, and $\sqrt{\Pi}'$. Instead, if a photon is lost before or during the beamsplitter, the resulting state becomes a weighted superposition of a photon loss occurring after the beamsplitter in one mode or the other. The PNMs then collapse this superposition into the states in either column 5 or 6.

Having addressed photon jumps, we now turn to the effect of no-jump evolution. As discussed in Sec. \ref{sec:4cats}, no-jump evolution simply reduces the size of the cat state without altering the photon number parity or 4-parity. Consequently, no-jump evolution after the beamsplitter preserves the Bell-state information that we aim to extract. However, unlike photon jumps, the fusions are not FT to no-jump evolution immediately before or during the beamsplitter. Fortunately, the beamsplitter is a fast operation \cite{chapman_2023, yao_lu_2023}, and errors during this brief window are several orders of magnitude less likely, contributing negligibly to the overall error budget.

After the beamsplitter, the fusion measurements are fully FT to dephasing and to evolution due to non-linearities in the bosonic modes, as these commute with the photon number. During the preparation of the 6-ring states, a SNAP gate was required to correct for the phase evolution caused by non-linearities, such as self-Kerr and $\chi'$. However, no such correction is needed during fusions, as these measurements occur at the end of the protocol, where any accumulated spurious phase does not affect the fusion outcomes. This robustness is a key advantage of encoding the Bell-state information into the photon number of each mode. As with no-jump evolution, such errors could in principle affect fusions if they occur before or during the beamsplitter; however, these events are comparatively rare.

We know from Sec. \ref{sec:core_ops_single} that errors in the ancilla during a PNM may propagate to the bosonic mode as random dephasing. Following the discussion above, dephasing in the bosonic mode may seem benign for the Bell measurement. However, these ancilla errors, including $T_1$, $T_\phi$, SPAM errors, and control pulse errors, can result in an incorrect measurement outcome. Nevertheless, these errors can be suppressed by realizing that fusions are end-of-the-line measurements and that the PNMs are quantum non-demolishing (QND) measurements on the photon-number space. Thus, we may reset the ancilla and repeat the PNMs to confirm the results. Appendix \ref{app:bell_msmt} provides complete details of the complete adaptive measurement scheme that uses a three-level ancilla to recover the correct Bell measurement result despite a single photon-loss or ancilla error. 

\begin{figure}[h]
	\includegraphics[width=\linewidth]{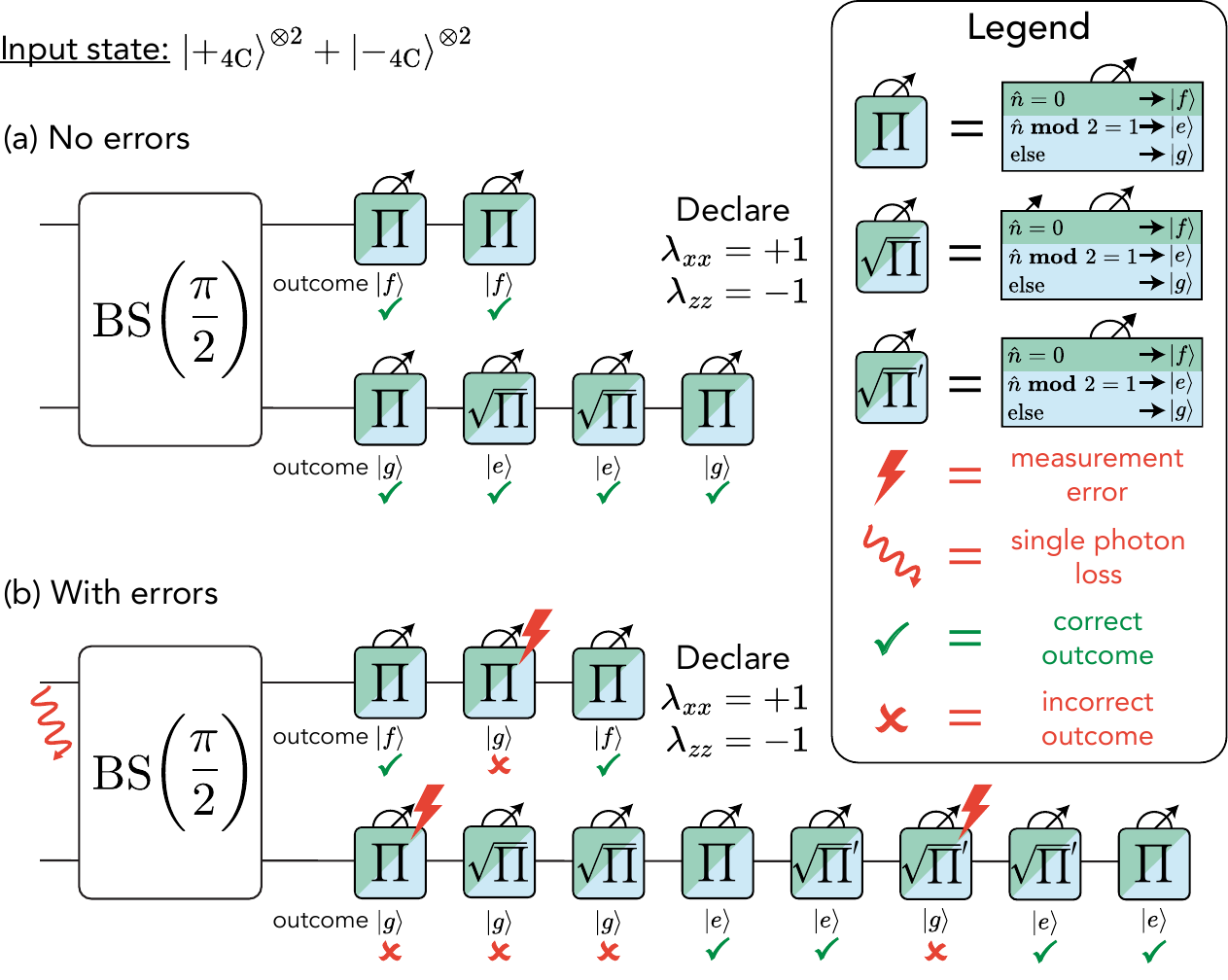}
	\caption{Examples of adaptive measurement sequences for the input state $\ket{\lambda_{xx}=+1, \lambda_{zz}=-1} = \frac{1}{\sqrt{2}}(\ket{+_\fC}^{\otimes 2} - \ket{-_\fC}^{\otimes 2})$. (a) In the ideal case, where no errors occur and repeated measurements yield consistent results. (b) Measurement errors and photon-loss cause disagreements between repeated measurements, which can be resolved through additional measurements. See App. \ref{app:bell_msmt} for a complete flowchart of the adaptive measurement sequence. }
    \label{fig:bell2}
\end{figure}

Fig. \ref{fig:bell2} shows an illustrative example of this adaptive measurement sequence for the input state $\ket{\lambda_{xx}=+1, \lambda_{zz}=-1} = \frac{1}{\sqrt{2}}(\ket{+_\fC}^{\otimes 2} - \ket{-_\fC}^{\otimes 2})$. In the ideal case, no errors occur and repeated measurements yield consistent results. Only one more measurement is needed to confirm the first result. In the example shown in Fig. \ref{fig:bell2}, a photon is lost from the first mode before the beamsplitter, putting it in an odd-parity state. A set of measurement errors result in disagreements between repeated measurements. Note that incorrect measurement outcomes may lead to incorrect choices for future measurements. However, our complete measurement scheme addresses these errors. 

%% file: tex/3.4.non_clifford.tex
\subsection{Non-Clifford resource states}
\label{sec:non_clifford}

While the cluster state natively supports Clifford operations via lattice surgery, non-Clifford operations are essential for universal quantum computation \cite{nielsen_chuang_2010,bravyi_2005,fowler_2012a}. Typically, logical non-Clifford operations can be implemented through magic state distillation and gate teleportation \cite{fowler_2009,fowler_2012a,litinski_2019,li_2015, singh_2022}. These schemes requires preparing a physical qubit in a non-Clifford state to distill a logical counterpart. In our protocol, this state is constructed by preparing a single bosonic mode in the $\ket{+_\fC}$ state, via displacements and parity measurements (see Appendix \ref{app:general_resource_prep}), followed by a $Z_\fC(\theta)$ using SNAP, 
\begin{subequations}
	\begin{align}
		Z_\fC(\theta) &= \sum_n e^{i \frac{\theta}{2}} \dyad{4n} + e^{-i \frac{\theta}{2}} \dyad{4n + 2}\\
		&= S\parens{\mqty[ \frac{\theta}{2} & 0 & -\frac{\theta}{2} & 0 & \dots ]^\rmT}.
	\end{align}
	\label{eqn:Z_theta}
\end{subequations}

The $ZZ_\fC(\theta)$ gate (see Sec. \ref{sec:core_ops_two} and Ref. \cite{tsunoda_2023}) can combined with the approach in Ref. \cite{singh_2022} to further improve the efficiency of magic state distillation. 

All operations used for preparing magic states are equipped with single-error detection, enabling attempts with flagged hardware errors to be discarded. Selecting only instances with no error flags ensures that the overall infidelity of the prepared magic states is quadratically suppressed, thereby reducing the overhead of the distillation protocol \cite{li_2015, singh_2022, jacoby_2025}.

%% file: tex/3.5.numerical_sims.tex
\subsection{Numerical Simulations}
\label{sec:numerical_sims}
We numerically simulate the Lindblad master equation in QuTiP \cite{qutip, qutip_2} to verify the FT properties of the six-ring preparation circuit and fusion measurements. Unfortunately, entirely simulating the preparation of the six-ring state is computationally infeasible as it involves a large Hilbert space, with multiple bosonic modes each with a three-level ancilla. Therefore, we separately simulate the PNMs and the $ZZ_\fC(\tfrac{\pi}{2})$ gate involved in the circuit. Then, we add the infidelities and failure probabilities associated with each operation to obtain estimates of these metrics for the complete circuit. For fusion measurements, we simulate the full adaptive measurement sequence with feed-forward. The simulations account for ancilla decay, ancilla dephasing, and photon loss in the bosonic modes (details in Appendix \ref{app:sims}). As shown in Fig. \ref{fig:all_sims}, we independently enable each error at a time, to verify first-order FT with each error source independently. These simulations also mimic experimental control pulses with a non-zero coherent pulse error, demonstrating first-order robustness these errors as well. 

For the resource state preparation, we evaluate the average gate failure probability, $p_{\text{fail}}$, and the average infidelity of successful gates, $\veps_{\text{pass}}$. As explained in Sec.\ref{sec:core_ops_two}, this circuit incorporates checks that detect all first-order jumps from each error channel, preventing any single error from propagating to the outer code. This is confirmed by the linear dependence of $p_{\text{fail}}$ on coherence times. Consequently, only second-order jump errors contribute to the infidelity of error-detected gates. These errors manifest as rare Pauli‑X errors, which are handled by the outer code. Our numerical simulations of the $ZZ$-gate confirm this first-order insensitivity, showing that $\veps_{\text{pass}}$ scales quadratically with each coherence time. 

We combine the simulations of the $ZZ$-gate with the parity measurements (implemented using $\mathrm{PNM}^{gf}$) to bound the fidelity and failure probability of preparing the 6-ring resource states. This is shown in the top row of Fig. \ref{fig:all_sims}. With realistic coherence times, we can achieve $p_{\text{fail}} \sim 1 \%$ and $\epsilon_{\text{pass}}$ below 0.01\%. The state preparation infidelity plateaus at $\sim 1 \times 10^{-6}$ which corresponds to the square of the pulse infidelities. Importantly, this highlights that the fusion measurements are first-order insensitive to control errors as well. 

\begin{figure*}[!ht]
	\includegraphics[width=\linewidth]{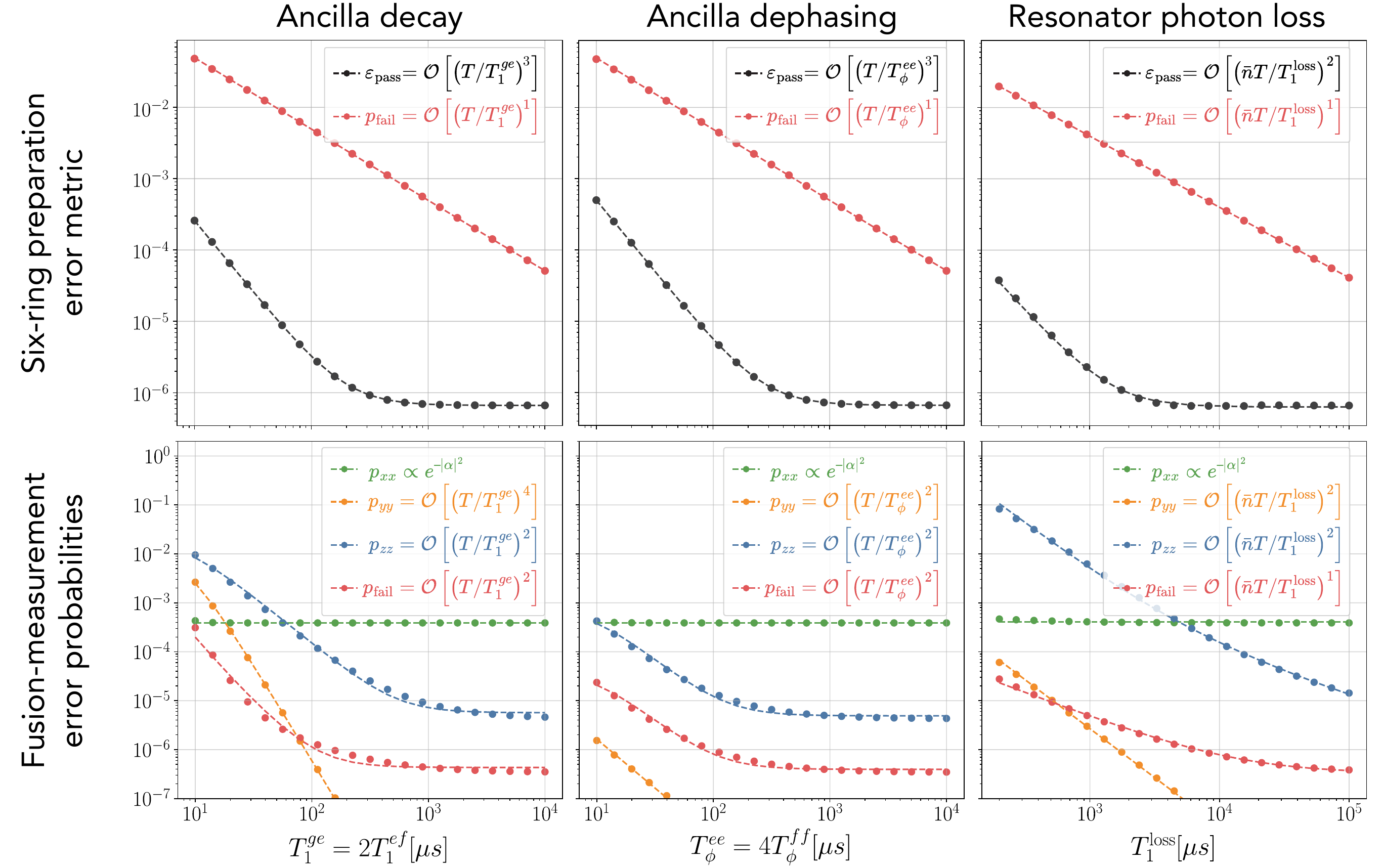}
	\caption{Numerical simulations (dots) and fits (dashed lines) of each step of our protocol, under three hardware error channels: ancilla dephasing, decay, and resonator photon loss. (Top) The failure probability $p_{\text{fail}}$ (red) and gate infidelity $\epsilon_{\text{pass}}$ (black) for the six-ring resource state preparation circuit. These values are to be interpreted only as an estimate, as they are computed by separately simulating the PNMs and $ZZ_\fC(\pi/2)$ gates, followed by adding the corresponding infidelities and probabilities. Notably, gate failure probability $p_{\text{fail}}$ scales linearly with coherence times, while gate infidelity $\epsilon_{\text{pass}}$ scales quadratically. (Bottom) Infidelities of the complete adaptive measurement sequence for a fusion measurement. The error in the fusion can be categorized into probabilities of misidentifying the $\lambda_{xx}$ outcome ($p_{xx}$, green), the $\lambda_{zz}$ outcome ($p_{zz}$, blue), or both outcomes ($p_{yy}$, yellow). $p_{xx}$ is dominated by a fundamental error in the measurement, $e^{-|\al|^2}$. In contrast, $p_{zz}$ and $p_{yy}$ scale quadratically (or better) with the coherence times. However, $p_{zz}$ eventually saturates to around $10^{-5}$, which is the square of the infidelity of the PNM pulses. The measurement sometimes yields an inconclusive result (i.e. an erased measurement), with probability $p_{\text{fail}}$ (red). This figure only provides a qualitative description of the fitted probabilities, in terms of the decoherence times and the execution time $T$. See App. \ref{app:sims} for the complete fitted expressions.}
	\label{fig:all_sims}
\end{figure*}

We simulate the complete adaptive measurement scheme to implement FT fusion measurements, where PNMs are repeated to resolve conflicting outcomes. We classify the measurement infidelity into the probabilities of misidentifying the $\lambda_{xx}$ outcome ($p_{xx}$), the $\lambda_{zz}$ outcome ($p_{zz}$), or both outcomes ($p_{yy}$). While $p_{xx}$ is fundamentally limited by the overlap of coherent states with vacuum $e^{-|\al|^2}$, both $p_{zz}$ and $p_{yy}$ demonstrate quadratic (or better) scaling with coherence times. Just as before, the probability of error plateaus at approximately $1 \times 10^{5}$, corresponding to the square of the PNM pulse infidelity. The fusion measurement sequence may lead to an inconclusive result, with probability $p_{\text{fail}}$. This can be treated by the outer QEC code as an erasure of the measurement outcome.

We simulate all operations on 4C states of size $|\al| = \sqrt{8} \approx 2.83$, and, where needed, also consider states with sizes $\sqrt{2}\alpha$ and $\sqrt{3}\alpha$. Note that unlike previous proposals \cite{rosenblum_2018, reinhold_2020, xu_2023b}, our protocol achieves fault-tolerance without matching $\chi_e$ to $\chi_f$. We explicitly show this by implementing an exaggerated difference in $\chi$ values, $\chi_{ge}/2\pi = -2$ MHz and $\chi_{gf}/2\pi = -1$ MHz. This results in a beamsplitter rate of approximately 2 MHz for the $ZZ_\fC(\tfrac{\pi}{2})$ gate. All simulations are implemented with realistic time-dependent pulses, with the error channels enabled in succession (details in Appendix \ref{app:sims}).

%% file: tex/4.discussion.tex
\section{Discussion and Conclusion}
\label{sec:discussion}

While single-mode bosonic codes have demonstrated impressive improvements in quantum memory lifetime, implementing error-corrected operations and achieving full fault tolerance with bosonic qubits has remained a persistent challenge. Our work addresses this challenge by introducing a novel framework for FT quantum computing by concatenating the 4C code with the $XZZX$ surface code through a fusion-based architecture. Crucially, all operations required for this concatenation are implemented using standard tools from the circuit-QED toolbox. Our numerical simulations demonstrate that the errors in these operations scale quadratically with the dominant hardware coherence times. This ensures that the concatenation retains the 4C code’s capability to suppress the most prevalent errors at the hardware level, leaving the $XZZX$ code responsible for addressing the rarer residual errors. Furthermore, we propose a new planar FBEC architecture that is fully compatible with superconducting circuits.

We simultaneously solve three key challenges which have plagued previous efforts to concatenate bosonic codes with circuit-based qubit codes: implementing FT entangling operations, correcting errors introduced by coupled ancillae, and mitigating state distortions caused by no-jump evolution and parasitic nonlinearities such as self-Kerr. Previous architectures have relied on complex techniques such as dissipation engineering \cite{mirrahimi_2014, xu_2023a}, multi-photon pumps and many-wave mixing couplers \cite{mirrahimi_2014, mundhada_2019, grimsmo_2020, vanselow_2025}, or intricate coupling methods such as $\chi$-matching \cite{rosenblum_2018, reinhold_2020, xu_2023b} to preserve the intrinsic error suppression of the bosonic codes. In contrast, our approach avoids these experimentally demanding techniques and instead employs conventional cQED tools: dispersively coupled transmons and beamsplitters. 

We opt for a fusion-based architecture for two reasons. First, fusion measurements can be easily implemented using conventional cQED tools. Second, FBEC naturally mitigates the no-jump evolution by repeatedly teleporting the logical 4C information, thereby circumventing the need for complex nonlinear interactions to stabilize the cat-code manifold. Taken together, our protocol substantially relaxes the hardware requirements for achieving fully FT QEC with the 4C code. The fusion-based protocol relies on two key steps: preparation of 6-ring resource states and the fusion of these states together using Bell measurements. Using current cQED hardware coherences, we project 6-ring preparation infidelities below 0.01\% with a 1\% probability of failure. Meanwhile, fusion measurements are expected to have a 1\% chance of an incorrect $ZZ$ outcomes and a 0.1\% chance of incorrect $XX$ outcome. 

Furthermore, our protocol addresses the limitations of conventional linear-optical FBEC. First, linear-optical FBEC requires a number of qubits that scales quadratically with the code distance and linearly with the length of the algorithm. In contrast, we propose a novel planar implementation of FBEC. This architecture efficiently reuses previously measured qubits and reduces the qubit overhead to only a quadratic dependence on code distance. Second, in linear optics, resource state preparation and fusions succeed only probabilistically. This necessitates multiplexing and a prohibitively large number of additional qubits to boost the success probability for preparing resource state above $\sim 1.5 \%$ and for recovering $XX$ fusion outcomes above $50 \%$. In contrast, all steps in our protocol are deterministic, limited only by rare hardware errors. Even with such errors, resource state preparation results in flagged failures just $\sim 1\%$ of the time, for realistic hardware coherences. Fusion measurements are even more robust, remaining entirely unaffected by single hardware errors.

In general, we expect that all operations can be performed below the threshold for FBEC \cite{claes_2023, sahay_2023a}. The $XZZX$ code is expected to use this noise bias to improve overall logical performance. \\

%% file: appendix.tex
\appendix
\onecolumngrid

\input{tex_appendix/four_cat.tex}
\input{tex_appendix/4c_bs.tex}

\input{tex_appendix/bell.tex}

\input{tex_appendix/resource_state_prep.tex}
\input{tex_appendix/fbec_no_jump.tex}
\input{tex_appendix/chi_prime.tex}
\input{tex_appendix/planar_fbec.tex}
\input{tex_appendix/sim_details.tex}

\input{tex_appendix/sim_kerr.tex}

%% file: tex_appendix/four_cat.tex
\section{The four-legged cat code}

\subsection{The four-legged cat states}
\label{app:four_cat_Z}

For a displacement $\al$, the orthonormal four-legged cat (4C) states $\{ \cat{n} | n = 0,1,2,3 \}$ are defined as an equal superposition of four coherent states displaced along the $q$ and $p$ quadratures of a bosonic mode $\{ \ket{i^m \al} | m = 0,1,2,3 \}$. As given in Eq. \ref{eqn:app_4cats}, the relative phases of these superpositions are chosen such that these states have disjoint support over the Fock states modulo four. 
\begin{subequations}    
    \begin{alignat}{3}
        \cat{0} &\equiv \frac{\coh{0} + \coh{1} + \coh{2} + \coh{3}}{2 \sqrt{2 e^{-|\al|^2} \parens{ \cosh |\al|^2 + \cos |\al|^2}} } &&= \sqrt{\frac{2}{\cosh |\al|^2 + \cos |\al|^2} } \sum_k \frac{\al^{4k}}{\sqrt{(4k)!}} \ket{4k} \\
        \cat{1} &\equiv \frac{\coh{0} - i \coh{1} - \coh{2} + i \coh{3}}{2 \sqrt{2 e^{-|\al|^2} \parens{ \sinh |\al|^2 + \sin |\al|^2}} } &&= \sqrt{\frac{2}{\sinh |\al|^2 + \sin |\al|^2} } \sum_k \frac{\al^{4k + 1}}{\sqrt{(4k + 1)!}} \ket{4k + 1} \\
        \cat{2} &\equiv \frac{\coh{0} - \coh{1} + \coh{2} - \coh{3}}{2 \sqrt{2 e^{-|\al|^2} \parens{ \cosh |\al|^2 - \cos |\al|^2}} } &&= \sqrt{\frac{2}{\cosh |\al|^2 - \cos |\al|^2} } \sum_k \frac{\al^{4k + 2}}{\sqrt{(4k + 2)!}} \ket{4k + 2} \\
        \cat{3} &\equiv \frac{\coh{0} + i \coh{1} - \coh{2} - i \coh{3}}{2 \sqrt{2 e^{-|\al|^2} \parens{ \sinh |\al|^2 - \sin |\al|^2}} } &&= \sqrt{\frac{2}{\sinh |\al|^2 - \sin |\al|^2} } \sum_k \frac{\al^{4k + 3}}{\sqrt{(4k + 3)!}} \ket{4k + 3}
    \end{alignat}
    \label{eqn:app_4cats}
\end{subequations}
In this paper we form a qubit by spanning the codespace with the two even states: $\ket{0_\fC} \equiv \cat{0}, \ket{1_\fC} \equiv \cat{2}$. Thus, the two odd states span the single-photon loss error-space: $\ket{0_{\text{4C, E}}} \equiv \cat{3} \sim \opa \cat{0}, \ket{1_{\text{4C, E}}} \equiv \cat{1} \sim \opa \cat{2}$. An undetected two-photon loss event results in an uncorrectable logical bit-flip, i.e. $\ket{0_\fC} \equiv \cat{0} \rightleftharpoons \ket{1_\fC} \equiv \cat{2}$ and $\ket{0_{\text{4C, E}}} \equiv \cat{3} \rightleftharpoons \ket{1_{\text{4C, E}}} \equiv \cat{1}$.

\subsection{Logical-X basis}
\label{app:four_cat_X}
We may notice that the symmetric superposition of the logical states results in a constructive interference of the $\ket{\pm \al}$ states and a destructive interference of the $\ket{\pm i \al}$ states. In general, these constructive/destructive interferences are imperfect, since the two logical states differ in their normalization factors. Regardless, it is helpful to define the \textit{two-legged cat states} as in Eq. \ref{eqn:2cats}. These are the famous Schr\"{o}dinger cat states with disjoint support over even or odd Fock states respectively. Note that the states $\cat{\pm}$ are aligned along the $q$ quadrature for purely real $\al$ and along the $p$ quadrature for purely imaginary $\al$.
\begin{subequations}
    \begin{alignat}{3}
        \tcat{0} &\equiv \frac{\coh{0} + \coh{2}}{2 \sqrt{e^{-|\al|^2} \cosh |\al|^2}} 
        && = \frac{1}{\sqrt{\cosh |\al|^2}} \sum_k \frac{\al^{2k}}{\sqrt{(2k)!}} \ket{2k} \\
        \tcat{1} &\equiv \frac{\coh{0} - \coh{2}}{2 \sqrt{e^{-|\al|^2} \sinh |\al|^2}}
        && = \frac{1}{\sqrt{\sinh |\al|^2}} \sum_k \frac{\al^{2k + 1}}{\sqrt{(2k + 1)!}} \ket{2k + 1} 
    \end{alignat}
    \label{eqn:2cats}
\end{subequations}

\begin{figure}[h]
    \centering
    \includegraphics[width=1\textwidth]{{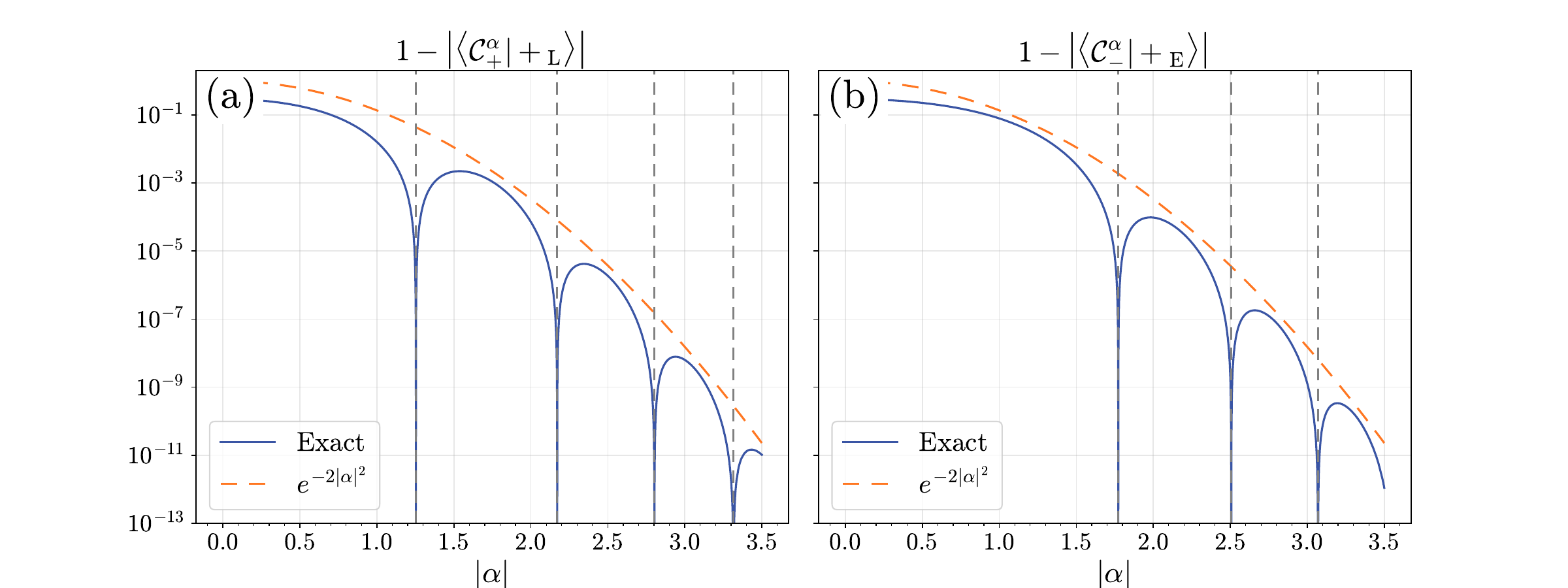}}
    \caption{The infidelity in defining the two-legged cat states as the X-basis states in the (a) logical and (b) error subspaces. This infidelity is defined as one minus the overlap between the corresponding X-basis states with the (a) even or (b) odd two-legged cat state. The exact infidelity (solid blue) is bounded by $e^{-2|\al|^2}$ (dashed orange). The dashed vertical gray lines indicate the values of $|\al|$ for which the X-basis states exactly equal the two-legged cat states. These values are given by (a) $|\al| = \sqrt{(n + \thalf) \pi}$ or (b) $|\al| = \sqrt{n \pi}$, respectively, for non-negative integers $n$.}
    \label{fig:cat4_xbasis_ovlp}
\end{figure}

We can express the logical-X basis states of the 4C code in terms of the two-legged cat states, as in Eq. \ref{eqn:4cat_even_xbasis}. 
\begin{align}
    \begin{split}
        \ket{\pm_\fC} &\equiv \frac{\cat{0} \pm \cat{2}}{\sqrt{2}} \\
        &= \half \parens{ \sqrt{\frac{\cosh |\al|^2}{\cosh |\al|^2 + \cos |\al|^2}} \pm \sqrt{\frac{\cosh |\al|^2}{\cosh |\al|^2 - \cos |\al|^2}} } \tcat{0}
        + \half \parens{ \sqrt{\frac{\cosh |\al|^2}{\cosh |\al|^2 + \cos |\al|^2}} \mp \sqrt{\frac{\cosh |\al|^2}{\cosh |\al|^2 - \cos |\al|^2}} } \ticat{0}
        \label{eqn:4cat_even_xbasis}
    \end{split}
\end{align}
Evidently, when $\cos |\al|^2 = 0$ (or $|\al| = \sqrt{\parens{n + \half} \pi}$ for non-negative integers $n$) the X-logical states $\ket{\pm_\rmL}$ exactly align with $\tcat{0}$ and $\ticat{0}$ respectively. For these values of $|\al|$ the $\tcat{0}$ and $\ticat{0}$ are perfectly orthogonal. Fig. \ref{fig:cat4_xbasis_ovlp}(a) shows the exact overlap error between the $\ket{+_\rmL}$ state and the $\tcat{0}$ two-legged cat state; we may observe that this error is always bounded by $e^{-2 {|\al|^2}}$.

Likewise, the X-basis of the error states are approximately the two-legged cats states $\tcat{1}$ and $\ticat{1}$, albeit with odd parity.  
\begin{align}
    \begin{split}
        \ket{\pm_{\text{4C, E}}} &\equiv \frac{\cat{3} \pm \cat{1}}{\sqrt{2}} \\
        &= \half \parens{ \sqrt{\frac{\sinh |\al|^2}{\sinh |\al|^2 - \sin |\al|^2}} \pm \sqrt{\frac{\sinh |\al|^2}{\sinh |\al|^2 + \sin |\al|^2}} } \tcat{1}
        + \frac{i}{2} \parens{ \sqrt{\frac{\sinh |\al|^2}{\sinh |\al|^2 - \sin |\al|^2}} \mp \sqrt{\frac{\sinh |\al|^2}{\sinh |\al|^2 + \sin |\al|^2}} } \ticat{1}
        \label{eqn:4cat_odd_xbasis}
    \end{split}
\end{align}
The error states in the X-basis $\ket{\pm_\rmE}$ exactly align with $\tcat{1}$ and $\ticat{1}$ when $\sin |\al|^2 = 0$ (or $|\al| = \sqrt{n \pi}$ for non-negative integers $n$). These are also the values of $|\al|$ for which $\tcat{1}$ and $\ticat{1}$ are perfectly orthogonal. Fig. \ref{fig:cat4_xbasis_ovlp}(b) shows the exact overlap between the $\ket{+_\rmE}$ state and the $\tcat{1}$ two-legged cat state; again we observe that this error is always bounded by $e^{-2 {|\al|^2}}$.

\subsection{Mean photon number}
\label{app:four_cat_nbar}
\begin{figure}[h]
    \centering
    \includegraphics[width=1\textwidth]{{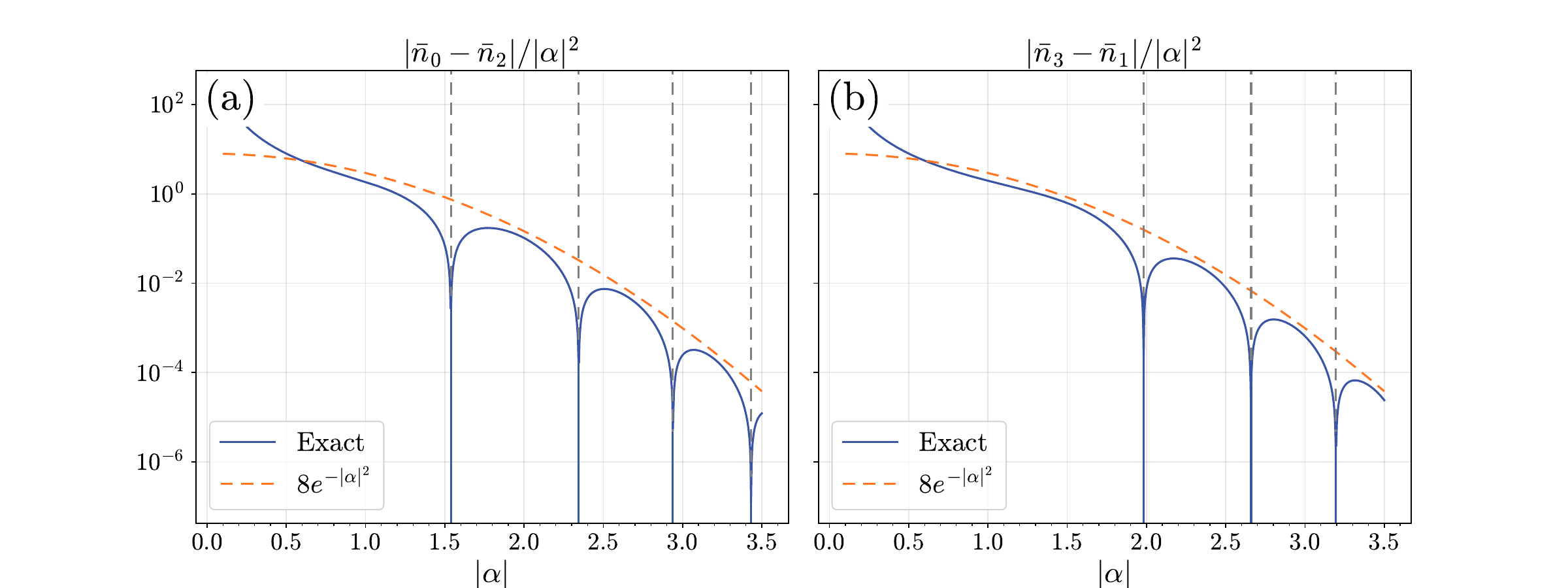}}
    \caption{The mean photon number for the (a) logical (i.e. even) and (b) error (i.e. odd) basis states, normalized by $|\al|^2$. Observe that for large $|\al|$, the mean photon number for all the states quickly approach $|\al|^2$. The vertical dashed gray lines indicate the values of $|\al|$ for which the (a) logical and (b) error basis states have exactly equal mean photon number. These values are numerical solutions to the transcendental equations in Eqn.\ref{4cat_nbar_equality_a} and \ref{4cat_nbar_equality_b} respectively.}
    \label{fig:cat4_nbar}
\end{figure}

Eq. \ref{eqn:4cat_nbar} gives the mean number of photons in each of the 4C states, as a function their ``size'', $|\al|$. 
\begin{subequations}
    \begin{alignat}{3}
        \nbar_0(\al) &\equiv \bracat{0} \opaDaga \cat{0} &&= |\al|^2 \ \, \frac{\sinh |\al|^2 - \sin |\al|^2}{\cosh |\al|^2 + \cos |\al|^2} \\
        \nbar_1(\al) &\equiv \bracat{1} \opaDaga \cat{1} &&= |\al|^2 \ \, \frac{\cosh |\al|^2 + \cos |\al|^2}{\sinh |\al|^2 + \sin |\al|^2} \\
        \nbar_2(\al) &\equiv \bracat{2} \opaDaga \cat{2} &&= |\al|^2 \ \, \frac{\sinh |\al|^2 + \sin |\al|^2}{\cosh |\al|^2 - \cos |\al|^2} \\
        \nbar_3(\al) &\equiv \bracat{3} \opaDaga \cat{3} &&= |\al|^2 \ \, \frac{\cosh |\al|^2 - \cos |\al|^2}{\sinh |\al|^2 - \sin |\al|^2}
    \end{alignat}
    \label{eqn:4cat_nbar}
\end{subequations}
In the limit of large $|\al|$, the mean photon number for all four cat states scales as $|\al|^2$. Fig. \ref{fig:cat4_nbar} shows the mean photon number for the (a) logical (i.e. even) and (b) error (i.e. odd) basis states, normalized by $|\al|^2$. This figure shows that the mean photon numbers exponentially approach each other with increasing $|\al|$.

In the presence of photon loss, the infidelity due to the no-jump evolution is minimized when the logical (error) states have an equal mean photon number. The values of $|\al|$ for which these are exactly equal is given by the following transcendental equations,
\begin{subequations}
    \begin{alignat}{3}
        \nbar_0(\al) - \nbar_2(\al) = 0 \qquad \Longleftrightarrow \qquad \tan |\al|^2 + \tanh |\al|^2 = 0 \label{4cat_nbar_equality_a}\\
        \nbar_1(\al) - \nbar_3(\al) = 0 \qquad \Longleftrightarrow \qquad \tan |\al|^2 - \tanh |\al|^2 = 0 \label{4cat_nbar_equality_b}
    \end{alignat}
\end{subequations}

\subsection{Choosing a cat size}
\label{app:alpha_choice}
For our numerical simulations, we $\al = \sqrt{2.5\pi} \approx 2.802$. Since $\cos(|\al|^2) = 0$, the logical-X states are exactly the two-legged cat states, $\ket{\pm_\rmL} = \cat{\pm}$. Furthermore, as we shall see in Appendix \ref{app:4c_bs}, $\cos(|\al|^2) = 0$ also implies that the true beamsplitter output is exactly the expected output in Table \ref{tbl:even_cat_bell}. 

The equal-mean photon number condition is approximately satisfied at this value of $|\al|$. Specifically, $\bar{n}_0 \approx 7.848$ and $\bar{n}_2 \approx 7.860$, such that $\bar{n}_2 - \bar{n}_0 \approx 0.012$.

%% file: tex_appendix/4c_bs.tex
\section{Evolving the 4C states through a beamsplitter}
\label{app:4c_bs}
Recall that the state preparation scheme probabilistically prepares any of the four 4C states. In the main text, we addressed fusions between two even 4C states. Sec. \ref{app:4c_bs_even_even} elaborates on this case by providing the exact expressions for the corresponding output states. In Sec. \ref{app:4c_bs_odd_odd}, we consider the case where the input states are odd 4C states. In both cases, we may distinguish between the Bell states by measuring properties of the photon number ($\op{n} = 0$)? and ($\op{n} \mod 4 = 0,1,2,$ or $3$)? on each resonator individually. To derive the expressions below, we use the simple fact that a symmetrized beamsplitter acting on a pair of coherent states results in a common and differential output: $\text{BS}(\tfrac{\pi}{2}) \Big[ \ket{\alpha} \otimes \ket{\alpha'} \Big] = \ket{\tfrac{1}{\sqrt{2}}(\alpha - \alpha')} \otimes \ket{\tfrac{1}{\sqrt{2}}(\alpha + \alpha')}$.

\subsection{Exact expressions for an even-even input}
\label{app:4c_bs_even_even}
Observe that choosing a cat size such that $\cos(|\al|^2) = 0$ implies that the true beamsplitter output is exactly the expected output, shown in Table \ref{tbl:even_cat_bell}.

\begin{subequations}
    \begin{align}
        \begin{split}
            \frac{\cat{0}\cat{0} + \cat{2}\cat{2}}{\sqrt{2}}  
            &\ce{->[BS]}
            \frac{\sqrt{2} \cosh(|\al|^2) \sqrt{\cosh(2 |\al|^2) + \cos(2|\al|^2)}}{\cosh(2|\al|^2) - \cos(2 |\al|^2)} \left( \frac{\ket{0}\bigcat{0} + \bigcat{0}\ket{0}}{\sqrt{2}} \right) \\
            &\qquad \qquad + \frac{\cos(|\al|^2)}{\sqrt{2}} \left[ \frac{\catB{0} \catB{0}}{\cos(|\al|^2) - \cosh(|\al|^2)} + \frac{\catB{2} \catB{2}}{\cos(|\al|^2) + \cosh(|\al|^2)}  \right] \\
            &\approx \left( \frac{\ket{0}\bigcat{0} + \bigcat{0}\ket{0}}{\sqrt{2}} \right) - 2 e^{-|\al|^2} \cos(|\al|^2) \left( \frac{\catB{0} \catB{0} + \catB{2} \catB{2}}{\sqrt{2}} \right) + \mcO\parens{e^{-2|\al|^2}}
        \end{split}\\[10pt]
        \begin{split}
            \frac{\cat{0}\cat{2} + \cat{0}\cat{2}}{\sqrt{2}}  
            &\ce{->[BS]}
            \frac{\ket{0}\bigcat{2} + \bigcat{2}\ket{0}}{\sqrt{2}}
        \end{split}\\[10pt]
        \begin{split}
            \frac{\cat{0}\cat{0} - \cat{2}\cat{2}}{\sqrt{2}}  
            &\ce{->[BS]}
            \frac{\sqrt{2} \cos(|\al|^2) \sqrt{\cosh(2 |\al|^2) + \cos(2|\al|^2)}}{\cosh(2|\al|^2) - \cos(2 |\al|^2)} \left( \frac{\ket{0}\bigcat{0} + \bigcat{0}\ket{0}}{\sqrt{2}} \right) \\
            &\qquad \qquad + \frac{\cosh(|\al|^2)}{\sqrt{2}} \left[ \frac{\catB{0} \catB{0}}{\cosh(|\al|^2) - \cos(|\al|^2)} - \frac{\catB{2} \catB{2}}{\cosh(|\al|^2) + \cos(|\al|^2)}  \right] \\
            &\approx \left( \frac{\catB{0} \catB{0} - \catB{2} \catB{2}}{\sqrt{2}} \right) + \sqrt{2} e^{-|\al|^2} \cos(|\al|^2) \left( \catB{0} \catB{0} + \catB{2} \catB{2} \right. \\
            &\qquad \qquad \left. - \ket{0}\bigcat{0} - \bigcat{0}\ket{0} \right) + \mcO\parens{e^{-2|\al|^2}}
        \end{split}\\[10pt]
        \begin{split}
            \frac{\cat{0}\cat{2} - \cat{2}\cat{0}}{\sqrt{2}}  
            &\ce{->[BS]}
            \frac{i \sqrt{2}}{\sqrt{\cosh(2|\al|^2) - \cos(2 |\al|^2)}} \left[ \sinh(|\al|^2) \frac{\catB{1} \catB{1} - \catB{3} \catB{3}}{\sqrt{2}} \right. \\
            &\qquad \qquad \left. + \sin(|\al|^2) \frac{\catB{1} \catB{1} + \catB{3} \catB{3}}{\sqrt{2}} \right] \\
            &\approx i \left( \frac{\catB{1} \catB{1} - \catB{3} \catB{3}}{\sqrt{2}} \right) + 2 i e^{-|\al|^2} \sin(|\al|^2) \left( \frac{\catB{1} \catB{1} + \catB{3} \catB{3}}{\sqrt{2}} \right) + \mcO\parens{e^{-2|\al|^2}}
        \end{split}
    \end{align}
\end{subequations}

\subsection{Exact expressions for an odd-odd input}
\label{app:4c_bs_odd_odd}
Evolving two odd cat states through a beamsplitter results in approximate states shown in Table \ref{tbl:odd_cat_bell}. The exact output states are given by Eq. \ref{eqn:odd_cat_bs}. Similar to the X-basis condition for a single odd-cat, the exact beamsplitter output coincides with expected output (Table \ref{tbl:odd_cat_bell}) when $\sin(|\al|^2) = 0$.

\begin{table}[h]
    \centering
    \begin{tabular}{||c | c || c | c | c | c ||} 
    \hline 
    $XX$ & $ZZ$ & Bell-states & After 50:50 B.S. & After 50:50 B.S. $+ \ \opa$ & After 50:50 B.S. $+ \ \opb$ \rule{0pt}{3ex} \\[3pt]
    \hline\hline 
    $+1$ & $+1$ & $\cat{3}\cat{3} + \cat{1}\cat{1}$ & $\ket{0}\bigcat{2} - \bigcat{2}\ket{0}$ & $\bigcat{1}\ket{0}$ & $\ket{0}\bigcat{1}$ \rule{0pt}{4ex} \\[15pt]
    $+1$ & $-1$ & $\cat{3}\cat{1} + \cat{1}\cat{3}$ & $\ket{0}\bigcat{0} - \bigcat{0}\ket{0}$ & $\bigcat{3}\ket{0}$ & $\ket{0}\bigcat{3}$ \\[15pt]
    $-1$ & $+1$ & $\cat{3}\cat{3} - \cat{1}\cat{1}$ & $\catB{0}\catB{2} - \catB{2}\catB{0}$ & $\catB{3}\catB{2} - \catB{1}\catB{0}$ & $\catB{0}\catB{1} - \catB{2}\catB{3}$\\[15pt]
    $-1$ & $-1$ & $\cat{3}\cat{1} - \cat{1}\cat{3}$ & $\catB{1}\catB{3} - \catB{3}\catB{1}$ & $\catB{0}\catB{3} - \catB{2}\catB{1}$ & $\catB{1}\catB{0} - \catB{3}\catB{0}$ \\[10pt]
    \hline
    \end{tabular}
    \caption{Bell measurements on the error (odd) cat states, $\ket{0_\rmE} \equiv \cat{3}, \ket{1_\rmE} \equiv \cat{1}$. The output states are approximately equal to the ideal output states, up to $\mcO\parens{e^{-|\al|^2}}$. Just as before, we may distinguish betweeen the Bell states by measuring properties of the photon number ($\op{n} = 0$)? and ($\op{n} \mod 4 = 0,1,2,$ or $3$)? on each resonator individually. }
    \label{tbl:odd_cat_bell}
\end{table}

\begin{subequations}
    \begin{align}
        \begin{split}
            \frac{\cat{3}\cat{3} + \cat{1}\cat{1}}{\sqrt{2}}  
            &\ce{->[BS]}
            \frac{\sqrt{2 \left[ \cosh(2 |\al|^2) - \cos(2|\al|^2) \right]}}{\cosh(2|\al|^2) + \cos(2 |\al|^2) - 2} \left[ \sinh(|\al|^2) \left( \frac{\ket{0}\bigcat{2} - \bigcat{2}\ket{0}}{\sqrt{2}} \right) \right. \\
            &\qquad \qquad \left. + i \sin(|\al|^2) \left( \frac{\cat{0}\cat{2} - \cat{2}\cat{0}}{\sqrt{2}} \right) \right] \\
            &\approx \left( \frac{\ket{0}\bigcat{2} - \bigcat{2}\ket{0}}{\sqrt{2}} \right) + i 2 e^{-|\al|^2} \sin(|\al|^2) \left( \frac{\cat{0}\cat{2} - \cat{2}\cat{0}}{\sqrt{2}} \right) + \mcO\parens{e^{-2|\al|^2}}
        \end{split}\\[10pt]
        \begin{split}
            \frac{\cat{3}\cat{1} + \cat{1}\cat{3}}{\sqrt{2}}  
            &\ce{->[BS]}
            \sqrt{1 + \frac{2}{\cos(2 |\al|^2) + \cosh(2 |\al|^2) - 2}} \left( \frac{\ket{0}\bigcat{0} - \bigcat{0}\ket{0}}{\sqrt{2}} \right)
        \end{split}\\[10pt]
        \begin{split}
            \frac{\cat{3}\cat{3} - \cat{1}\cat{1}}{\sqrt{2}}  
            &\ce{->[BS]}
            \frac{\sqrt{2 \left[ \cosh(2 |\al|^2) - \cos(2|\al|^2) \right]}}{\cosh(2|\al|^2) + \cos(2 |\al|^2) - 2} \left[ \sin(|\al|^2) \left( \frac{\ket{0}\bigcat{2} - \bigcat{2}\ket{0}}{\sqrt{2}} \right) \right. \\
            &\qquad \qquad \left. + i \sinh(|\al|^2) \left( \frac{\cat{0}\cat{2} - \cat{2}\cat{0}}{\sqrt{2}} \right) \right] \\
            &\approx i \left( \frac{\cat{0}\cat{2} - \cat{2}\cat{0}}{\sqrt{2}} \right) + 2 e^{-|\al|^2} \sin(|\al|^2) \left( \frac{\ket{0}\bigcat{2} - \bigcat{2}\ket{0}}{\sqrt{2}} \right) + \mcO\parens{e^{-2|\al|^2}}
        \end{split}\\[10pt]
        \begin{split}
            \frac{\cat{0}\cat{2} - \cat{2}\cat{0}}{\sqrt{2}}  
            &\ce{->[BS]}
            \frac{\cat{3}\cat{1} - \cat{1}\cat{3}}{\sqrt{2}}
        \end{split}
    \end{align}
    \label{eqn:odd_cat_bs}
\end{subequations}

%% file: tex_appendix/bell.tex
\section{Adaptive sequence for fusion measurements}
\label{app:bell_msmt}
\begin{figure}[h]
    \centering
    \includegraphics[width=0.78\textwidth]{{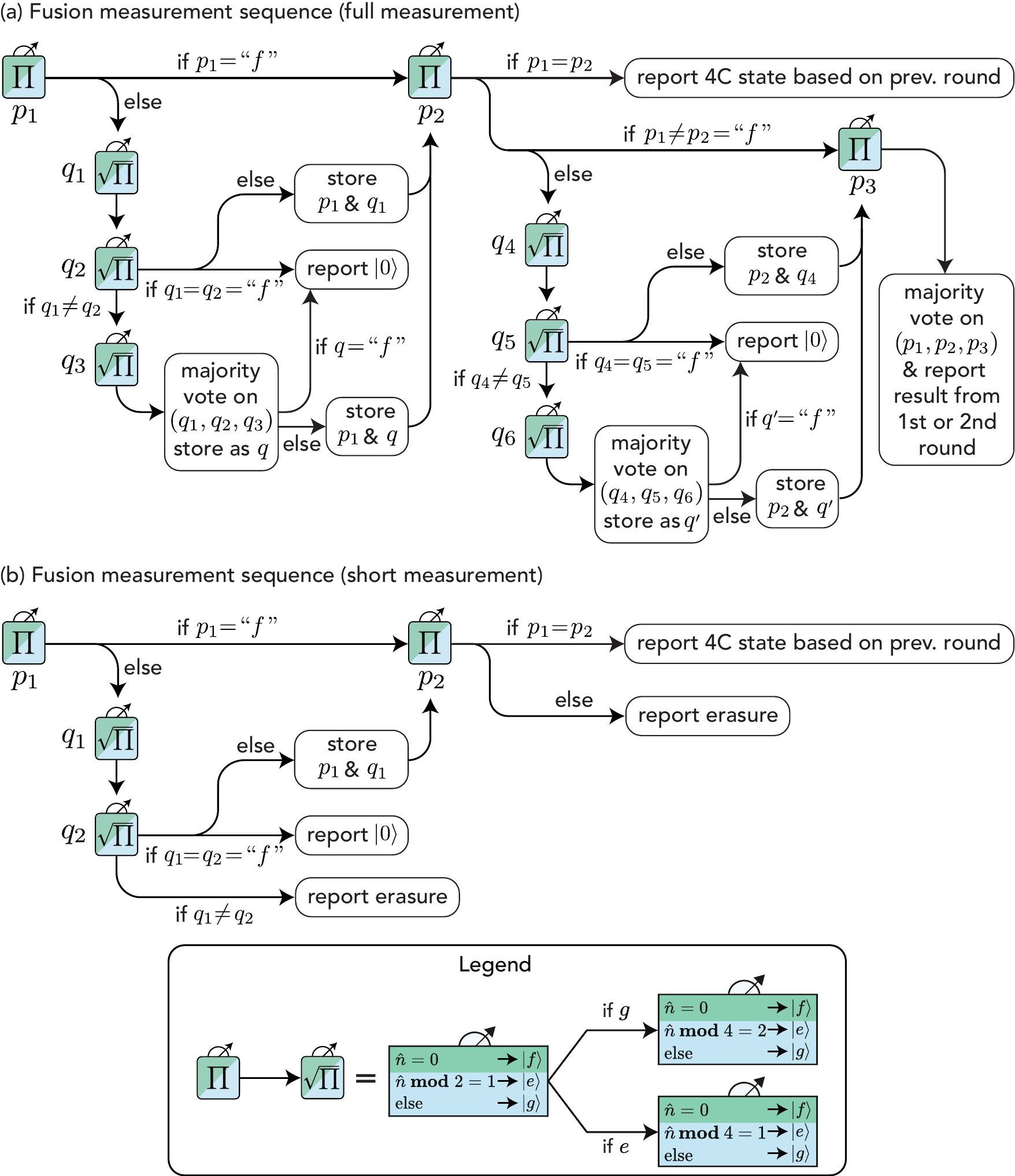}}
    \caption{Flowchart of the adaptive sequence for fusion measurement. The sequence starts with a parity measurement $\hat{\Pi}$, followed by a 4-parity measurement $\hat{4\Pi}$. These measurements are repeated, and results are only reported if two measurements agree with each other. (a) shows the complete sequence where a third measurement is performed to break ties between disagreeing measurements, while (b) shows a shorter sequence where disagreeing measurements are reported as erasures.}
    \label{fig:fusions_scheme}
\end{figure}

Fusion measurements on the 4C code involve evolving two 4C states through a beamsplitter, followed by a sequence of PNMs. Each PNM is performed by dispersively coupling a bosonic mode to a three-level ancilla, measuring the ancilla in the computational basis, and inferring the parity or photon number modulo four (4-parity) of the bosonic mode. Figure \ref{fig:fusions_scheme} illustrates the adaptive sequences used to identify whether each bosonic mode contains a 4C state or the vacuum. Since the sequences are identical for both modes, only the sequence for one of the modes is shown. These sequences consist of parity ($\Pi$) and 4-parity ($\sqrt{\Pi}$) measurements, where each step depends on prior results. Measurements are repeated for consistency, with two approaches for handling disagreements: a tie-breaking third measurement or reporting disagreements as erasures.


%% file: tex_appendix/resource_state_prep.tex
\section{Circuits to prepare resource states}
\label{app:general_resource_prep}
\begin{figure}[h]
    \centering
    \includegraphics[width=0.9\textwidth]{{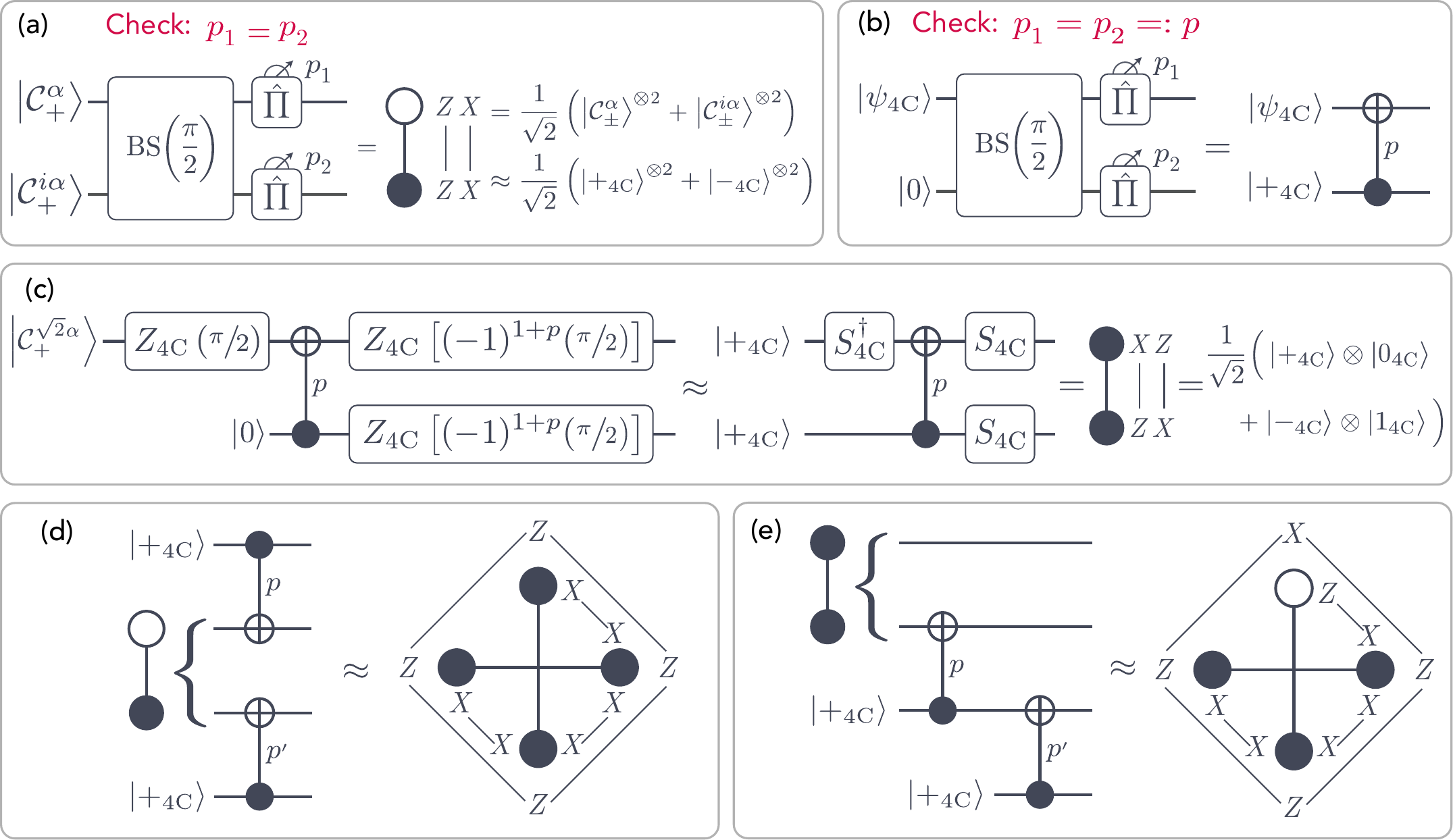}}
    \caption{Circuits for Bell-state preparation and resource state construction. (a) Circuit to approximately prepare a canonical Bell state, stabilized by $Z \otimes Z$ and $X \otimes Z$. This circuit, like all others in this figure, can detect single-photon loss through the parity check $p_1 = p_2$. (b) CNOT gadget where the control is always initialized in $\ket{+_\fC}$ and the target is an arbitrary 4C state. (c) Circuit to prepare a Hadamard-deformed Bell state, stabilized by $X \otimes Z$ and $Z \otimes X$. (d-e) Examples of circuits that combine the gadgets described above, to form four-body GHZ states. The respective stabilizers for both states are also shown.}
    \label{fig:general_resource_prep}
\end{figure}

In this section we look at circuits (shown in Figure \ref{fig:general_resource_prep}) to prepare entangled 4C states, that are not covered in the main text. Each subsection derives these circuits in detail. 

\subsection{Preparing cat states}
\label{app:prep_4cats}

The cat state preparation begins with displacing the bosonic mode to a coherent state, $\ket{\al}$. Next, we measure the parity operator, $\op{\Pi} = \exp(-i \pi \opaDag \opa)$, using a PNM conditioned on the odd photon numbers, i.e., $\text{PNM}^{gf}(\{1, 3, \dots, 2n+1\})$ \citeapp{ofek_2016}. Let the parity measurement result be $p \in \{0, 1\}$. The post-measurement state becomes:
\begin{align}
    \frac{\opI_{\text{bm}} + (-1)^p \, \op{\Pi}}{2} \ket{\al} = \frac{\opI_{\text{bm}} + (-1)^p \, e^{-i \pi \opaDag \opa}}{2} \ket{\al} \propto \ket{\al} + (-1)^p \ket{-\al} = \cat{\pm} \approx \ket{+_\fC} \text{ OR } \ket{+_\fCE}
\end{align}
This forms an even or odd cat state $\cat{\pm}$ based on $p$, which approximates the $\ket{+_\fC}$ (codeword) or $\ket{+_\fCE}$ (error-word) in the 4C code. Similarly, $\icat{\pm}$ states can be prepared by initially displacing the bosonic mode to $\ket{i\al}$.

Next, we form the 4C states by measuring the 4-parity operator, $\op{\sqrt{\Pi}} = \exp(-i \thalf \pi \opaDag \opa)$ \citeapp{ofek_2016}, the measurement is conditioned on the prior parity result. For even ($p=0$) or odd ($p=1$) outcomes, the respective measurement uses $\sqrt{\Pi} = \text{PNM}^{ge}(\{2, 6, \dots, 4n+2\})$ or $\sqrt{\Pi}' = \text{PNM}^{ge}(\{1, 5, \dots, 4n+1\})$. Let the 4-parity result be $q \in \{0, 1\}$. Assuming both measurements yield even results ($p=0, q=0$), the post-measurement state is:
\begin{align}
    \frac{\opI_{\text{bm}} + \op{\sqrt{\Pi}}}{2} \left[ \ket{\al} + \ket{-\al} \right] = \frac{\opI_{\text{bm}} + e^{-i \thalf \pi \opaDag \opa}}{2} \left[ \ket{\al} + \ket{-\al} \right]  \propto \ket{\al} + \ket{i \al} + \ket{- \al} + \ket{-i \al} \propto \cat{0} = \ket{0_\fC}
\end{align}
Other measurement outcomes correspond to different cat states, $\cat{{2q + p}}$. In this section, the entangled 4C states are prepared using even cat states as inputs, but similar results hold for odd inputs, as the parity can be tracked and adjusted in software \citeapp{ofek_2016}.

To ensure 1st-order FT, we repeat the measurement sequence twice, i.e., $\op{\Pi} \rightarrow \op{\sqrt{\Pi}} \rightarrow \op{\Pi} \rightarrow \op{\sqrt{\Pi}}$. To proceed, we require the parity and 4-parity measurements to agree; otherwise, the bosonic mode is reset, and the process is retried. This protocol mitigates ancilla errors and photon loss, ensuring 1st-order FT, similar to the GHZ state preparation circuit (see \ref{sec:resource_states}).

\subsection{Bell-state preparation}
\label{app:prep_bell}
Fig. \ref{fig:general_resource_prep}(a) illustrates the circuit used to approximately prepare the canonical Bell state, $\tfrac{1}{\sqrt{2}} \left( \ket{+_\fC}^{\otimes 2} + \ket{-_\fC}^{\otimes 2} \right)$. While this state is not directly used in our protocol, the circuit demonstrates three key principles underlying our approach. 

Firstly, we apply a symmetrized beamsplitter acting on a pair of coherent states. 
\begin{align}
    \tilde{\text{BS}}\parens{\frac{\pi}{2}} \Big[ \ket{\al} \ket{\al'} \Big] = \exp[-i \frac{\pi}{4} \parens{\opaDag \opb + \opbDag \opa}] \Big[ \ket{\al} \ket{\al'} \Big] \propto \ket{\tfrac{1}{\sqrt{2}} (\al - i \al')} \ket{-i \tfrac{1}{\sqrt{2}} (\al + i \al')}.
\end{align}
Note that we have used a beamsplitter Hamiltonian with a different phase compared to the definition in the main text. However, this phase can easily be adjusted in hardware \citeapp{yao_lu_2023, chapman_2023}.

Second, parity measurements on coherent states form 2C states, as shown in App. \ref{app:prep_4cats}. For simplicity, we assume even parity measurement outcomes ($p_1 = p_2 = 0$), as odd outcomes result in odd-parity states that can be tracked and adjusted in software. Finally, as discussed in Sec. \ref{sec:resource_states}, the beamsplitter conserves the joint-photon number parity. This enables the check $p_1 = p_2$, which can detect a single-photon loss. If this check is violated, the modes are reset, and the circuit is retried.
\begin{subequations}
\begin{align}
    \cat{+} \cat{+} &\propto \Big[ \ket{\al} + \ket{-\al} \Big] \Big[ \ket{\al} + \ket{-\al} \Big] \\
    &\ce{->[$\tilde{\text{BS}}$]} \ket{\tfrac{1}{\sqrt{2}} (1-i) \al} \ket{\tfrac{1}{\sqrt{2}} (1-i) \al} + \ket{- \tfrac{1}{\sqrt{2}} (1-i) \al} \ket{- \tfrac{1}{\sqrt{2}} (1-i) \al} \nonumber \\ 
    &\qquad \qquad + \ket{-i \tfrac{1}{\sqrt{2}} (1-i) \al} \ket{i \tfrac{1}{\sqrt{2}} (1-i) \al} + \ket{i \tfrac{1}{\sqrt{2}} (1-i) \al} \ket{-i \tfrac{1}{\sqrt{2}} (1-i) \al}\\
    &= \ket{\beta}\ket{\beta} + \ket{-\beta}\ket{-\beta} + \ket{i \beta}\ket{-i\beta} + \ket{-i \beta}\ket{i\beta} \\
    &\ce{->[Parity meas.]} \Big( \ket{\beta} + \ket{-\beta} \Big)^{\otimes 2} + \Big( \ket{i \beta} + \ket{- i\beta} \Big)^{\otimes 2}
\end{align}
\end{subequations}
Here we have defined $\beta = \tfrac{1}{\sqrt{2}} (1+i) \al$, and ignored constant prefactors. To complete the derivation, we recall that resulting 2C states ($\cat{+}$ and $\icat{+}$) are approximately equal to the logical-X eigenstates ($\ket{\pm_\fC}$ respectively), up to an exponentially small error $\veps \propto e^{-|\al|^2}$. 

The Bell state can also be prepared using the CNOT gadget discussed in the next subsection.

\subsection{CNOT gadget}
\label{app:prep_cnot}
This section discusses the CNOT gadget shown in Fig. \ref{fig:general_resource_prep}(b). This gadget implements a CNOT gate with a control qubit initialized in the $\ket{+_\fC}$ state and a target qubit in an arbitrary 4C code state, $\ket{\psi_\fC}$. The input $\ket{\psi_\fC}$ must be initialized with twice the desired energy (i.e. with a cat size $\sqrt{2}\alpha$) since the beamsplitter evenly distributes the energy between the two modes.

In this circuit, we revert to the beamsplitter phase convention used in the main text, where $\text{BS}(\tfrac{\pi}{2}) \Big[ \ket{\alpha} \ket{\alpha'} \Big] = \ket{\tfrac{1}{\sqrt{2}} (\alpha - \alpha')} \ket{\tfrac{1}{\sqrt{2}} (\alpha + \alpha')}$. 

To see how this gadget works, let us consider the input $\ket{\psi_\fC} = \ket{0_\fC}$. 
\begin{align}
    \bigcat{0} \ket{0} &\propto \Big[ \ket{\sqrt{2} \al} + \ket{i \sqrt{2} \al} + \ket{-\sqrt{2} \al} + \ket{-i \sqrt{2} \al} \Big] \ket{0} \\
    &\ce{->[BS]} \ket{\al}\ket{\al} + \ket{i \al}\ket{i \al} + \ket{-\al}\ket{-\al} + \ket{-i \al}\ket{-i \al} \\
    &\ce{->[Parity meas.]} \Big[ \ket{\alpha} + (-1)^p \ket{-\alpha} \Big]^{\otimes 2} + \Big[ \ket{i \alpha} + (-1)^p \ket{- i\alpha} \Big]^{\otimes 2},
\end{align}
where $p$ denotes the parity measurement result. The output is approximately equal to $\tfrac{1}{\sqrt{2}} \left( \ket{+_\fC}^{\otimes 2} + \ket{-_\fC}^{\otimes 2} \right)$, providing an alternate way of creating a Bell state.

We can similarly analyze the input $\ket{\psi_\fC} = \ket{1_\fC}$:
\begin{align}
    \bigcat{2} \ket{0} &\propto \Big[ \ket{\sqrt{2} \al} - \ket{i \sqrt{2} \al} + \ket{-\sqrt{2} \al} - \ket{-i \sqrt{2} \al} \Big] \ket{0} \\
    &\ce{->[BS]} \ket{\al}\ket{\al} - \ket{i \al}\ket{i \al} + \ket{-\al}\ket{-\al} - \ket{-i \al}\ket{-i \al} \\
    &\ce{->[Parity meas.]} \Big[ \ket{\alpha} + (-1)^p \ket{-\alpha} \Big]^{\otimes 2} - \Big[ \ket{i \alpha} + (-1)^p \ket{- i\alpha} \Big]^{\otimes 2}.
\end{align}
This approximately corresponds to the output $\tfrac{1}{\sqrt{2}} \parens{\ket{+_\fC}^{\otimes 2} - \ket{-_\fC}^{\otimes 2}}$. Together, these results confirm the implementation of the desired CNOT operation, with the control qubit set to $\ket{+_\fC}$. Just as before. this gadget enables a single photon loss to be detected, by confirming $p_1 = p_2 =: p$. As shown in Fig. \ref{fig:general_resource_prep}(b), we record the measurement $p$ next to the CNOT symbol because it could be used in future operations. 

In Sec. \ref{sec:resource_states}, this gadget is utilized to generate GHZ states. As discussed, the beamsplitting angle can be modified to unequally divide the energy between the modes while preserving the logical operation on the 4C code.

\subsection{Hadamard-deformed Bell-state preparation}
\label{app:prep_hadamard}
As illustrated in Fig. \ref{fig:general_resource_prep}(c), the CNOT gadget can be used to prepare a deformed Bell state. This state differs from the canonical Bell state by the application of a Hadamard gate on one of the qubits, resulting in a state stabilized by the operators $X \otimes Z$ and $Z \otimes X$. 

The circuit can be understood as the CNOT gadget conjugated by $S_\fC$ gates. In the 4C code, the $S_\fC$ gate, defined as $Z_\fC(- \tfrac{\pi}{2})$, is implemented using SNAP gates, as detailed in Sec. \ref{sec:non_clifford}. A key feature of the CNOT gadget is that the output parity is random; however, since the parity measurement result, $p$, is known, the subsequent SNAP gates can be adjusted accordingly. Finally, note that the approximation in Fig. \ref{fig:general_resource_prep}(c) arises from the fact that the 2C states are approximately equal to the X-logical states in the 4C code.

As discussed in Sec. \ref{sec:core_ops_single}, \ref{sec:resource_states}, and \ref{sec:non_clifford}, SNAP gates can detect first-order hardware errors, ensuring consistency with the capabilities of other gadgets in the circuit.

\subsection{Preparing four-body GHZ states}
\label{app:prep_other}
Figs. \ref{fig:general_resource_prep}(d) and (e) depict circuits that use Bell-state preparation and CNOT gadgets to create four-body GHZ states. These examples highlight the importance of carefully managing the energy distribution among modes to ensure the final states have the desired size. While it is natural to conceptualize state preparation as logical operations on the 4C code, adjustments to the beamsplitting angles and initial state sizes are crucial for achieving the correct photon distribution.

The circuit in Fig. \ref{fig:general_resource_prep}(d) generates the four-body GHZ state, $\tfrac{1}{\sqrt{2}} \parens{\ket{+_\fC}^{\otimes 4} + \ket{-_\fC}^{\otimes 4}}$. The initial Bell state can be prepared using the method shown in either Fig. \ref{fig:general_resource_prep}(a) or (b), provided the state constitutes cat states of double the required size, $\sqrt{2} \alpha$. Symmetrized beamsplitters in the subsequent CNOT gadgets ensure equal photon distribution across all modes in the output state.

Fig. \ref{fig:general_resource_prep}(e) presents a circuit to prepare the state $\tfrac{1}{\sqrt{2}} \parens{\ket{+_\fC}\ket{+_\fC}\ket{+_\fC} \ket{0_\fC} + \ket{-_\fC}\ket{-_\fC}\ket{-_\fC} \ket{1_\fC}}$. In this case, the initial Bell state is prepared with an asymmetric beamsplitter to allocate three times as many photons to the second mode as the first. These photons are subsequently distributed evenly among the bottom three modes via CNOT gadgets. Specifically, the beamsplitting angles are configured as follows: the initial Bell state uses an angle of $\frac{2 \pi}{3}$, the first CNOT gadget uses an angle of $2 \arccos(\frac{1}{\sqrt{3}})$, and the second CNOT employs a symmetric beamsplitter with an angle of $\frac{\pi}{2}$.

\subsection{Resource states for 4-star FBEC}
\label{app:prep_4star}
In the main text, we examined FBEC using 6-ring resource states. An alternative approach involves utilizing four-body resource states, commonly referred to as 4-star FBEC \citeapp{bartolucci_2023, sahay_2023a}. The 4-star protocol is conventionally less favored compared to the 6-ring protocol for two interconnected reasons. First, implementing the same code distance with 4-star requires more physical qubits. Second, it demands a greater number of fusions per unit cell compared to the 6-ring approach \citeapp{bartolucci_2023, sahay_2023a}. These factors have led previous studies to report a lower QEC threshold for the 4-star protocol, albeit under different noise modes \citeapp{bartolucci_2023, sahay_2023a}.

\begin{figure}[h]
    \centering
    \includegraphics[width=0.6\textwidth]{{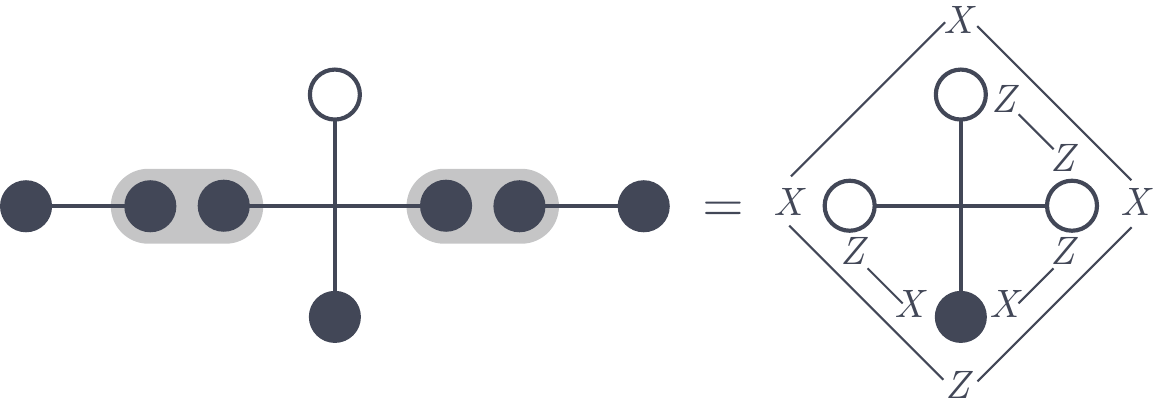}}
    \caption{Construction of one of the 4-star resource states using known resource states and fusions (gray ovals).}
    \label{fig:4star_resource}
\end{figure}

Nonetheless, we propose a method to construct the 4-star resource states, for the sake of completeness. One of these states is exactly the GHZ state depicted in Fig. \ref{fig:general_resource_prep}(d). The other can be generated by fusing the states shown in Fig. \ref{fig:general_resource_prep}(c) and (e), as illustrated by Fig. \ref{fig:4star_resource}. It is worth noting that constructing this resource state requires 8 physical qubits, further emphasizing the qubit inefficiency of this approach.

%% file: tex_appendix/fbec_no_jump.tex
\section{Accounting for the no-jump evolution}
\label{app:no_jump}
As discussed in Sec. \ref{sec:4cats}, the no-jump evolution of a lossy bosonic mode shrinks the size of the 4C states: $|\al(t)| = e^{-\half \bar{n} \kappa t} \alpha_0$, where $\kappa$ is the single-photon loss rate in the bosonic mode and $\alpha_0$ is the initial size. Although this effect contributes only a small correction to the overall error budget, it can be further suppressed by appropriately adjusting the beamsplitter angles during resource state preparation. 

To see how this works, consider the action of a beamsplitter on two coherent states: $\text{BS}(\theta) \Big[ \ket{\al} \ket{\beta} \Big] =  \ket{\cos(\tfrac{\theta}{2}) \al - \sin(\tfrac{\theta}{2})\beta} \ket{\sin(\tfrac{\theta}{2})\al + \cos(\tfrac{\theta}{2}) \beta}$. This is consistent with the phase convention used in the main text. 

In our resource state preparation, we typically interfere a larger cat state (e.g., $\Bigcat{0}$) in one mode with vacuum in the other. To compensate for amplitude decay over a short timescale $\kappa t < \ln(\sqrt{2})$, we can tune the beamsplitter angle to $\theta = 2 \arccos(e^{\kappa t}/\sqrt{2})$, ensuring that one of the output modes retains amplitude $\alpha$, while the other is slightly reduced to approximately $(1 - 2\kappa t)\alpha$.

This approach enables the construction of a 6-ring resource state using three ``shrunk'' cat states and three with full amplitude. Conveniently, the shrunk states can be fused with leftover cat states from the previous round of preparation, which will also have shrunk due to no-jump decay. As a result, the amplitudes of both states involved in the fusion are closely matched, improving the fidelity of the fusion step.

Meanwhile, the other half of the resource state (the cats with full amplitude) continue to evolve and shrink slightly before the next round. In the subsequent round they are fused with the newly prepared shrunk cats, maintaining the same cadence.

Assuming the bosonic modes have similar loss rates, this strategy produces a steady rhythm during which the beamsplitter angles remain largely fixed, minimizing the need for dynamic recalibration. 

%% file: tex_appendix/chi_prime.tex
\section{Accounting for \texorpdfstring{$\chi'$}{chi'}}
\label{app:chi-prime}
The dispersive coupling between a bosonic mode and an ancilla, such as a transmon, induces several undesired non-linearities. These include the self-Kerr $\half K \opa^{\dagger 2} \opa^2$ as well as a second-order correction to the dispersive shift, $\chi' \opa^{\dagger 2} \opa^2$. The self-Kerr term is a photon-number dependant phase shift, which is addressed in the main text. In this section, we address the $\chi'$ term, which is a photon-number dependant correction to the dispersive shift. 

Recall that the PNMs were implemented as weak drives on the ancilla at a set of frequencies $\{\omega_{ge} + n \chi_e | n \in \mcN \}$, where $\omega_{ge}$ is the ancilla's $g$-$e$ frequency and $\mcN$ is a set of photon-numbers. These drives conditionally flip the ancilla's state, depending on the number of photons in the resonator. The $\chi'_e$ term additionally shifts the ancilla's $g$-$e$ frequency by an amount proportional to the square of the photon-number. This shift can be accounted for by driving the ancilla at $\{\omega_{ge} + n \chi_e + \chi_e' n(n + 1) | n \in \mcN \}$ instead. Similarly, the drives on the $g$-$f$ manifold can be shifted by $\chi_f' n(n + 1)$. Since the PNMs a projective measurements on the mode's photon number, the photon-number dependant phase-shift induced by $\chi'$ is irrelevant.

SNAP gates, being two consecutive photon-number-selective rotations, are similarly modified. The additional photon-number-dependent phase shift is absorbed into the phases, $\vec{\phi}$ imparted the SNAP gate.

Similar to self-Kerr, the beamsplitter operation BS$(\frac{\pi}{2})$ is not FT to $\chi'$. However, the beamsplitter is a relatively fast operation \citeapp{chapman_2023,yao_lu_2023}, thus these non-linearities negligible contributes to the overall error. In contrast, we suspect that $\chi'$ during the slower $ZZ_\fC(\theta)$ gate may result in a meaningfully undetectable error. A thorough analysis of this error is left for future work.

%% file: tex_appendix/planar_fbec.tex
\section{Planar FBEC}
\label{app:planar_fbec}
In this Appendix, we outline a foliation scheme to construct an FBEC protocol from a 2D stabilizer code. While the final architecture is not inherently planar, it can often be adapted to achieve planarity. As an illustrative example, we focus on the $XZZX$ code, which is relevant to the main text.

\subsection{Teleportation chain}
\label{app:teleportation_chain}
\begin{figure}[ht]
    \centering
    \includegraphics[width=\textwidth]{{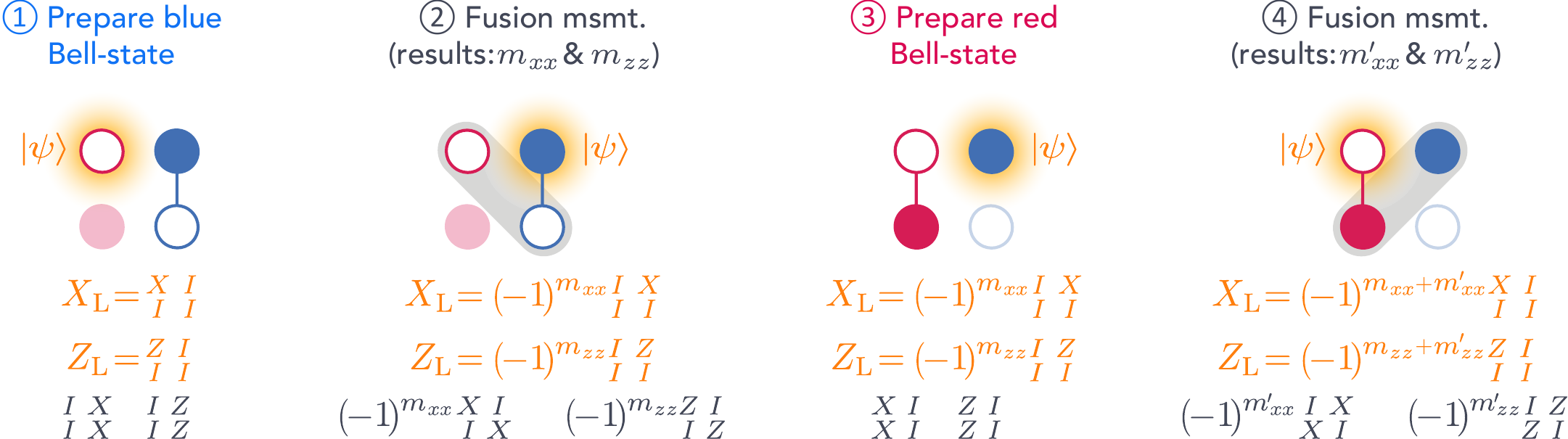}}
    \caption{A four-qubit teleportation chain can be understood by tracking the logical Pauli operators, $X_\rmL$ and $Z_\rmL$ (orange), and stabilizers (black) across four physical qubits. Initially, a single-qubit state $\ket{\psi}$ (orange glow) is encoded in a red qubit, with the blue qubits initialized in a Bell state. A fusion measurement between a red and blue qubit teleports $\ket{\psi}$ to the remaining blue qubit. Then both red qubits are reinitialized in a Bell state. This enables another fusion measurement to teleport $\ket{\psi}$ back to a red qubit, completing one cycle. These steps can be repeated to teleport $\ket{\psi}$ back-and-forth between the red and blue qubits. Note that, at each step, one qubit (faded) remains inactive, hinting that the scheme could be optimized (see App.\ref{app:fbec_optimization}).}
    \label{fig:teleportation}
\end{figure}

To introduce the foliation of a QEC code, we first examine the four qubit teleportation chain. As shown in Fig.~\ref{fig:teleportation}, a single-qubit state is teleported back-and-forth between red and blue Bell states via Bell measurements (fusions). This setup is a symmetrized version of the textbook 3-qubit teleportation circuit \citeapp{nielsen_chuang_2010}. Instead of tracking the quantum state, we analyze the stabilizers (black) and logical Pauli-$X$ and $Z$ operators (orange) of the 4-qubit system, demonstrating that the state is indeed teleported, up to with Pauli corrections dependent on the measurement outcomes.

The suboptimal four qubit version of teleportation is pedagogically convenient, when understanding the foliation protocol in the next subsection. However, as shown in App. \ref{app:fbec_optimization}, the number of qubits in the resulting FBEC architecture can be optimized, analogously to how this teleportation can be optimized from four to three qubits. 

\subsection{Quasi-planar foliation of the \texorpdfstring{$XZZX$}{XZZX} code}
\label{app:quasi_planar_foliation}

\begin{figure}[ht]
    \centering
    \includegraphics[width=1\textwidth]{{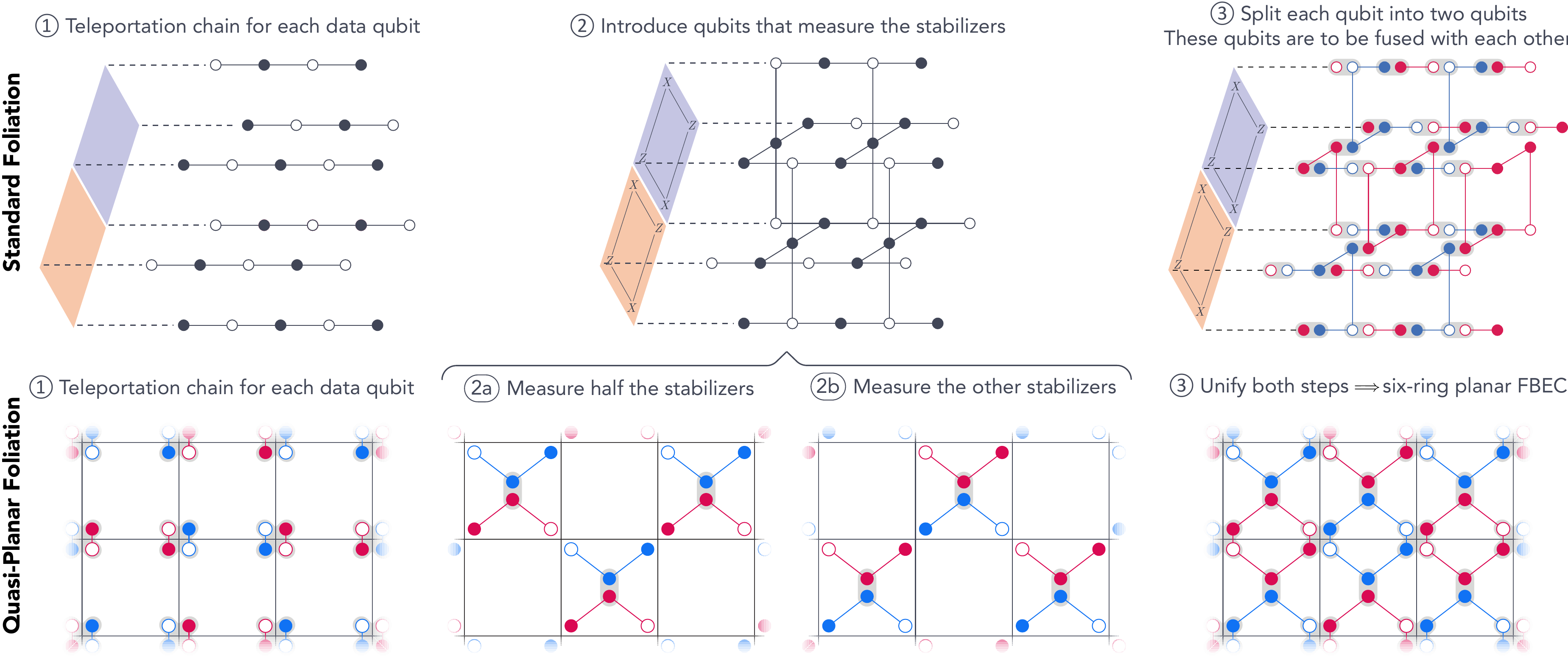}}
    \caption{Two methods for foliating the $XZZX$ code into a fusion-based architecture: (top row) the standard procedure~\protect\citeapp{claes_2023, sahay_2023a}, and (bottom row) our quasi-planar procedure. In the top row, the circuit-based $XZZX$ surface code is depicted as a shadow (on the left) of the cluster state, with data qubits on vertices and stabilizers on plaquettes. The standard foliation replaces each data qubit with a teleportation chain (steps 1 and 2), coupling these chains to measure stabilizers. The resulting cluster state is transformed into a fusion-based implementation by splitting each qubit into two (step 3)~\protect\citeapp{bartolucci_2023, sahay_2023a, bombin_2024}. In contrast, our procedure directly constructs a fusion-based architecture. The bottom row shows the circuit-based $XZZX$ surface code in the plane. First, each data qubit is replaced with a four-qubit teleportation chain (step 1). The Bell states in the each chain are coupled to measure plaquette stabilizers (step 3). This directly produces the 6-ring FBEC implementation.}
    \label{fig:planar_foliation}
\end{figure}

Foliation is a well-established method for creating cluster states from stabilizer codes~\citeapp{raussendorf_2003, raussendorf_2006, brown_2020, bolt_2016}. Following foliation, each qubit in the resulting lattice may be ``split'' into two, as described in~\citeapp{bartolucci_2023, sahay_2023a, bombin_2024}, to form the fusion-based implementation of the cluster state. Here, we review the traditional foliation approach and introduce an optimized procedure. As an example, in Fig. \ref{fig:planar_foliation}, we show how both procedures are implemented on the $XZZX$ code. 

The standard foliation procedure constructs the Raussendorf-Harrington-Goyal (RHG) cluster state~\citeapp{raussendorf_2001, raussendorf_2003, raussendorf_2006} from the CSS surface code. The RHG cluster state is defined by initializing all qubits in the $\ket{+}$ state, entangling them with CZ gates. The foliation process specifies the placement of CZ gates, which determines the cluster state’s connectivity. Following Brown and Roberts~\citeapp{brown_2020}, the RHG cluster state can be derived from the CSS surface code in two steps: (1) replacing each data qubit with a 1D teleportation chain, and (2) coupling these chains to repeatedly measure the stabilizers during teleportation. Once the cluster state is formed, the qubits are measured out in the $X$-basis, which enacts the computation.

However, Brown and Roberts' procedure disrupts the noise bias critical to the $XZZX$ code~\citeapp{claes_2023}. Furthermore, when applied to the XZZX code, their procedure results in a non-nearest-neighbor cluster state. Claes et al.~\citeapp{claes_2023} addressed these issues by proposing a bias-preserving cluster state for the $XZZX$ code. This cluster state is defined by qubits that are initialized in either $\ket{+}$ (denoted ``$X$-type'' and depicted by $\sbullet $) or $\ket{0}$ (denoted ``$X$-type'' and depicted by {\large $\circ$}). These qubits are entangled using CZ gates (between $X$-types) or CX gates (between $X$- and $Z$-types, with the $X$-type as the target). Measurements are performed in the $X$- or $Z$-basis, respectively. As shown in the top row of Fig.~\ref{fig:planar_foliation}, this similarly results in a cluster state with 1D teleportation chains for each data qubit, which are subsequently coupled to enact stabilizer measurements. Note that adjacent stabilizer measurements are staggered in a checkerboard-like pattern to avoid conflicts.

The FBEC equivalent of these cluster states is obtained by splitting each qubit into two~\citeapp{bartolucci_2023, sahay_2023a, bombin_2024}. Each $X$-type ($Z$-type) qubit is represented by a pair of qubits, fused via joint $Z \otimes Z$ ($X \otimes X$) measurements to create an effective degree of freedom. The effective qubits are then measured with a joint $X \otimes X$ ($Z \otimes Z$) measurement, enacting a measurement on the effective cluster state. Together these two measurements complete the fusion. This splitting partitions the lattice into individual six-qubit states, corresponding to the hexagonal 6-ring resource states required for FBEC~\citeapp{bartolucci_2023, sahay_2023a}.

Our approach builds directly on these concepts but avoids an intermediate cluster state. Instead, we construct an FBEC implementation from the outset, as shown in the bottom row of Fig.~\ref{fig:planar_foliation}. We begin by replacing each data qubit with a 4-qubit teleportation chain (see App. \ref{app:teleportation_chain}). Next, we introduce a pair of ancilla qubits to measure each stabilizers. One of these qubits is coupled to the top half of the data qubits involved in the stabilizer, and the other is coupled to bottom half. This results in a quasi-planar architecture for 6-ring FBEC, exactly as depicted in Fig.~\ref{fig:qec_schedule}(b). However, the cross-link connections within the teleportation chain prevent full planarity, leading us to term this approach ``quasi-planar foliation''. The next subsection resolves these issues to achieve a fully planar architecture.

\subsection{Optimizing the number of qubits}
\label{app:fbec_optimization}
\begin{figure}[h]
    \centering
    \includegraphics[width=0.9\textwidth]{{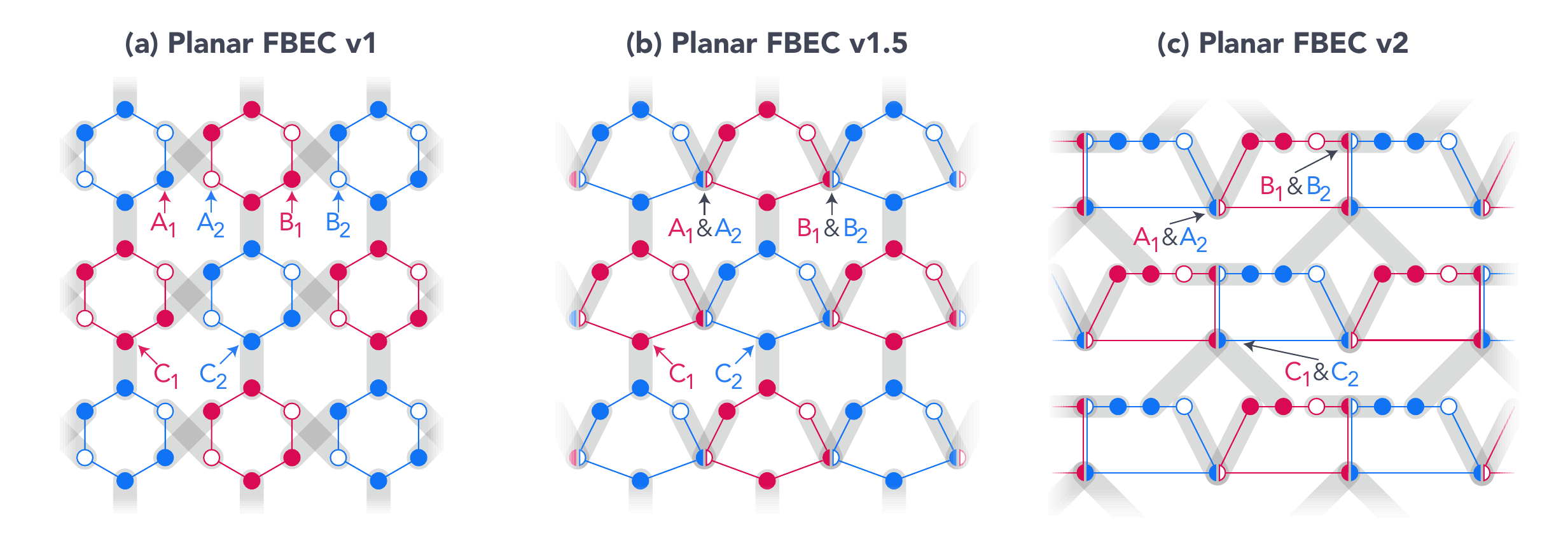}}
    \caption{Optimizing the planar FBEC layouts. (a) Quasi-planar layout derived from the quasi-planar foliation of the XZZX code. This layout is inefficient since some qubits (denoted as pairs $\{\rmA_1, \rmA_2\}$ and so on), are never active simultaneously. (b) Planar FBEC v1.5 layout, achieved by merging $\rmA_1$ with $\rmA_2$ and $\rmB_1$ with $\rmB_2$, resulting in a fully planar configuration with primarily degree-3 and some degree-4 connectivity. (c) Planar FBEC v2.0 layout, the most efficient design, obtained by further merging $\rmC_1$ with $\rmC_2$ at the cost of introducing degree-5 connections.}
    \label{fig:all_planar_fbec_versions}
\end{figure}

As explained in Sec. \ref{sec:fbqc_overview}, certain qubits sometimes remain idle during the stabilizer cycle. These qubits neither store logical information nor participate in imminent operations for logical teleportation. Instead, they these qubits have recently been measured and are awaiting reinitialization.

Fig. \ref{fig:all_planar_fbec_versions} illustrates how these idle qubits can be repurposed to reduce the total qubit count. Fig. \ref{fig:all_planar_fbec_versions}(a) shows the quasi-planar layout derived the previous subsection. Notably, qubits labeled $\rmA_1$ and $\rmA_2$ never participate simultaneously. Therefore, they can be replaced by a single qubit that assumes the role of $\rmA_1$ or $\rmA_2$ as required. Similarly, $\rmB_1$ and $\rmB_2$ can be merged into one qubit. These mergers result in the layout depicted in Fig. \ref{fig:all_planar_fbec_versions}(b), referred to as ``Planar FBEC v1.5''. This layout is fully planar with no cross-links, featuring primarily degree-3 connectivity and some degree-4 connections. Overall this v2 reduces the qubit count by 16.7\%, compared to v1.

We may further optimize the layout at the cost of introducing degree-5 connections. As shown Fig \ref{fig:all_planar_fbec_versions}(c), qubits $\rmC_1$ and $\rmC_2$ can be merged into a single qubit as well. This yields the ``Planar FBEC v2.0''. This is the most efficient configuration, with 25\% fewer qubits compared to the original. 

%% file: tex_appendix/sim_details.tex
\section{Simulation details}
\label{app:sims}

\subsection{Photon-Number Measurements (PNM)}
\label{app:sims_pnm}
We model PNMs using a single mode coupled to a three-level ancilla, given by the Hamiltonian $\opH(t) = \opH_{\chi} + \opH_d(t)$. Here, $\opH_{\chi}$ describes static dispersive coupling between the bosonic mode and the ancilla, while $\opH_d(t)$ represents a time-dependent drive on the ancilla. These are given by,
\begin{align}
    \opH_{\chi} &= - \chi_{e} \opaDaga \dyad{e} - \chi_{f} \opaDaga \dyad{f}, \\
    \opH_d(t) &= \Omega_{e}(t) \big( \dyad{g}{e} + \dyad{e}{g} \big) + \Omega_{f}(t) \big( \dyad{g}{f} + \dyad{f}{g} \big),
    \label{eqn:app_dispersive}
\end{align}
Here, $\ket{g}$, $\ket{e}$, and $\ket{f}$ denote the ground, first-excited, and second-excited states of the ancilla, respectively, and $\opa$ is the bosonic annihilation operator. As defined in the main text, $\chi_{e}$ ($\chi_{f}$) is the strength of the dispersive interaction to $\ket{e}$ ($\ket{f}$) state, and $\Omega_{e}(t)$ [$\Omega_{f}(t)$] is a time-dependant drive that triggers the $\ket{g} \leftrightarrow \ket{e}$ ($\ket{g} \leftrightarrow \ket{f}$) transition. Note that we have written this Hamiltonian in a frame rotating at the qubit's transition frequency. 

To implement a PNM, the ancilla is initialized in $\ket{g}$, photon-number information (e.g., $\op{\Pi}$ or $\op{V}$) is encoded into the ancilla, and the ancilla state is measured. As explained in the main text, we encode any binary information about the bosonic mode's photon-number into the $\ket{e}$ state by selectively exciting the ancilla for any photon number within a set $\mcN_e$. This is done with a pulse, $\Omega_{e}(t) = A(t) \left[ \sum_{n \in \mcN_e} \cos(n \chi_e t) \right]$. This pulse predominantly contains frequency components at the shifted transition frequencies $n \chi_e$. However, the envelope $A(t)$ broadens the spectral components around the desired frequencies. To minimize the infidelity caused by this spectral leakage, we chose $A(t)$ to be a Kaiser window \citeapp{kaiser_1966}. The Kaiser window is known to maximize the spectral content around a frequency while simulatenously minimizing the overall pulse length \citeapp{kaiser_1966,scipy}. This window is computed using the scientific computing package \texttt{scipy} \citeapp{scipy}. Similarly, $\Omega_{f}(t)$ can be tailored to selectively trigger the $\ket{g} \leftrightarrow \ket{f}$ conditioned on the bosonic mode having a photon number within a disjoint set $\mcN_f$. 

We use a total pulse duration of $2 , \mu$s with a Kaiser shape parameter $\beta = 4$ \citeapp{kaiser_1966,scipy}. The coupling strengths are set to $\chi_e/2\pi = -2$ MHz and $\chi_f/2\pi = -1$ MHz, demonstrating that our scheme does not require $\chi$-matching. The complete time-dependent Hamiltonian is simulated using \texttt{QuTiP} \citeapp{qutip,qutip_2}. Table \ref{tab:app_pnm_fidelity} presents the measurement fidelities of PNMs relevant to our scheme for select states. Although these measurements exhibit non-negligible error, even in the absence of decoherence in the ancilla or bosonic mode, this error is rendered insignificant as it is quadratically suppressed by our 4C-FBEC protocol.

\begin{table*}[tbh]
    \centering
    \begin{tblr}{
        colspec = {X[c,m, wd=3.5cm]X[c,m]X[c,m]X[c,m]X[c,m]X[c,m]X[c,m]},
        stretch = 0,
        rowsep = 3pt,
        vlines = {black, .6pt},
      }
        \hline
        \SetCell[r=2]{c,m}{Measurement} & \SetCell[r=2]{c,m}{Measured ancilla state} & \SetCell[c=5]{c,m}{Measurement probabilities for each input state} \\ 
        \hline[dashed]
        & & {$\cat{0}$} & {$\cat{1}$} & {$\cat{2}$} & {$\cat{3}$} & {Vacuum, $\ket{0}$} \\
        \hline \hline
        $\Pi = \text{PNM}^{ge}(\{ 2n+1 \})$ & $\rmP(g)$ &  $0.998$ & $2.15 \times 10^{-3}$ & $0.99998$ & $2.11 \times 10^{-3} $ & $2.43 \times 10^{-4}$\\
        and & $\rmP(e)$ & $2.20 \times 10^{-7}$ & $0.998$ & $1.87 \times 10^{-7}$ & $0.998$ & $1.33 \times 10^{-5}$\\
        $V = \text{PNM}^{gf}(\{ 0 \})$ & $\rmP(f)$ & $1.56 \times 10^{-3}$ & $4.09 \times 10^{-5}$ & $1.12 \times 10^{-5}$ & $2.40 \times 10^{-5}$ & $0.9997$\\
        \hline
        $\sqrt{\Pi} = \text{PNM}^{ge}(\{ 4n+2 \})$ & $\rmP(g)$ & $0.998$ & $0.9997$ & $4.73 \times 10^{-4}$ & $0.9998$ & $1.49 \times 10^{-5}$\\
        and & $\rmP(e)$ & $4.11 \times 10^{-9}$ & $1.50 \times 10^{-4}$ & $0.9995$ & $1.53 \times 10^{-4}$ & $1.66 \times 10^{-7}$\\
        $V = \text{PNM}^{gf}(\{ 0 \})$ & $\rmP(f)$ &  $1.56 \times 10^{-3}$ & $7.35 \times 10^{-5}$ & $2.76 \times 10^{-6}$ & $5.10 \times 10^{-6}$ & $0.99998$\\
        \hline
        $\sqrt{\Pi}' = \text{PNM}^{ge}(\{ 4n+1 \})$ & $\rmP(g)$ & $0.998$ & $9.22 \times 10^{-4}$ & $0.9998$ & $0.99996$ & $1.50 \times 10^{-4}$\\
        and & $\rmP(e)$ & $1.50 \times 10^{-4}$ & $0.9991$ & $1.53 \times 10^{-4}$ & $3.55 \times 10^{-5}$ & $1.40 \times 10^{-5}$\\
        $V = \text{PNM}^{gf}(\{ 0 \})$ & $\rmP(f)$ &  $1.56 \times 10^{-3}$ & $2.49 \times 10^{-5}$ & $1.09 \times 10^{-5}$ & $4.74 \times 10^{-6}$ & $0.9998$\\
        \hline
    \end{tblr}
    \caption{Simulated measurement probabilities for the parity measurement $\op{\Pi}$ and the 4-parity measurements $\sqrt{\Pi}$ and $\sqrt{\Pi}'$ for the 4C states and the vacuum state. In addition to encoding the corresponding parity (or 4-parity) information into the $\ket{e}$, these measurements selectively excite the ancilla to $\ket{f}$ if the bosonic mode is in the vacuum state. The ancilla qutrit is measured at the end of the pulse. All of these measurements are 2$\mu$s long.}
    \label{tab:app_pnm_fidelity}
\end{table*}

As highlighted in the main text, we use PNMs for both resource state generation and fusions. When simulating PNMs in those contexts, we include decoherence in the bosonic mode and the ancillae. This is done by simulating the Lindblad master equation in \texttt{QuTiP} \citeapp{qutip,qutip_2}:
\begin{align}
    \dv{\rho}{t} = -i \comm{\opH}{\rho} + \frac{1}{T_1^{\text{loss}}} \superOp{\opa} \rho + \frac{1}{T_1^{ge}} \superOp{\op{t}} \rho + \frac{1}{T_{\phi}^{ee}} \superOp{\op{t}^\dagger \op{t}} \rho,
    \label{eqn:app_lme}
\end{align}
where $\superOp{\op{L}}\rho = \op{L} \rho \op{L}^\dagger - \half \op{L}^\dagger \op{L} \rho - \half \rho \op{L}^\dagger \op{L}$ is the usual Lindblad dissipator and $\op{t} = \ket{g}_1 \bra{e}_1 + \sqrt{2} \ket{e}_1 \bra{f}_1$ is the annihilation operator for the three-level ancilla. $T_1^{\text{loss}}$ is the bosonic mode's lifetime, $T_1^{ge} = 2 T_1^{ef}$ is the ancilla's characteristic lifetime, and $T_{\phi}^{ee} = 4 T_{\phi}^{ff}$ is the ancilla's characteristic dephasing time. 

\subsection{Beamsplitter}
\label{app:sims_bs}
The beamsplitter is described by the following Hamiltonian, 
\begin{align}
    \opH_{\text{BS}} &= \half \left[ g(t) \opaDag \opb + g^*(t) \opa \opbDag \right] + \Delta \opaDaga
    \label{eqn:app_beamsplitter}
\end{align}

The $\text{BS}(\theta)$ gate requires $\Delta = 0$, $g = -i |g|$, and $T = \theta/|g|$. Unlike the $cZZ_\fC$, imposes no strict constraints on the amplitude $|g|$. In experiment, high-fidelity beamsplitters have been implemented with high-fidelity and operating much faster than any coherence time in the system~\citeapp{chapman_2023,yao_lu_2023}. Therefore, we may assume that the $\text{BS}(\theta)$ are implemented instantaneously. 

The complete Hamiltonian also includes the static dispersive interaction between each mode and its corresponding ancilla, as described in Eq.~\ref{eqn:app_dispersive}. We set $\chi_{e}/2\pi = -2$ MHz and $\chi_{f}/2\pi = -1$ MHz. The dispersive coupling to the second-mode’s ancilla can be ignored, as this ancilla remains in its ground state during both $\text{BS}(\theta)$ and $ZZ_\fC(\frac{\pi}{2})$ operations. However, the dispersive interaction with the first mode’s ancilla is critical for implementing the $ZZ_\fC$ gate.

To realize the ancilla-controlled $ZZ_\fC$, we apply the Hamiltonian $\opH = \opH_{\chi, 1} + \opH_{\text{BS}}$ for a time $T= \pi/\chi_{f,1}$, with $g=\sqrt{15}/2 \, \chi_{f,1}$, $\Delta = -\chi_{f,1}/2$~\citeapp{tsunoda_2023}.  As described in Sec. \ref{sec:core_ops_two}, this operation is used to implement the $ZZ_\fC(\theta)$ gate. 

We simulate a similar Lindblad master equation to Eq. \ref{eqn:app_lme}, but now include loss in both modes. 
\begin{align}
    \dv{\rho}{t} = -i \comm{\opH}{\rho} + \frac{1}{T_1^{\text{loss}}} \superOp{\opa} \rho + \frac{1}{T_1^{\text{loss}}} \superOp{\opb} \rho + \frac{1}{T_1^{ge}} \superOp{\op{t}} \rho + \frac{1}{T_{\phi}^{ee}} \superOp{\op{t}^\dagger \op{t}} \rho,
    \label{eqn:app_lme2}
\end{align}
Note that we have implicitly neglected the no-jump evolution by imposing equal decay rates for both bosonic modes. When the decay rates are not equal, the entangled states gradually become biased toward the longer-lived cavity \citeapp{tsunoda_2023}. To mitigate this effect, periodic SWAP operations can be performed between the two bosonic modes, effectively echoing out the no-jump evolution \citeapp{weiss_2024}.


\subsection{\texorpdfstring{$ZZ_\fC(\frac{\pi}{2})$}{ZZ(pi/2)}  gate}
\label{app:sim_zz}
From Eq. \ref{eqn:ZZ_theta} (in the main text), we observe that the beamsplitter and transmon pulses do not overlap, allowing us to assume a piecewise evolution. The entire $ZZ_\fC(\frac{\pi}{2})$ gate is simulated as a composition of quantum channels, $\rho_f = \left[ \mcC_{T_3} \circ \mcC_{\text{BS}_2} \circ \mcC_{T_2} \circ \mcC_{\text{BS}_1} \circ \mcC_{T_1} \right] \parens{\rho_i}$. Here, $\mcC_{T_i}$ and $\mcC_{\text{BS}_j}$ denote the ancilla rotations and beamsplitter-mediated $cZZ_\fC$ gates respectively. Since these rotations are unselective and significantly faster than the beamsplitter, they negligibly contribute to the overall error budget and can be approximated as ideal unitary evolutions. Only the ancilla-controlled $cZZ_\fC$ gates are simulated. During these gate, the beamsplitter amplitude modeled as a rectangular pulse with a cosine ramp, an amplitude of $g/2\pi = \sqrt{15} (\chi_{f,1}/2\pi)/2  \approx  1.93$ MHz, and a duration of $T = \pi/\chi_{f,1} = 0.5 \mu$s. The beamsplitter detuning is fixed to $\Delta/2\pi = - (\chi_{f,1}/2 \pi)/2 = 0.5$ MHz.

We aim to compute the postselected fidelity of the $ZZ_\fC(\frac{\pi}{2})$ gate. To achieve this, we define the measurement operators $\op{M}_\checkmark$ and $\op{M}_\times$, which correspond to detecting no error and detecting an error, respectively \citeapp{weiss_2024}. If the system state before measurement is $\rho$, the post-measurement state, conditioned on detecting no errors, is given by $\rho' = \op{M}_\checkmark \rho \op{M}_\checkmark^\dagger/P_\checkmark(\rho)$, where $P_\checkmark(\rho) = \tr[ \rho \op{M}_\checkmark^\dagger \op{M}_\checkmark]$ the probability of not detecting an error. 

Recall that measuring the transmon in the state $\ket{g}$ indicates no detected errors, while measuring $\ket{e}$ or $\ket{f}$ signals a bit-flip or a phase-flip error, respectively. However, the gate may still fail due to other undetected mechanisms, such as two ancilla errors. Additionally, the 4C code can detect a single-photon loss in the bosonic modes via parity measurements on each mode, at the end of the sequence. Thus, the measurement operator corresponding to no errors is:
\begin{align}
    \op{M}_\checkmark = \parens{\frac{\opI + e^{-i \pi \opaDaga}}{2}} \otimes \parens{\frac{\opI + e^{-i \pi \opbDagb}}{2}} \otimes \ket{g}_1 \bra{g}_1,
\end{align}
Here we have written the ideal projector onto the even photon number space, $\thalf \parens{\opI + e^{-i \pi \opaDaga}}$. However, in our simulations, the parity measurements are simulated explicitly (as explained in App. \ref{app:sims_pnm}). Therefore, our simulation includes coherent and incoherent errors during the parity operation into $\op{M}_\checkmark$. 

While normalizing by $P_\checkmark(\rho)$ ensures that the density matrix $\rho{\prime}$ has unit trace, this nonlinear mapping complicates the use of standard fidelity metrics \citeapp{weiss_2024}, which are designed for linear quantum channels \citeapp{nielsen_2002, dankert_2005, pedersen_2007, weiss_2024}. To address this, we temporarily ignore the normalization and treat the process as a linear map producing a subnormalized state. We can then use Nielsen's formula for the entanglement fidelity for a gate $\op{U}$ \citeapp{nielsen_2002, weiss_2024}. 
\begin{align}
    \tilde{F}_e(\op{U}) = \frac{\sum_{j} \Tr( \op{U} \op{V}_j^\dagger \op{U}^\dagger \left[ \op{M}_\checkmark \veps_U \{ \op{V}_j \} \op{M}_\checkmark^\dagger \right] )}{d^3}
\end{align}
where $\veps_U$ is the noisy quantum channel (before the error-detecting measurements) derived from simulating the Lindblad master equation (Eq. \ref{eqn:app_lme2}) in QuTiP \citeapp{qutip,qutip_2}. $d$ is the dimension of the Hilbert space (here, $d = 4$) and the $\{ \op{V}_j \}_{j=1}^{16}$ are an orthonormal basis of operators on the space of two 4C states (i.e. $\op{V}_j = \sqrt{d} \dyad{m,n}{p,q}$, where $m,n,p,q \in \{ 0_\fC, 1_\fC \big\}$). Since these $\op{V}_j$'s are not pure density matrices, we can decompose them as $\dyad{m,n}{p,q} = \dyad{+} + i \dyad{+i} - \half (1 + i) \Big[ \dyad{m,n} + \dyad{p,q} \Big]$, where $\ket{+} = \tfrac{1}{\sqrt{2}}(\ket{m,n} + \ket{p,q})$ and $\ket{+i} = \tfrac{1}{\sqrt{2}}(\ket{m,n} + i \ket{p,q})$ \citeapp{lidar_2008, weiss_2024}. We can simulate these pure density matrices under the noisy quantum channel, and compute $\veps_U\{ \op{V}_j \}$ by linearity \citeapp{nielsen_2002, dankert_2005, pedersen_2007, weiss_2024}.

Following Ref. \citeapp{weiss_2024}, we divide $\tilde{F}_e(\op{U})$ by the average success probability, $P_U = \sum_j \tr[\veps_U\{ \op{V}_j \}]/(2d^2 - d)$ to obtain the average postselected entanglement fidelity, $F_e(U) = \tilde{F}_e(\op{U})/P_U$. The average postselected gate fidelity $F(U)$ is then given by the standard formula
\begin{align}
    F(U) = \frac{d F_e(U) + 1}{d+1}.
\end{align}

\subsection{Fusion measurements}
As outlined in Appendix \ref{app:bell_msmt}, fusion measurements are implemented through an adaptive sequence of local PNMs. Simulating this sequence concurrently for both modes is computationally expensive but unnecessary since the sequence is identical for each mode. Instead, we simulate the adaptive sequence for a representative set of input states in one mode, specifically the states emerging from the beamsplitter: $\Psi \equiv \left\{ \ket{0}, \cat{j}, \bigcat{k} \big| j,k = 0,1,2,3 \right\}$. 

As shown in Table \ref{tbl:even_cat_bell}, there is a set of acceptable product states $\ket{\psi_1} \ket{\psi_2}$ for each 4C Bell state, where $\ket{\psi_1}, \ket{\psi_2} \in \Psi$. For instance, the $\ket{\lambda_{xx} = +1, \lambda_{zz} = +1}$ Bell state corresponds to $\Phi_{++} = \left\{ \cat{0} \cat{0}, \cat{2} \cat{2}, \ket{0} \bigcat{0}, \bigcat{0} \ket{0}, \ket{0} \bigcat{3}, \bigcat{3} \ket{0} \right\}$. If the adaptive measurement identifies any of these states, we declare $\ket{\lambda_{xx} = +1, \lambda_{zz} = +1}$. Analogous sets $\Phi_{\pm\pm}$ are constructed for the other Bell states.

However, PNM infidelity and decoherence may lead to measurement errors. To estimate Bell measurement infidelity, we compute the probability $\rmP(R_1 \land R_2 | I_1 \land I_2)$ of reporting $\ket{R_1}\ket{R_2}$ for a given input $\ket{I_1}\ket{I_2}$, where $\ket{I_i}, \ket{R_i} \in \Psi$. We can use the standard Bayes identity and the independence of the measurements on each mode to decompose this probability. 
\begin{align}
    \rmP(R_1 \land R_2 | I_1 \land I_2) = \frac{\rmP \big[(R_1 \land R_2) \land (I_1 \land I_2) \big]}{\rmP(I_1 \land I_2)} = \frac{\rmP(R_1 \land I_1)}{\rmP(I_1)} \cdot \frac{\rmP(R_2 \land I_2)}{\rmP(I_2)} = \rmP(R_1 | I_1) \cdot \rmP(R_2 | I_2)
\end{align}

This simiplification lets us construct $\rmP(\phi_{\text{out}} | \phi_{\text{in}})$, the probability of measuring Bell state $\ket{\phi_{\text{out}}}$ given input Bell state $\ket{\phi_{\text{in}}}$:
\begin{gather}
    \rmP(\phi_{\text{out}} | \phi_{\text{in}}) = \sum_{ \ket{R_1} \ket{R_2} \in \Phi_{\text{out}}} \ \sum_{ \ket{I_1}, \ket{I_2}} |c_{ {I_1}, {I_2} } |^2 \ \rmP(R_1 | I_1) \ \rmP(R_2 | I_2), \\[10pt]
    \text{where } \qquad \ket{\phi_{\text{in}}} = \sum_{ \ket{I_1}, \ket{I_2}} c_{ {I_1}, {I_2} } \ket{I_1} \ket{I_2},
\end{gather}
and $\Phi_{\text{out}}$ is the set of acceptable states for $\ket{\phi_{\text{out}}}$, and $\ket{\phi_{\text{in/out}}}$ are the Bell states after the beamsplitter. Averaging these probabilities over the four Bell state inputs yields the measurement infidelities $p_{xx}$, $p_{yy}$, and $p_{zz}$, shown in Fig.~\ref{fig:all_sims}.

Fig.~\ref{fig:all_sims} also plots the probability of inconclusive measurements. These occur if either mode’s sequence fails to provide a definitive result:
\begin{gather}
    \rmP(\text{inconclusive} | \phi_{\text{in}}) = \sum_{ \ket{I_1}, \ket{I_2}} |c_{ {I_1}, {I_2} } |^2 \ \big[ \rmP( \text{inconclusive} | I_1) + \rmP(\text{inconclusive} | I_2) \big].
\end{gather}

\begin{figure}[h]
    \centering
    \includegraphics[width=\textwidth]{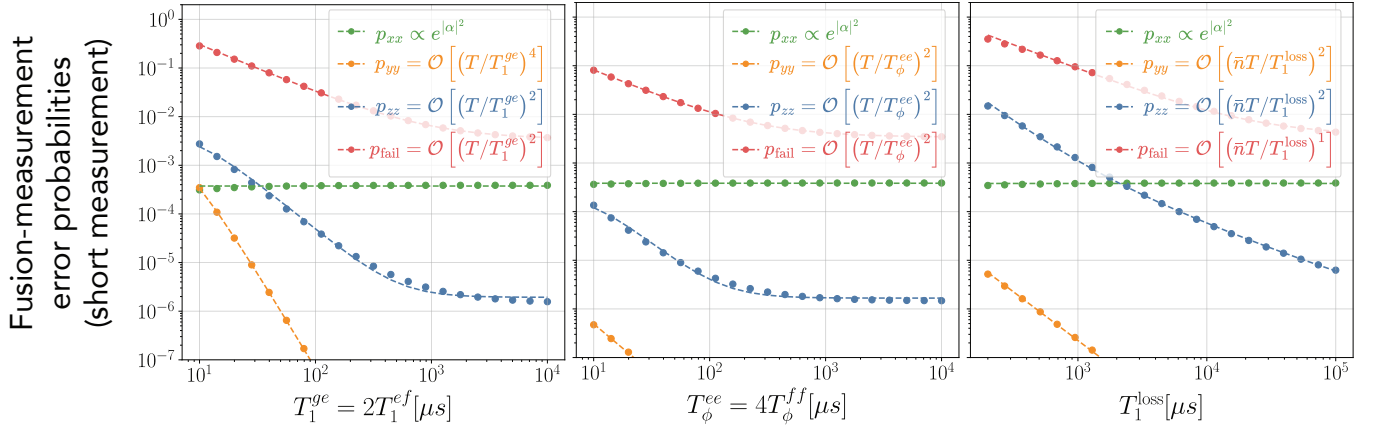}
    \caption{Numerical simulation of the shorter fusion measurement sequence.}
    \label{fig:short_msmt_sims}
\end{figure}

Fig.~\ref{fig:short_msmt_sims} shows the measurement error probabilities for the shorter measurement sequence discussed in App. \ref{app:bell_msmt}.

\subsection{Fits to the numerical simulations}
The numerical results can be fit to low order polynomials, as shown by the dashed lines in Fig.\ref{fig:all_sims} and \ref{fig:short_msmt_sims}. These resulting fits are shown in Table \ref{tab:fit_params}. 
\begin{table*}[tbh]
    \centering
    \begin{tblr}{
        colspec = {X[c,m, wd=1.5cm]X[c,m]X[c,m]X[c,m]},
        stretch = 0,
        rowsep = 5pt,
        vlines = {black, .6pt},
      }
        \hline 
        Operation & Varying ancilla $T_1^{ge} = 2 T_1^{ef}$ & Varying ancilla $T_\phi^{ee} = 2 T_1^{ef}$ & Varying single-photon loss rate $T_1^{\text{loss}}$\\
        \hline \hline
        \SetCell[r=2]{c,m}{Six-ring preparation} & $p_{\text{fail}} = 0.496 x$ & $p_{\text{fail}} = 0.494 x$ & $p_{\text{fail}} = 3.98 x$ \\
        & $\veps_{\text{pass}} = 0.026 x^2 + (6.67 \times 10^{-7}) $ & $\veps_{\text{pass}} = 0.050 x^2 + (6.64 \times 10^{-7}) $ & $\veps_{\text{pass}} = 1.413 x^2 + (6.32 \times 10^{-7})$\\
        \hline
        \SetCell[r=4]{c,m}{Fusion (complete sequence)} & $p_{xx} = 3.90 \times 10^{-4}$ & $p_{xx} = 3.95 \times 10^{-4}$ & $p_{xx} = 3.88 \times 10^{-4}$ \\
        & $p_{yy} = 3.95 x^4 + 4.00 x^5$ & $p_{yy} = (4.08 \times 10^{-5}) x^2$ & $p_{yy} = 0.01 x^2$ \\
        & $p_{zz} = 0.38 x^2 - 0.86 x^3 + (5.68 \times 10^{-6})$ & $p_{zz} = 0.02 x^2 - 0.05 x^3 + (4.90 \times 10^{-6})$ & $p_{zz} = 16.29 x^2 + (1.83 \times 10^{-6})$ \\
        & $p_{\text{fail}} = (1.36 \times 10^{-3})x^2 + 0.02 x^3 + (4.31 \times 10^{-7})$ & $p_{\text{fail}} = (1.08 \times 10^{-3})x^2 + (3.94 \times 10^{-7})$ & $p_{\text{fail}} = (2.91 \times 10^{-4})x + (3.26 \times 10^{-7})$ \\
        \hline
        \SetCell[r=4]{c,m}{Fusion (shorter sequence)} & $p_{xx} = 3.80 \times 10^{-4}$ & $p_{xx} = 3.88 \times 10^{-4}$ & $p_{xx} = 3.82 \times 10^{-4}$ \\
        & $p_{yy} = 0.44 x^4 + 4.00 x^5$ & $p_{yy} = (1.23 \times 10^{-5}) x^2$ & $p_{yy} = (9.09 \times 10^{-4})x^2$ \\
        & $p_{zz} = 0.12x^2 - 0.32 x^3 + (1.93 \times 10^{-6})$ & $p_{zz} = (6.30 \times 10^{-3})x^2 - 0.02 x^3 + (1.67 \times 10^{-6})$ & $p_{zz} = 2.40 x^2 + (8.09 \times 10^{-7})$ \\
        & $p_{\text{fail}} = 1.5x + (3.43 \times 10^{-3})$ & $p_{\text{fail}} = 0.39 x + (3.39 \times 10^{-3})$ & $p_{\text{fail}} = 5.37 x + (3.66 \times 10^{-3})$ \\
        \hline[dashed]
        where & $x = (T/T_1^{ge})$ & $x = (T/T_\phi^{ee})$ & $x = (\bar{n} T/T_1^{\text{loss}})$ \\
        \hline
    \end{tblr}
    \caption{Fitted expressions for how the error probabilities vary with each of the decoherence times.}
    \label{tab:fit_params}
\end{table*}